\documentclass[amssymb, notitlepage,amsmath, pra, 10pt, aps]{revtex4-1}

\usepackage{amsmath,amssymb,amsfonts} 
\usepackage{graphicx}
\usepackage{bm}
\usepackage{psfrag}
\usepackage[caption=false]{subfig}

\allowdisplaybreaks[1]

\begin{document}

\title{Homogeneous shear turbulence as a second-order cone program}

\author{Luoyi Tao}
\email[]{luoyitao@iitm.ac.in}
\affiliation{Department of Aerospace Engineering\\
             Indian Institute of Technology Madras\\
             Chennai 600 036, India}

\date{\today}

\def\s{\!}
\def\ss{\!\!}
\def\sss{\!\!\!}
\def\l{\left}
\def\r{\right}
\def\bo{\mathbf{0}}
\def\bx{{\bf x}}
\def\by{{\bf y}}
\def\bz{{\bf z}}
\def\ba{{\bf a}}
\def\bw{{\bf w}}
\def\tbw{\tilde{\bf w}}
\def\bV{{\bf V}}
\def\Vi{V_i}
\def\Vj{V_j}
\def\Vk{V_k}
\def\Vij{\Vi,_j}
\def\Vji{\Vj,_i}
\def\Vkl{\Vk,_l}
\def\bv{{\bf v}}
\def\vi{v_i}
\def\vj{v_j}
\def\tw{\tilde{w}}
\def\wi{w_i}
\def\twi{\tilde{w}_i}
\def\wj{w_j}
\def\twj{\tilde{w}_j}
\def\wk{w_k}
\def\twk{\tilde{w}_k}
\def\wl{w_l}
\def\twl{\tilde{w}_l}
\def\Bm{{\bf m}}
\def\bmp{\Bm^{\prime}}
\def\bmpp{\Bm^{\prime\prime}}
\def\mi{m_i}
\def\bbm{(\Bm)}
\def\bbmp{\s\l(\bmp\r)}
\def\bM{{\bf M}}
\def\bbM{\s\l(\bM;\bx\r)}
\def\bn{{\bf n}}
\def\bnp{\bn^{\prime}}
\def\bnpp{\bn^{\prime\prime}}
\def\bbn{(\bn)}
\def\bbnp{\s\l(\bnp\r)}
\def\bk{{\bf k}}
\def\bkp{\bk^{\prime}}
\def\bbk{(\bk)}
\def\bbkp{\s\l(\bkp\r)}
\def\bK{{\bf K}}
\def\bbK{\s\l(\bK;\bx\r)}
\def\bl{{\bf l}}
\def\blp{\bl^{\prime}}
\def\bbl{(\bl)}
\def\bblp{\s\l(\blp\r)}
\def\bi{{\bf i}}
\def\bip{\bi^{\prime}}
\def\bbi{\s\l(\bi\r)}
\def\bbip{\s\l(\bip\r)}
\def\bj{{\bf j}}
\def\bjp{\bj^{\prime}}
\def\bbj{(\bj)}
\def\bbjp{\s\l(\bjp\r)}
\def\bL{{\bf L}}
\def\bp{{\bf p}}
\def\bpp{\bp^{\prime}}
\def\bbp{(\bp)}
\def\bbpp{(\bpp)}
\def\bq{{\bf q}}
\def\bqp{\bq^{\prime}}
\def\p{p}
\def\bbq{(\bq)}
\def\bbqp{\s\l(\bqp\r)}
\def\P{P}
\def\q{q}
\def\tq{\tilde{q}}
\def\barwiwj{\overline{\wi \wj}}
\def\barwiwjwk{\overline{\wi \wj \wk}}
\def\barwixwix{\overline{\wi(\bx)\, \wi(\bx)}}
\def\barwixwjx{\overline{\wi(\bx)\, \wj(\bx)}}
\def\barwixwjy{\overline{\wi(\bx)\, \wj(\by)}}
\def\barwjywkz{\overline{\wj(\by)\, \wk(\bz)}}
\def\barwjywjy{\overline{\wj(\by)\, \wj(\by)}}
\def\barwkzwkz{\overline{\wk(\bz)\, \wk(\bz)}}
\def\barwixwjywkz{\overline{\wi(\bx)\,\wj(\by)\,\wk(\bz)}}
\def\barwixwjxwkx{\overline{\wi(\bx)\,\wj(\bx)\,\wk(\bx)}}
\def\barwipxwjpywkpz{\overline{\wi,_{i'}\ss(\bx)\,\wj,_{j'}\ss(\by)\,\wk,_{k'}\ss(\bz)}}
\def\barwixwjywkzwla{\overline{\wi(\bx,t)\,\wj(\by,t)\,\wk(\bz,t)\,\wl(\ba,t)}}
\def\barbwbw{\overline{\bw\bw}}
\def\bartwimtwjn{\overline{\twi\bbm\twj\bbn}}
\def\bartwimtwln{\overline{\twi\bbm\twl\bbn}}
\def\bartwiktwjl{\overline{\twi\bbk\,\twj\bbl}}
\def\bartwititwjtj{\overline{\twi(t;\bi)\twj(t;\bj)}}
\def\bartwitktwjtl{\overline{\twi(t;\bk)\twj(t;\bl)}}
\def\bartwititwjtjtwktk{\overline{\twi(t;\bi)\,\twj(t;\bj)\,\twk(t;\bk)}}
\def\B{{\cal B}}
\def\DD{{\cal D}}
\def\DDi{\DD\s\l(i\r)}
\def\DDj{\DD\s\l(j\r)}
\def\DDk{\DD\s\l(k\r)}
\def\n{{\rm N}}
\def\nO{{\cal O}\s\l(\n\r)}
\def\b{b}
\def\Aj{A_j}
\def\Al{A_l}
\def\Ak{A_k}
\def\Bi{B_i}
\def\Bj{B_j}
\def\Bk{B_k}
\def\Bl{B_l}
\def\Cji{C_{ji}}
\def\Cli{C_{li}}
\def\Clk{C_{lk}}
\def\Cjk{C_{jk}}
\def\Clm{C_{lm}}
\def\Cnm{C_{nm}}
\def\D{D}
\def\E{E}
\def\EE{{\cal E}}
\def\H{H}
\def\Hij{\H_{ij}}
\def\Wi{W_i}
\def\Wk{W_k}
\def\Wl{W_l}
\def\W{W}
\def\Wij{W_{ij}}
\def\Wil{W_{il}}
\def\Wjl{W_{jl}}
\def\Wkl{W_{kl}}
\def\WiHOMS{\overline{W}_i}
\def\WkHOMS{\overline{W}_k}
\def\WlHOMS{\overline{W}_l}
\def\W{W}
\def\WHOMS{\overline{\W}}
\def\Wij{W_{ij}}
\def\Wil{W_{il}}
\def\Wjl{W_{jl}}
\def\Wkl{W_{kl}}
\def\WijHOMS{\overline{W}_{ij}}
\def\WilHOMS{\overline{W}_{il}}
\def\WjlHOMS{\overline{W}_{jl}}
\def\WklHOMS{\overline{W}_{kl}}
\def\soc{\beta}
\def\socR{\soc^{(R)}}
\def\socI{\soc^{(I)}}
\def\socDim{\soc^{(R)}}
\def\socDimO{\soc^{(\infty)}}
\def\socS{\Pi}
\def\toc{\gamma}
\def\tocR{\toc^{(R)}}
\def\tocI{\toc^{(I)}}
\def\tocDim{\toc}
\def\tocDimR{\toc^{(R)}}
\def\tocDimI{\toc^{(I)}}
\def\tocDimIO{\toc^{(\infty)}}
\def\tocDimIOpm{\toc^{(Ia)\pm}}
\def\tocDimIOmp{\toc^{(Ia)\mp}}
\def\tocDimIOp{\toc^{(Ia)+}}
\def\tocDimIOm{\toc^{(Ia)-}}
\def\foc{\delta}
\def\focDim{\hat{\foc}}
\def\foocG{\delta^G}
\def\siocG{\zeta^G}
\def\eiocG{\theta^G}
\def\bN{{\bf N}}
\def\bH{{\bf H}}
\def\bR{\mathbb{R}}
\def\bX{{\bf X}}
\def\barbwbwbw{\overline{\bw\bw\bw}}
\def\G{{\cal G}}
\def\xi{x_i}
\def\xj{x_j}
\def\xk{x_k}
\def\xl{x_l}
\def\balpha{{{\bf \alpha}}}
\def\mbi{b_i}
\def\mbj{b_j}
\def\mbk{b_k}

\def\inform{I}
\def\pdf{f}
\def\pdfG{f_G}
\def\pdfD{f_D}
\def\pdfL{f^{(L)}}
\def\pdfH{f^{(H)}}
\def\varvec{\hat{\hat{\bw}}}
\def\Reprevarvec{\hat{\hat{\bw}}}
\def\ReprevarvecL{\hat{\hat{\bw}}^{(L)}}
\def\ReprevarvecH{\hat{\hat{\bw}}^{(H)}}
\def\bkH{{\bf k}^{(H)}}
\def\LagMultiplier{\lambda}
\def\bK{{\bf K}}
\def\hhw{\hat{\hat{w}}}
\def\Det{\inform_D}
\def\K{\inform_T}
\def\vort{\omega}
\def\BETA{\pmb{\beta}}
\def\ip{i^{\prime}}
\def\jp{j^{\prime}}
\def\kp{k^{\prime}}
\def\lp{l^{\prime}}
\def\twip{\tilde{w}_{\ip}}
\def\twjp{\tilde{w}_{\jp}}
\def\twkp{\tilde{w}_{\kp}}
\def\twlp{\tilde{w}_{\lp}}
\def\imaginary{\imath}
\def\Real{\text{RE}}
\def\Imag{\text{IM}}
\def\VGrad{V}
\def\twl{\tilde{w}_l}
\def\twm{\tilde{w}_m}
\def\twn{\tilde{w}_n}
\def\twmp{\tilde{w}_{m'}}
\def\twnp{\tilde{w}_{n'}}
\def\twmpp{\tilde{w}_{m''}}
\def\twnpp{\tilde{w}_{n''}}
\def\Bm{{\bf m}}
\def\bmp{\Bm^{\prime}}
\def\bmpp{\Bm^{\prime\prime}}
\def\mi{m_i}
\def\bbm{(\Bm)}
\def\bbmp{\s\l(\bmp\r)}
\def\bM{{\bf M}}
\def\bbM{\s\l(\bM;\bx\r)}
\def\bn{{\bf n}}
\def\bnp{\bn^{\prime}}
\def\bnpp{\bn^{\prime\prime}}
\def\bbn{(\bn)}
\def\bbnp{\s\l(\bnp\r)}
\def\bk{{\bf k}}
\def\bkp{\bk^{\prime}}
\def\bbk{(\bk)}
\def\bbkp{\s\l(\bkp\r)}
\def\bK{{\bf K}}
\def\bbK{\s\l(\bK;\bx\r)}
\def\bl{{\bf l}}
\def\blp{\bl^{\prime}}
\def\bbl{(\bl)}
\def\bblp{\s\l(\blp\r)}
\def\bi{{\bf i}}
\def\bip{\bi^{\prime}}
\def\bbi{\s\l(\bi\r)}
\def\bbip{\s\l(\bip\r)}
\def\bj{{\bf j}}
\def\bjp{\bj^{\prime}}
\def\bbj{(\bj)}
\def\bbjp{\s\l(\bjp\r)}
\def\bL{{\bf L}}
\def\bp{{\bf p}}
\def\bpp{\bp^{\prime}}
\def\bbp{(\bp)}
\def\bbpp{(\bpp)}
\def\bq{{\bf q}}
\def\bqp{\bq^{\prime}}
\def\p{p}
\def\bbq{(\bq)}
\def\bbqp{\s\l(\bqp\r)}
\def\P{P}
\def\q{q}
\def\tq{\tilde{q}}
\def\barqq{\overline{\q \q}}
\def\barwiwj{\overline{\wi \wj}}
\def\barwiwjwk{\overline{\wi \wj \wk}}
\def\barwiwjwkwl{\overline{\wi \wj \wk \wl}}
\def\barwixwjy{\overline{\wi(\bx)\, \wj(\by)}}
\def\barwkxwkx{\overline{\wk(\bx)\, \wk(\bx)}}
\def\barwixwjywkz{\overline{\wi(\bx)\,\wj(\by)\,\wk(\bz)}}
\def\barwixwjywkzwla{\overline{\wi(\bx)\,\wj(\by)\,\wk(\bz)\,\wl(\ba)}}
\def\barbwbw{\overline{\bw\bw}}
\def\bartwimtwjn{\overline{\twi\bbm\twj\bbn}}
\def\bartwimtwln{\overline{\twi\bbm\twl\bbn}}
\def\bartwiktwjl{\overline{\twi\bbk\,\twj\bbl}}
\def\bartwititwjtj{\overline{\twi(t;\bi)\twj(t;\bj)}}
\def\bartwitktwjtl{\overline{\twi(t;\bk)\twj(t;\bl)}}
\def\bartwititwjtjtwktk{\overline{\twi(t;\bi)\,\twj(t;\bj)\,\twk(t;\bk)}}
\def\bartwiitwjj{\overline{\twi(\bi)\twj(\bj)}}
\def\bartwiktwjl{\overline{\twi(\bk)\twj(\bl)}}
\def\bartwiitwjjtwkk{\overline{\twi(\bi)\,\twj(\bj)\,\twk(\bk)}}
\def\ReynoldsNo{\text{Re}}
\def\tDim{\hat{t}}
\def\WNS{{\cal W}}
\def\transformedsoc{\hat{\soc}}
\def\transformedsocDim{\hat{\soc}}
\def\transformedsocDimO{\hat{\soc}^{(\infty)}}
\def\transformedtocDim{\hat{\toc}}
\def\transformedtocDimR{\hat{\toc}^{(R)}}
\def\transformedtocDimI{\hat{\toc}^{(I)}}
\def\transformedtocDimRO{\hat{\toc}^{(Ra)}}
\def\transformedtocDimIO{\hat{\toc}^{(I\infty)}}
\def\transformedbk{\hat{\mathbf{k}}}
\def\transformedbm{\hat{\mathbf{m}}}
\def\transformedbn{\hat{\mathbf{n}}}
\def\transformedbl{\hat{\mathbf{l}}}
\def\transformedbj{\hat{\mathbf{j}}}
\def\transformedk{\hat{k}}
\def\transformedm{\hat{m}}
\def\transformedn{\hat{n}}
\def\transformedl{\hat{l}}

\def\betaT{\sqrt{\s\s\beta}}
\def\g{\gamma}
\def\gO{\g^{(\infty)}}
\def\bO{\betaT^{(\infty)}}
\def\B{B}
\def\G{G}

\def\br{\mathbf{r}}
\def\bs{\mathbf{s}}
\def\wm{w_m}
\def\wn{w_n}
\def\W{U}
\def\Wij{\W_{ij}}
\def\Wkj{\W_{kj}}
\def\Wji{\W_{ji}}
\def\Wikj{\W_{ikj}}
\def\Wkij{\W_{kij}}
\def\Wjki{\W_{jki}}
\def\Wkji{\W_{kji}}
\def\Wjlk{\W_{jlk}}
\def\Wlkj{\W_{lkj}}
\def\Q{Q}
\def\Qi{\Q_i}
\def\Qj{\Q_j}
\def\Qk{\Q_k}
\def\tW{\tilde{U}}
\def\tWij{\tW_{ij}}
\def\tWkj{\tW_{kj}}
\def\tWji{\tW_{ji}}
\def\tWikj{\tW_{ikj}}
\def\tWkij{\tW_{kij}}
\def\tWjki{\tW_{jki}}
\def\tWkji{\tW_{kji}}
\def\tWjlk{\tW_{jlk}}
\def\tWjli{\tW_{jli}}
\def\tWlkj{\tW_{lkj}}
\def\tWlki{\tW_{lki}}
\def\tQ{\tilde{Q}}
\def\tQi{\tQ_i}
\def\tQj{\tQ_j}
\def\tQk{\tQ_k}

\def\tWOneOneDim{\beta^{(R)}}
\def\tWOneOneOneDim{\gamma^{(I)}_{1}}
\def\tWOneTwoOneDim{\gamma^{(I)}_{2}}
\def\ok{\overline{k}}
\def\obk{\overline{\bk}}
\def\obr{\overline{\br}}
\def\ot{\overline{t}}

\def\socAsy{\beta^{(\infty)}}
\def\tocAsy{\gamma^{(\infty)}}
\def\focAsy{\delta^{(\infty)}}
\def\foc{\delta}
\def\fociAsy{\foc^{(\infty)}_i}
\def\focIAsy{\foc^{(\infty)}_1}
\def\focIIAsy{\foc^{(\infty)}_2}
\def\focIIIAsy{\foc^{(\infty)}_3}
\def\foci{\foc_i}
\def\focI{\foc_1}
\def\focII{\foc_2}
\def\focIII{\foc_3}
\def\QAsy{Q^{(\infty)}}
\def\tQAsy{\tQ^{(\infty)}}
\def\WAsy{\W^{(\infty)}}
\def\tWAsy{\tW^{(\infty)}}
\def\kp{k^{\prime}}
\def\lp{l^{\prime}}
\def\bkp{\bk^{\prime}}
\def\blp{\bl^{\prime}}
\def\Kinfty{K^{\infty}}
\def\psiAsy{\psi^{(\infty)}}
\def\tWtocAsy{\tW^{(I\infty)}}

\def\gW{\dot{\W}}
\def\gQ{\dot{\Q}}
\def\gsoc{\dot{\soc}}
\def\gtoc{\dot{\toc}}
\def\gfoc{\dot{\foc}}
\def\gfocI{\dot{\focI}}
\def\gfocII{\dot{\focII}}
\def\gsocAsy{\gsoc^{(\infty)}}
\def\gtocAsy{\gtoc^{(\infty)}}
\def\gfocAsy{\dot{\foc}^{(\infty)}}
\def\gfocIAsy{\dot{\foc}^{(\infty)}_1}
\def\gfocIIAsy{\dot{\foc}^{(\infty)}_2}
\def\ggfoc{\ddot\foc}

\def\dsoc{\dot{\soc}}
\def\dtoc{\dot{\toc}}
\def\dfoc{\dot{\foc}}
\def\dfocI{\dot{\foc}_1}
\def\dfocII{\dot{\foc}_2}
\def\dsocAsy{\dot{\soc}^{(\infty)}}
\def\dtocAsy{\dot{\toc}^{(\infty)}}
\def\dfocAsy{\dot{\foc}^{(\infty)}}
\def\dfocIAsy{\dot{\foc}^{(\infty)}_1}
\def\dfocIIAsy{\dot{\foc}^{(\infty)}_2}

\def\w{w}
\def\tw{\tilde{\w}}
\def\tw{\tilde{\w}}
\def\underlinei{\underline{i}}
\def\underlinej{\underline{j}}
\def\underlinek{\underline{k}}
\def\underlinel{\underline{l}}
\def\underlinem{\underline{m}}
\def\underlinen{\underline{n}}

\def\tk{\tilde{k}}
\def\tbk{\tilde{\bk}}

\def\CharacteristicFunctionTriangleSNPosi{\chi_{\TriangleSNPos_i}}
\def\CharacteristicFunctionTriangleSNPosj{\chi_{\TriangleSNPos_j}}
\def\CharacteristicFunctionTriangleSNPosk{\chi_{\TriangleSNPos_k}}
\def\CharacteristicFunctionTriangleSNPosl{\chi_{\TriangleSNPos_l}}
\def\CharacteristicFunctionTriangleSNPosm{\chi_{\TriangleSNPos_m}}

\def\CharacteristicFunctionTriangleSNNegi{\chi_{\TriangleSNNeg_i}}
\def\CharacteristicFunctionTriangleSNNegj{\chi_{\TriangleSNNeg_j}}
\def\CharacteristicFunctionTriangleSNNegk{\chi_{\TriangleSNNeg_k}}
\def\CharacteristicFunctionTriangleSNNegl{\chi_{\TriangleSNNeg_l}}
\def\CharacteristicFunctionTriangleSNNegm{\chi_{\TriangleSNNeg_m}}

\def\dW{\dot{\W}}
\def\dWImag{\dW^{(I)}}
\def\dWImagInit{\dW^{(I0)}}
\def\tWImag{\tW^{(I)}}
\def\tWImagInit{\dW^{(I0)}}

\def\tWImagAsy{\tW^{(I\infty)}}

\def\CVgamma{\tilde{\gamma}}
\def\dCVgamma{\dot{\gamma}}
\def\CVgammaAsy{\CVgamma^{(\infty)}}
\def\dCVgammaAsy{\dot{\gamma}^{(\infty)}}

\def\bzero{\mathbf{0}}
\def\tWImag{\tW^{(I)}}
\def\tQImag{\tQ^{(I)}}

\def\Vorticity{\omega}

\def\dWAsy{\dot{\W}^{(\infty)}}
\def\dWImag{\dot{\W}^{(I)}}
\def\dWImagAsy{\dot{\W}^{(I\infty)}}
\def\pWImagAsy{\dot{\W}^{(I\infty)\prime}}
\def\tWImagInit{\tW^{(I0)}}

\def\bo{\mathbf{0}}

\def\deltak{\delta{k_2}}
\def\ThresholdK{K}
\def\Expal{E}
\def\DomainbkGreat{{\cal D}^c_0}

\def\EstimatedSupportForgammaij{{\cal D}_{\Gamma}}
\def\EstimatedSupportForWijk{{\cal D}_{\W_{(ij)k}}}
\def\EstimatedSupportForWij{{\cal D}_{\W}}
\def\LowerSupportBoundForgammaij{K^{\Gamma}_{2L}}
\def\UpperSupportBoundForgammaij{K^{\Gamma}_{2U}}
\def\LowerSupportBoundForWij{K^{\W}_{2L}}
\def\UpperSupportBoundForWij{K^{\W}_{2U}}

\def\mangled{\langle\Bm\rangle}

\def\hw{\hat\w}
\def\hv{\hat{\Vorticity}}
\def\tv{\tilde{\Vorticity}}

\def\pdCVgammaAsy{\dot{\gamma}^{(\infty)\prime}}

\def\fa{f_a}
\def\fb{f_b}
\def\fc{f_c}
\def\ta{\tilde{a}}
\def\tb{\tilde{b}}
\def\tc{\tilde{c}}
\def\fap{f^{\prime}_a}
\def\fbp{f^{\prime}_b}
\def\fcp{f^{\prime}_c}

\def\Contractedtoc{\tilde\Gamma}
\def\dContractedtoc{\dot{\Gamma}}
\def\dContracted{\dot{\Gamma}^{(0)}}

\def\ContractedtocAsy{\Contractedtoc^{(\infty)}}
\def\dContractedtocAsy{\dContractedtoc^{(\infty)}}
\def\pdContractedtocAsy{\dContractedtoc^{(\infty)\prime}}

\def\argmin{\text{argmin}}
\def\bS{\mathit{S}}
\def\bX{\mathit{X}}
\def\bY{\mathit{Y}}
\def\bZ{\mathit{Z}}
\def\bU{\mathit{U}}
\def\bbR{\mathbb{R}}
\def\cA{{\cal A}}
\def\cB{{\cal B}}
\def\cC{{\cal C}}
\def\cD{{\cal D}}
\def\cE{{\cal E}}
\def\cF{{\cal F}}
\def\cG{{\cal G}}
\def\cH{{\cal H}}
\def\cI{{\cal i}}
\def\cK{{\cal K}}
\def\cL{{\cal L}}
\def\cP{{\cal P}}
\def\cQ{{\cal Q}}
\def\cR{{\cal R}}
\def\dist{\textbf{dist}}
\def\dom{\textbf{dom}}
\def\Indicator{I}
\def\maximize{\text{maximize}}
\def\minimize{\text{minimize}}
\def\overX{\bar{X}}
\def\overY{\bar{Y}}
\def\Projection{\Pi}
\def\prox{\textbf{prox}}
\def\subjectto{\text{subject to}}
\def\Lik{\Lambda_i^k}
\def\Ljk{\Lambda_j^k}
\def\NCV{N_{\text{CV}}}

\def\normk{||\mathbf{k}||}
\def\norml{||\mathbf{l}||}
\def\normkl{||\mathbf{k}+\mathbf{l}||}

\def\DensitydWAsybb{\rho^{(\infty)}_{22}}
\def\DensitydWAsyab{\rho^{(\infty)}_{12}}
\def\DensitydWAsyaa{\rho^{(\infty)}_{11}}
\def\DensitydWAsy{\rho^{(\infty)}}

\def\CharacteristicFunction{\chi}
\def\tphi{\hat{\phi}}

\def\hatw{\hat{\w}}
\def\hatq{\hat{\q}}
\def\bc{\mathbf{c}}

\def\RAsy{R^{(\infty)}}

\begin{abstract}
To help resolve issues of non-realizability and restriction to homogeneity 
faced by analytical theories of  turbulence, 
we explore  three-dimensional homogeneous shear turbulence
of an incompressible Newtonian fluid
within the context of optimal control and convex optimization.
The framework is composed of multi-point spatial correlations
of velocity and pressure fluctuations up to the degenerate fourth order,
their evolution equations and constraints.
 It is argued that the integral of the trace of the second order correlations is 
 the objective functional to be maximized.
 The sources of the constraints are discussed,
 such as the Cauchy-Schwarz inequality and the non-negativity of variance of products.
 Two models are introduced:   the second-order model
 uses the contracted and degenerate third order correlations as the control variables;
 the third-order model takes the degenerate fourth order correlations as the control variables.
Both model are  second-order cone programs when discretized.
 The nature of large-scale and huge-scale computations and the link to big data are commented on.
 It is shown that the exponential growth rates of the asymptotic states are bounded 
 from above by zero.
The asymptotic steady state of the second-order model is solved numerically.
 Three bounded characteristic macro length scales are predicted, 
 streamwise, vertical, and spanwise, beyond which the two-point
 correlations are negligible.
The predicted values of the anisotropy tensor are consistent with experimental data  qualitatively
(in regard to the relative numerical order pattern of the diagonal components),
albeit with significant quantitative differences.
 Such differences are attributed to the non-enforceability
 of the non-negativity of variance of products within the second-order model.
 Compared with DNS data, 
 the predicted second order correlation functions contain flawed features
 of local minima either too large in magnitude or present spuriously.
 The third-order model is expected to improve predictions,
 because of its ability to include constraints generated by the non-negativity
  of variance of products
 and its mathematical structure formulated in an enlarged space of control variables.
   The issue of how to solve this huge-scale problem is yet open.
\end{abstract}

\maketitle

\section{Introduction}

As regards the analytical theories of  
 turbulence for an incompressible Newtonian fluid
\cite{Davidson2004, Lesieur2008, Orsazg1970, Tatsumi1980, SagautCambon2008},
especially the quasi-normal model (QN) and a few of its variants concerned,
there are  known issues of non-realizability 
(a large part of the turbulent energy spectrum becoming
negative in isotropic turbulence \cite{OBrienFrancis1962,Ogura1963,Davidson2004})
and restriction to homogeneity
(``Two-point models are restricted, more or less, to homogeneous turbulence
      (unlike one-point models), but nevertheless they have proved very popular.''
\cite{Davidson2004}).
In this study, we explore a framework of optimal control and convex optimization
to help resolve these issues.
We choose three-dimensional
homogeneous shear turbulence as a test problem,
considering the relative
simplicity of the motion and 
the availability of  experimental and
direct numerical simulation (DNS) data 
  \cite{Piquet1999, RogersMoin1987, Sekimotoetal2016}.

To determine the statistically averaged structure
of three-dimensional homogeneous shear turbulence,
we start with the relations resulting from the Navier-Stokes equations,
   the Reynolds decomposition, 
and the ensemble average operation,
 following conventional practice. 
Our attention is restricted to a framework accounting for the multi-point spatial 
correlations of velocity and pressure fluctuations 
up to the degenerate fourth order. It is
guided by the general structure of
  analytical theories, the consideration of  mathematical and computational demands, 
 and the availability of  experimental  data \cite{TaoRamakrishna2010}.
 One part of the relations  concerns the dynamical equations of evolution 
 for the correlations and is derived in  conventional fashion.
The other part involves the constraints of equality and inequality 
for  the correlations
that are the essence of realizability and whose comprehensive enforcement
in the proposed framework
  is a departure from the practice of  analytical theories.
 Satisfaction of these constraints provides a solution to
 the issue of non-realizability. Alternatively,
one may pursue a comprehensive statistical formulation of turbulence 
with a probability density functional
which produces functional equations governing the evolution
of the characteristic functional \cite{Hopf1952, McComb1990}.
Such a formalism, however, faces the challenge of solving the functional equations
and in going beyond homogeneous turbulence.

There are several sources, physically and mathematically,
that give rise to  constraints in ensemble 
averaged turbulence modeling.
The constraints of equality come from the self-consistency of the definitions of correlations,
 the divergence-free fluctuating velocity field from the supposed incompressibility,
 and certain symmetries that arise from the mean flow characteristics.
The constraints of inequality are constructed mathematically
  from the applications of the Cauchy-Schwarz inequality and
the non-negativity of variance of products.
These inequalities include, as special cases,
 physically motivated and known constraints, 
such as the positive semi-definiteness of the Reynolds stress tensor and
 the non-negativity of the viscous dissipation, that
 are the realizability conditions usually addressed
 in engineering turbulence modeling methodology \cite{Schumann1977}.
However, the vast majority of these inequalities 
are ignored or even nonenforceable
within  conventional schemes of turbulence modeling.

Here, a conventional scheme means a closure approach which approximates
the highest order correlations in terms of lower ones,
in the form of equalities, within a turbulence model.
In QN, for example, the fourth order correlations are represented 
by certain combinations of
the products of the second order correlations in the Fourier wave-number space;
the specific numerical simulation of isotropic turbulence
has demonstrated the emergence of  nonphysical negative energy spectrum
in this specific model
\cite{OBrienFrancis1962,Ogura1963}.
It is thus inferred that in general
  such a scheme cannot satisfy
  all the constraints aforementioned.
This observation
motivates us to pursue an optimal control approach:
the realization of the constraints
is taken as essential;
 an objective is optimized, 
subject to the dynamical equations and the constraints.
One possible choice for the objective is Shannon entropy and its maximization
\cite{Shannon1948, Jaynes1957, Jaynes1983, CoverThomas2006, EdwardsMcComb1969},
which poses difficulties of formidable computational size and
numerical integration in very high dimensions.
Shannon entropy, however, provides guidance on selection of the integral
associated with the second order correlations, $\int_{\mathbb{R}^3} d\br\,\W_{kk}(\br)$,
as the  objective functional to be maximized.
This form is also substantiated from the strong mixing perspective
of physical turbulence.

Two models are introduced,
motivated by the structure of the dynamical equations of evolution.
The second-order model takes 
 the contracted and degenerate third order correlations as the control variables.
The third-order model takes
 the degenerate fourth order correlations as the control variables.
In their discretized forms, both models are second-order cone programs (SOCP)
according to the criteria defined in 
\cite{Loboetal1998, AlizadehGoldfarb2003, BoydVandenberghe2009}.
 There is a constitutive difference between them:
 The second-order model cannot include any constraints 
 from the non-negativity of variance of products,
 while the third-order model
  includes those constraints from the non-negativity requirement
  involving only the second, third, and fourth order correlations.
  The significance of this difference lies in that such
  constraints  provide a mechanism to bound 
  the correlations  and their exponential growth rate  in the asymptotic states.
  It is noted that these constraints 
  demand that  the maximum growth rate of the correlations be zero
  ($\max\sigma=0$),
 contradicting  experimental and DNS data
 \cite{Piquet1999, IsazaCollins2009, Sekimotoetal2016}.

Restricted by  computational feasibility at present,
we focus on the numerical solution of the second-order model 
in the asymptotic steady state in this paper.
As part of the basis  to evaluate the framework,
we compute the non-trivial components of the  anisotropy tensor $b^{(\infty)}_{ij}$, 
 against  experimental data summarized in \cite{Piquet1999}.
Both our prediction and the experimental data
are given here:
(a) $b^{(\infty)}_{11}=0.5031$ $> b^{(\infty)}_{33}=-0.2080$ 
       $> b^{(\infty)}_{22}=-0.2951$ (Prediction) versus
$b^{(\infty)}_{11}=0.203$ $> b^{(\infty)}_{33}=-0.06$ $> b^{(\infty)}_{22}=-0.143$
 (Experiment);
(b) $b^{(\infty)}_{12}=-0.1161$ (Prediction) versus 
$b^{(\infty)}_{12}=-0.156$ (Experiment).
The comparison shows that  the model
  produces values  matching qualitatively with
the order pattern of the experimental data,
albeit with  significant quantitative differences.
Such differences are attributed to the 
non-enforceability of the non-negativity requirement
within the second-order model.
With regard to the prediction of the second order correlation functions,
three bounded characteristic length scales are obtained, 
streamwise, spanwise, and vertical, beyond which the two-point 
correlations are effectively negligible;
the relative numerical order of the length scales
 is compatible with the physical flow.
Compared with the DNS data of \cite{Sekimotoetal2016}, however,
the predicted spatial distributions of the correlation functions
are flawed in that the local negative minima are
either too large in magnitude or present  spuriously.

The third-order model is
expected to improve predictions,
because it is formulated in an enlarged control variable space
and it contains a bounding
mechanism  produced by the non-negativity requirement.
 However, the computational scale of the model is huge
  according to the criteria listed in \cite{Nesterov2014},
 a great effort is required with regard to
  the mathematical formulation, algorithms, codes, etc.
The third-order model has a close link to big data \cite{Cevheretal2014}
and  research in the field of big data may help 
 explore and implement the model.

The present paper is organized as follows.
 In Sec.~\ref{sec:HomogeneousTurbulence},
 we develop the dynamical equations of  evolution for the 
 correlations, define the two models,
  and discuss the constraints.
 The issue of closure and choice of objective functional is addressed,
 and the basic features are summarized.
 In Sec.~\ref{sec:AsymptoticStates},
we consider the  asymptotic states of two models at large time and
  present one consequence of the non-negativity of variance of products
  --- the maximum exponential growth rate of the correlations is zero.
Also, we comment on the issues of  huge-scale computation and DNS.
 In Sec.~\ref{Sec:AsymptoticStatesofSOM},
 we study numerically the asymptotic steady state of the second-order model. 
  The predicted values of the anisotropy tensor and other quantities
   are compared against  experimental data,
   the role of the non-negativity of variance constraints is examined,
   and the predictions of the second order correlation functions
   are evaluated against DNS data.
   The main results of the present work are summarized
  in Sec.~\ref{sec:Conclusion}.
  Appendices provide details of
 certain analytical, numerical, and computational aspects.

\section{\label{sec:HomogeneousTurbulence}Basic framework}

Consider homogeneous shear turbulence of an incompressible
Newtonian fluid of mass density $\rho$ and kinematic viscosity $\nu$
in $\mathbb{R}^3$ and in the Cartesian coordinate system $x_i$ 
with the mean velocity field, 
\begin{align}
V_i=\delta_{i1} S x_2,
\label{AverageVelocityField}
\end{align}
where $\delta_{ij}$ is the Kronecker delta and $S$ is the spatially uniform mean shear rate,
positive and constant.
The equations of motion governing the fluctuation fields of velocity $\wi(\bx,t)$ and
 scaled pressure $\q(\bx,t)=p(\bx,t)/\rho$ can be
  derived from the Navier-Stokes equations, using Eqs.~\eqref{AverageVelocityField}
 and the Reynolds decomposition.
Next, dimensionless quantities (accented) are defined via
\begin{align}
\label{DimensionlessQuantityDefn}
&
t=t^{\prime}/S,\quad
x_i=(\nu/S)^{1/2} x^{\prime}_i,\quad
w_i=(\nu S)^{1/2} w^{\prime}_i,\quad
\q=\nu S q^{\prime}.
\end{align}
The aforementioned equations of fluctuation fields are non-dimensionalized to give
\begin{align}
\label{NSEqns}
 &
\frac{\partial\wk}{\partial x_k}=0,
\quad
\frac{\partial \wi}{\partial t}
+ x_2 \frac{\partial \wi}{\partial x_1}
+ \delta_{i1} w_2
+\frac{\partial (\wi \wk)}{\partial x_k}
=-\frac{\partial \q}{\partial x_i}
+ \frac{\partial^2 \wi}{\partial x_k \partial x_k},
\end{align}
where the accent $'$ is removed for brevity.

Some clarification is required regarding the above treatment.
Within the present formulation, the infinite domain $\mathbb{R}^3$ is necessary 
to realize homogeneity throughout the whole domain.
It is conventionally known 
that this homogeneous shear flow has the macro timescale $S^{-1}$,
but it has no characteristic outer length scale,
except for the Kolmogorov micro length scale whose value is yet to be determined.
In Eqs.~\eqref{DimensionlessQuantityDefn}, both $S$ and $\nu$ are combined
to produce a length scale $(\nu/S)^{1/2}$ \cite{RogersMoin1987},
and consequently, Eqs.~\eqref{NSEqns} are 
free of controlling parameters.
Our numerical simulation suggests that in homogeneous shear turbulence
there are statistically averaged characteristic macro length scales,
streamwise, vertical, and spanwise, as presented 
in Subsec.~\ref{subsec:NumericalResultsandDiscussion}.

As done in \cite{MoninYaglom1975}, 
Eqs.~\eqref{NSEqns} are used to construct
the dynamical equations governing the evolution of the second order correlations
$\overline{\wi(\bx) \wj(\by)}$ and
the third order correlations $\overline{\wi(\bx)\wj(\by)\wk(\bz)}$,
along with those governing
 the spatial correlations 
$\overline{\q(\bx) \wj(\by)}$, $\overline{\q(\bx)\wj(\by)\wk(\bz)}$,
and $\overline{\q(\bx)\,\q(\by)}$,
where the dependence of the correlations on $t$ is suppressed.
Further,  homogeneity is adopted to
 simplify the mathematical treatment. 
The structure of the dynamical equations
and  homogeneity lends grounds to define
\begin{align}
\label{Homogeneity}
&
\Wij(\br)=\overline{\wi(\bx) \wj(\by)},\quad
\W_{(ij)k}(\br)=\overline{\wi(\bx) \wj(\bx) \wk(\by)},\quad
\W_{ijk}(\br,\bs)=\overline{\wi(\bx) \wj(\by) \wk(\bz)},
\notag\\[3pt]&
\W_{(ij)kl}(\br,\bs)=\overline{\wi(\bx) \wj(\bx) \wk(\by) \wl(\bz)},\quad
\W_{ijkl}(\bs',\br,\bs)=\overline{\wi(\bx) \wj(\bz') \wk(\by) \wl(\bz)},
\notag\\[-15pt]&
\\[3pt]&
\Q(\br)=\overline{\q(\bx) \q(\by)},\quad
\Qj(\br)=\overline{\q(\bx)\wj(\by)},\quad
\Q_{jk}(\br,\bs)=\overline{\q(\bx)\wj(\by)\wk(\bz)},
\notag
\end{align}
where $\br=\by-\bx$, $\bs=\bz-\bx$, $\bs'=\bz'-\bx$.
The dynamical equations are
\begin{align}
\label{EquationsOfCorrelationsInPhysSpace}
&
\frac{\partial \W_{kj}(\br)}{\partial r_k}=0,\quad
\bigg\{\frac{\partial}{\partial s_l},\quad
       \frac{\partial}{\partial r_k}+\frac{\partial}{\partial s_k}
      \bigg\}\W_{kjl}(\br,\bs)=0,\quad
\frac{\partial \W_{(ij)kl}(\br,\bs)}{\partial r_k}=0,\quad
\frac{\partial \Q_{k}(\br)}{\partial r_k}=0\quad
\frac{\partial \Q_{kl}(\br,\bs)}{\partial r_k}=0,
\notag\\&
\bigg(\frac{\partial}{\partial s'_i}+\frac{\partial}{\partial r_i}
        +\frac{\partial}{\partial s_i}\bigg)\W_{ijkl}(\bs',\br,\bs)=0,
\notag\\&
\bigg(\frac{\partial}{\partial t}+ r_2 \frac{\partial }{\partial r_1}\bigg) \W_{ij}(\br)
+ \delta_{i1} \W_{2j}(\br)
+ \delta_{j1} \W_{i2}(\br)
-\frac{\partial \W_{ikj}(\mathbf{0},\br)}{\partial r_k}
+\frac{\partial \W_{jki}(\mathbf{0},-\br)}{\partial r_k}
=
 \frac{\partial \Q_j(\br)}{\partial r_i}
-\frac{\partial \Q_i(-\br)}{\partial r_j}
+2  \frac{\partial^2 \W_{ij}(\br)}{\partial r_k\partial  r_k},
\notag\\&
\bigg(\frac{\partial }{\partial t} + r_2 \frac{\partial }{\partial r_1}
      + s_2 \frac{\partial }{\partial s_1}\bigg) \W_{ijk}(\br,\bs)      
+ \delta_{i1}   \W_{2jk}(\br,\bs)    
+ \delta_{j1}   \W_{i2k}(\br,\bs)    
+ \delta_{k1}   \W_{ij2}(\br,\bs)  
-\bigg(\frac{\partial }{\partial r_l}+\frac{\partial }{\partial s_l}\bigg) \W_{(il)jk}(\br,\bs)   
\notag\\[-10pt]&
\\[-3pt]&\ \
+\frac{\partial \W_{(jl)ik}(-\br,\bs-\br)}{\partial r_l}  
+\frac{\partial \W_{(kl)ij}(-\bs,\br-\bs)}{\partial s_l} 
=
\bigg(\frac{\partial }{\partial r_i}+\frac{\partial }{\partial s_i}\bigg) \Q_{jk}(\br,\bs)  
-\frac{\partial \Q_{ik}(-\br,\bs-\br)}{\partial r_j} 
-\frac{\partial \Q_{ij}(-\bs,\br-\bs)}{\partial s_k} 
\notag\\&\ \
+2 \bigg(
 \frac{\partial^2 }{\partial r_l\partial  r_l}
+\frac{\partial^2 }{\partial s_l\partial  s_l}
+\frac{\partial^2}{\partial r_l \partial s_l}
\bigg) \W_{ijk}(\br,\bs), 
\notag\\&
\frac{\partial^2 \Q(\br)}{\partial r_k \partial r_k}
=-2 \frac{\partial \Q_2(\br)}{\partial r_1}
-\frac{\partial^2 \Q_{kl}(\br,\br)}{\partial r_k \partial r_l},\quad
\frac{\partial^2 \Qj(\br)}{\partial r_k\partial  r_k}  
= 2 \frac{\partial \W_{2j}(\br)}{\partial r_1}   
-\frac{\partial^2 \W_{lkj}(\mathbf{0},\br)}{\partial r_k \partial r_l},
\notag\\&
\bigg(\frac{\partial}{\partial r_l}+\frac{\partial}{\partial s_l}\bigg)
\bigg(\frac{\partial}{\partial r_l}+\frac{\partial}{\partial s_l}\bigg)\Q_{jk}(\br,\bs)
=
2 \bigg(\frac{\partial }{\partial r_1}+\frac{\partial }{\partial s_1}\bigg)  \W_{2jk}(\br,\bs)
-\bigg(\frac{\partial}{\partial r_m}+\frac{\partial}{\partial s_m}\bigg)
\bigg(\frac{\partial}{\partial r_l}+\frac{\partial}{\partial s_l}\bigg) \W_{(lm)jk}(\br,\bs).
 \notag
\end{align}
These dynamical equations of evolution act as constraints of equality
and can be viewed as such. They represent mathematically
the core physics of homogeneous shear turbulence.

To ensure self-consistency,
definitions~\eqref{Homogeneity} themselves
impose the equality constraints of symmetry,
\begin{align}
\label{HomogeneitySymmetry}
&
\Wij(\br)=\Wji(-\br),\quad
\W_{(ij)k}(\br)=\W_{(ji)k}(\br),\quad
\W_{(ij)k}(\bo)=\W_{(ik)j}(\bo),
\notag\\&
\W_{ijk}(\br,\bs)=\W_{ikj}(\bs,\br)=\W_{jik}(-\br,\bs-\br)
=\W_{kij}(-\bs,\br-\bs),\quad
\W_{(ij)kl}(\br,\bs)=\W_{(ji)kl}(\br,\bs)=\W_{(ij)lk}(\bs,\br),
\\&
\W_{(ij)kl}(\br,\br)=\W_{(kl)ij}(-\br,-\br),\quad
\W_{(ij)kl}(\bo,\br)
=\W_{(ik)jl}(\bo,\br),\quad
\Q(\br)=\Q(-\br),\quad
\Q_{ij}(\br,\bs)=\Q_{ji}(\bs,\br).
\notag
\end{align}
Further, motivated by experimental data \cite{Piquet1999, SagautCambon2008}
that give
$\W_{13}(\bo)=\W_{23}(\bo)=0$
and the need to reduce computational size of the problem,
we consider 
the geometric and kinematic symmetries underlying 
the mean velocity field~\eqref{AverageVelocityField}.
They suggest inversion and mirror  symmetries described by
\begin{align}
\label{InversionMirror}
&
\W_{ij}(\br)=\W_{ij}(-\br)
=(-1)^{\delta_{\underlinei 3}+\delta_{\underlinej 3}}\,\W_{\underlinei \underlinej}(\br'),
\quad
\W_{(ij)k}(\br)=-\W_{(ij)k}(-\br)
=(-1)^{\delta_{\underlinei 3}+\delta_{\underlinej 3}
           +\delta_{\underlinek 3}}\W_{(\underlinei\underlinej)\underlinek}(\br'),
\notag\\&
\W_{ijk}(\br,\bs)=-\W_{ijk}(-\br,-\bs)
=(-1)^{\delta_{\underlinei 3}+\delta_{\underlinej 3}
        +\delta_{\underlinek 3}}\W_{\underlinei\underlinej\underlinek}(\br',\bs'),
\notag\\[-6.5pt]&
\\[-6.5pt]&
\W_{(ij)kl}(\br,\bs)=\W_{(ij)kl}(-\br,-\bs)
=(-1)^{\delta_{\underlinei 3}+\delta_{\underlinej 3}
       +\delta_{\underlinek 3}+\delta_{\underlinel 3}}
           \W_{(\underlinei\underlinej)\underlinek\underlinel}(\br',\bs'),
\quad
\Q(\br)=\Q(-\br)=\Q(\br'),
\notag\\&
\Q_i(\br)=-\Q_i(-\br)=(-1)^{\delta_{\underlinei 3}}\Q_{\underlinei}(\br'),
\quad
\Q_{ij}(\br,\bs)=\Q_{ij}(-\br,-\bs)=(-1)^{\delta_{\underlinei 3}
                 +\delta_{\underlinej 3}}\Q_{\underlinei\underlinej}(\br',\bs').
\notag
\end{align}
Here, $\br'=(r_1,r_2,-r_3)$ and $\bs'=(s_1,s_2,-s_3)$.
The summation rule is suspended for the subscripts underlined.
The derivation of the  equalities
  is outlined in Appendix~\ref{Appensec:InversionAndMirror}.

Owing to the convenience of explicit representation
of the constraints and objective in terms of the control variables
and the need to generate standard conic forms
for numerical simulation,
 the problem is formulated in the Fourier wave-number space
with the help of  Fourier transforms,
\begin{align*}
&
\phi(\br)=\int_{\mathbb{R}^3} d\bk\,
     \tilde{\phi}(\bk) \exp(\imaginary\,\bk\cdot\br),
\quad
\phi(\br,\bs)
=\int_{\mathbb{R}^3\times\mathbb{R}^3} d\bk d\bl\,  \tilde{\phi}(\bk,\bl)
   \exp[\imaginary\, (\bk\cdot\br+\bl\cdot\bs)],
\notag\\&
\phi(\bs',\br,\bs)
=\int_{\mathbb{R}^3\times\mathbb{R}^3\times\mathbb{R}^3}
 d\Bm\, d\bk\, d\bl\,\tilde{\phi}(\Bm,\bk,\bl)
\exp[\imaginary\, (\Bm\cdot\bs'+\bk\cdot\br+\bl\cdot\bs)].
\end{align*}
In  the wave-number space,
 $\tW_{ij}$, $\tW_{(ij)kl}$,  $\tW_{ijkl}$, $\tQ$, 
and $\tQ_{ij}$ are real, 
 $\tW_{(ij)k}$,
 $\tW_{ijk}$, and $\tQ_j$ are purely imaginary and 
 are transformed according to
\begin{align}
\label{wiwjwkPurelyImaginaryfs}
\tW_{(ij)k}(\bk)=\imaginary\,\tWImag_{(ij)k}(\bk),\quad
\tW_{ijk}(\bk,\bl)=\imaginary\,\tWImag_{ijk}(\bk,\bl),
\quad
\tQ_i(\bk)=\imaginary\,\tQImag_i(\bk).
\end{align}
Equations~\eqref{EquationsOfCorrelationsInPhysSpace}
through \eqref{InversionMirror} are transformed as
\begin{subequations}
\label{EquationsOfMotionfs}
\begin{align}
\label{DivergenceFreeInPhysicalSpace_qandw_fs}
&
k_k\tW_{kj}(\bk) =0,\quad
k_k\tWImag_{(ij)k}(\bk)=0,\quad
k_j\tWImag_{kjl}(\bk,\bl)=0,\quad
(k_k+l_k)\tWImag_{kij}(\bk,\bl)=0,\quad
k_k\tW_{(ij)kl}(\bk,\bl)=0,
\notag\\[-7pt]&
\\[-7pt]&
(m_i+k_i+l_i)\tW_{ijkl}(\Bm,\bk,\bl)=0,
\notag
\end{align}
\vspace{-1.em}
\begin{align}
 \tQ(\bk) 
=
-\frac{2 k_1 \tQ^{(I)}_2(\bk)}{|\bk|^2}  
-\frac{k_k k_l}{|\bk|^2}\int_{\mathbb{R}^3}d\bl\tQ_{kl}(\bk-\bl,\bl),
\label{PressureInPhysicalSpace_qq_fs}
\end{align}
\vspace{-1.em}
\begin{align}
\tQImag_j(\bk) = 
 -\frac{2 k_1 \tW_{2j}(\bk) }{|\bk|^2} 
-\frac{k_k k_l}{|\bk|^2} \int_{\mathbb{R}^3}d\bl \tWImag_{lkj}(\bl,\bk),  
\label{PressureInPhysicalSpace_qw_fs}
\end{align}
\begin{align}
\tQ_{jk}(\bk,\bl) =
\frac{2 (k_1+l_1) \tWImag_{2jk}(\bk,\bl)}{|\bk+\bl|^2}
 -\frac{(k_m+l_m)(k_l+l_l) \tW_{(lm)jk}(\bk,\bl)}{|\bk+\bl|^2},
\label{PressureInPhysicalSpace_qww_fs}
\end{align}
\begin{align}
\label{CLMInPhysicalSpace_ww_fs}
&
\bigg(
      \frac{\partial}{\partial t}
      -k_1 \frac{\partial}{\partial k_2}
      +2|\bk|^2\bigg)\tWij(\bk)
+ \delta_{i1} \tW_{2j}(\bk)+\delta_{j1} \tW_{i2}(\bk) 
\notag\\&
=
-k_i\tQImag_j(\bk)
+k_j\tQImag_i(-\bk) 
-k_k \int_{\mathbb{R}^3} \Big(\tWImag_{ijk}(\bk,\bl)-\tWImag_{jik}(-\bk,\bl)\Big) d\bl ,
\end{align}
\begin{align}
\label{CLMInPhysicalSpace_www_fs}
&
 \bigg(\frac{\partial }{\partial t}         
       -k_1\frac{\partial}{\partial k_2}
       -l_1\frac{\partial}{\partial l_2}
       +2(|\bk|^2+|\bl|^2+\bk\cdot\bl)\bigg)\tWImag_{ijk}(\bk,\bl)
+\delta_{i1}  \tWImag_{2jk}(\bk,\bl)
+\delta_{j1} \tWImag_{i2k}(\bk,\bl)
+\delta_{k1}\tWImag_{ij2}(\bk,\bl)   
\notag\\ &     
=
 k_i\tQ_{jk}(\bk,\bl)
+l_i\tQ_{jk}(\bk,\bl)
-k_j\tQ_{ik}(-\bk-\bl,\bl)
-l_k\tQ_{ij}(-\bk-\bl,\bk)         
+(k_l+l_l)\tW_{(il)jk}(\bk,\bl)  
\notag\\&\hskip4mm
-k_l\tW_{(jl)ik}(-\bk-\bl,\bl)  
-l_l\tW_{(kl)ij}(-\bk-\bl,\bk),
\end{align}
\end{subequations}
and
\begin{align}
\label{HomogeneitySymmetryInversionMirrorfs}
&
\tW_{ij}(\bk)
=\tW_{ji}(\bk)=\tW_{ij}(-\bk)
=(-1)^{\delta_{\underlinei 3}+\delta_{\underlinej 3}}\,
       \tW_{\underlinei\underlinej}(\bk'),
\quad
\tWImag_{(ij)k}(\bk)=\tWImag_{(ji)k}(\bk)=-\tWImag_{(ij)k}(-\bk)
=(-1)^{\delta_{\underlinei 3}+\delta_{\underlinej 3}+\delta_{\underlinek 3}}
    \,\tWImag_{(\underlinei\,\underlinej)\underlinek}(\bk'),
\notag\\&
\tWImag_{ijk}(\bk,\bl)=\tWImag_{ikj}(\bl,\bk)=\tWImag_{jik}(-\bk-\bl,\bl)
=\tWImag_{kij}(-\bk-\bl,\bk)=-\tWImag_{ijk}(-\bk,-\bl)
=(-1)^{\delta_{\underlinei 3}+\delta_{\underlinej 3}+\delta_{\underlinek 3}}
    \,\tWImag_{\underlinei\,\underlinej\,\underlinek}(\bk',\bl'),
\notag\\&
 \tW_{(ij)kl}(\bk,\bl)=\tW_{(ji)kl}(\bk,\bl)=\tW_{(ij)lk}(\bl,\bk)
=\tW_{(ij)kl}(-\bk,-\bl)
=(-1)^{\delta_{\underlinei 3}+\delta_{\underlinej 3}+\delta_{\underlinek 3}+\delta_{\underlinel 3}}
 \tW_{(\underlinei\underlinej)\underlinek\underlinel}(\bk',\bl'),
\notag\\[-7pt]&
\\[-7pt]&
\int_{\mathbb{R}^3} d\bk\,
 \Big[\tW_{(ij)kl}(\bk,\bl)-\tW_{(ik)jl}(\bk,\bl)\Big]=0,
\quad
\int_{\mathbb{R}^3} d\bl\,
 \Big[\tW_{(ij)kl}(\bk+\bl,-\bl)-\tW_{(kl)ij}(\bk+\bl,-\bl)\Big] =0,
\notag\\&
\tQ(\bk)=\tQ(-\bk)=\tQ(\bk'),
\quad
\tQImag_i(\bk)=-\tQImag_i(-\bk)
=(-1)^{\delta_{\underlinei 3}} \tQImag_{\underlinei}(\bk'),
\notag\\&
\tQ_{ij}(\bk,\bl)=\tQ_{ij}(-\bk,-\bl)
=\tQ_{ji}(\bl,\bk)
=(-1)^{\delta_{\underlinei 3}+\delta_{\underlinej 3}}
     \tQ_{\underlinei\underlinej}(\bk',\bl'),
\notag
\end{align}
where $\bk'=(k_1,k_2,-k_3)$, $\bl'=(l_1,l_2,-l_3)$.

The structure of Eqs.~\eqref{EquationsOfMotionfs}
provides the ground to introduce  
the second-order model and the third-order model.

\subsection{\label{subsec:PrimaryEquationsEvolutionInSOM}
    The second-order model}

Equations~\eqref{PressureInPhysicalSpace_qw_fs}
and \eqref{CLMInPhysicalSpace_ww_fs} indicate that
 $\tWImag_{ijk}$
affect $\tWij$ only through the degenerate and  contracted
$k_k \tWImag_{(ki)j}(\bk)$. Therefore,
the second-order model is formulated with 
\begin{align}
\Contractedtoc_{ij}(\bk)=k_k \tWImag_{(ki)j}(\bk)
\label{ContractedTOC}
\end{align}
as the control variables.
The constraint of $k_j\Contractedtoc_{ij}(\bk)=0$ 
from Eqs.~\eqref{DivergenceFreeInPhysicalSpace_qandw_fs} yields
\begin{align}
\Contractedtoc_{j3}(\bk)=
-\frac{(k_1 \Contractedtoc_{j1}+k_2 \Contractedtoc_{j2})(\bk)}{k_3},
\label{vHST_DivergenceFreeInPhysicalSpace_wiwjwk_fs_SOM}
\end{align}
which implies that
 $\Contractedtoc_{ij}$, $i=1,2,3$, $j=1,2$,
act as the primary control variables.
    
Correlations $\tW_{ij}$ are state variables.
 Equations~\eqref{DivergenceFreeInPhysicalSpace_qandw_fs} are solved to obtain
\begin{align}
\label{DivergenceFreeInPhysicalSpace_wiwj_fs}
\tW_{j3}(\bk) =-\frac{(k_1\tW_{j1}+k_2\tW_{j2})(\bk)}{k_3},\ \ j=1,2,
\quad 
\tW_{33}(\bk) 
=\frac{((k_1)^2 \tW_{11}+(k_2)^2 \tW_{22}+2 k_1 k_2\tW_{12})(\bk)}{(k_3)^2}.
\end{align}
That is, $\tW_{11}$, $\tW_{12}$, and $\tW_{22}$ 
are treated as the primary components of $\tW_{ij}$. 
Their equations of evolution are derived from
Eqs.~\eqref{CLMInPhysicalSpace_ww_fs},
\begin{align}
\label{vEvolutionOftWPrimarySOM}
&
\frac{\exp[-2H(\bk)]}{2|\bk|^4}
\bigg(\frac{\partial}{\partial t}-k_1\frac{\partial}{\partial k_2}\bigg)
  \Big(|\bk|^4\exp\big[2 H(\bk)\big] \tW_{22}(\bk)\Big)  
=
 \frac{k_2 k_l \Contractedtoc_{l2}(\bk)}{|\bk|^2}
-\Contractedtoc_{22}(\bk),
\notag\\[8pt]&
\frac{\exp[-2H(\bk)]}{|\bk|^2}
\bigg(\frac{\partial}{\partial t}-k_1 \frac{\partial}{\partial k_2}\bigg)
   \Big(|\bk|^2 \exp\big[2 H(\bk)\big]\tW_{12}(\bk)\Big)
+\bigg(1-\frac{2(k_1)^2}{|\bk|^2}\bigg) \tW_{22}(\bk)
\notag\\[-7pt]&
\\[-7pt]&\ \
=
 \frac{k_l [k_1\Contractedtoc_{l2}(\bk)+k_2\Contractedtoc_{l1}(\bk)]}{|\bk|^2}
-\Contractedtoc_{12}(\bk)
-\Contractedtoc_{21}(\bk),  
\notag\\[4pt]&
\frac{\exp[-2H(\bk)]}{2}
 \bigg(\frac{\partial}{\partial t}-k_1\frac{\partial}{\partial k_2}\bigg)
         \Big(\exp\big[2 H(\bk)\big]\tW_{11}(\bk)\Big)
+\bigg(1-\frac{2 (k_1)^2}{|\bk|^2}\bigg)\tW_{12}(\bk) 
=
       \frac{k_1 k_l \Contractedtoc_{l1}(\bk)}{|\bk|^2}   
      -\Contractedtoc_{11}(\bk),
\notag
\end{align}
where
\begin{align*}
H(\bk)=
-\frac{k_2}{k_1}\bigg((k_1)^2+\frac{1}{3}(k_2)^2+(k_3)^2\bigg).
\end{align*}
The remaining component equations of
Eqs.~\eqref{CLMInPhysicalSpace_ww_fs} are redundant.
These differential equations may be solved  under
initial conditions and the boundary conditions of
\begin{align}
\label{BoundaryConditionOftWij}
\lim_{k_2\rightarrow \pm\infty}\tW_{ij}(\bk)=0,
\end{align}
owing to the physical boundedness of $\W_{ij}(\br)$.

Four sets of constraints are present for  
 $\{\Contractedtoc_{ij}: i=1,2,3$, $j=1,2\}$.
(a) The symmetries of $\tW_{ij}$ in
Eqs.~\eqref{HomogeneitySymmetryInversionMirrorfs},
linked by Eqs.~\eqref{DivergenceFreeInPhysicalSpace_wiwj_fs} 
and \eqref{vEvolutionOftWPrimarySOM}.
(b) The symmetries of $\tW_{(ij)k}$ in
Eqs.~\eqref{HomogeneitySymmetryInversionMirrorfs} require that
\begin{align}
&
\Contractedtoc_{ij}(\bk)=\Contractedtoc_{ij}(-\bk)
=
 (-1)^{\delta_{\underlinei 3}+\delta_{\underlinej 3}}
    \Contractedtoc_{\underlinei\underlinej}(\bk'),
\ \ 
\bk'=(k_1,k_2,-k_3).
\label{ContractedSymmetryInversionMirrorfs}
\end{align}
(c) The conditions $(k_k+l_k)\tWImag_{kij}(\bk,\bl)$ $=0$ from
Eqs.~\eqref{DivergenceFreeInPhysicalSpace_qandw_fs} result in 
(details in Appendix~\ref{Appensec:DetailsSecondOrderModel})
\begin{align}
\int_{-\infty}^0 dk_1 \int_{\mathbb{R}^2} dk_2 dk_3
\Big(\Contractedtoc_{ij}(\bk)+\Contractedtoc_{ji}(\bk)\Big)
=0,\ \ i\leq j.
\label{vIntrinsicEquality_kk_jjk_Half_SOM}
\end{align}
(d)  Constraints of inequality~\eqref{twiwj_CSInWNS} formulated in 
Subsec.~\ref{Subsec:ConstraintsInvolvingSecondOrderCorrelations}.

Correlations $\tQImag_j$ are state variables.
 Equations~\eqref{PressureInPhysicalSpace_qw_fs},
    \eqref{vHST_DivergenceFreeInPhysicalSpace_wiwjwk_fs_SOM},
 and \eqref{DivergenceFreeInPhysicalSpace_wiwj_fs} yield
\begin{align}
k_j\tQImag_j(\bk)=0.
\label{vHST_DivergenceFreeInPhysicalSpace_LowestOrderOnq}
\end{align}
Combining the contracted Eqs.~\eqref{CLMInPhysicalSpace_ww_fs} ($i=j$) 
and Eq.~\eqref{vHST_DivergenceFreeInPhysicalSpace_LowestOrderOnq} produces
\begin{align}
\bigg(\frac{\partial}{\partial t}
           -k_1 \frac{\partial}{\partial k_2}
            +2 |\bk|^2\bigg)\tW_{jj}(\bk)
+2\tW_{12}(\bk)
=-2\Contractedtoc_{jj}(\bk).
\label{vEvolutionOftWkk_SOM} 
\end{align}
Next, combination of Eqs.~\eqref{vIntrinsicEquality_kk_jjk_Half_SOM}
and \eqref{vEvolutionOftWkk_SOM}
leads to the known relationship,
\begin{align}
\frac{\partial \W_{jj}(\bo)}{\partial t}
+2\W_{12}(\bo)
-2\frac{\partial^2 \W_{jj}(\br)}{\partial r_k\partial  r_k}\bigg|_{\br=\bo}
=0,
\label{vIntrinsicRelationForUkk(0)Evolution}
\end{align}
which is satisfied automatically.
The structures of 
Eqs.~\eqref{vEvolutionOftWkk_SOM} and \eqref{vIntrinsicRelationForUkk(0)Evolution}
indicate that
Eqs.~\eqref{vIntrinsicEquality_kk_jjk_Half_SOM}
and \eqref{vHST_DivergenceFreeInPhysicalSpace_LowestOrderOnq}
play the known role of distributing $\tW_{kk}$
among its components $\tW_{\underlinei \underlinei}$
and distributing  $\W_{kk}(\bzero)$ among its
  components $\W_{\underlinei \underlinei}(\bzero)$.
  
It is noticed that the second-order model has 
$\{\tW_{ij}, \W_{ij},\tQImag_j, \Q_j\}$ 
as its state variables
and $\Contractedtoc_{ji}$ as the control variables.
  
  \vspace{-4.5mm}
  
\subsection{\label{subsec:PrimaryEvolutionEquationsoftWijk}
The third-order model}

The third-order model contains correlations up to 
the degenerate fourth order  $\tW_{(ij)kl}$.
 $\tW_{ij}$, $\tWImag_{ijk}$,
$\tQ$, $\tQImag_i$, $\tQ_{ij}$, and their correspondences in the 
physical space are the state variables,
 $\tW_{(ij)kl}$ and  $\W_{(ij)kl}$ act as the control variables.

The evolution of $\tW_{ij}$ is governed by
Eqs.~\eqref{DivergenceFreeInPhysicalSpace_wiwj_fs}
and \eqref{vEvolutionOftWPrimarySOM}.
$\tQ$, $\tQImag_i$, and $\tQ_{ij}$ are given by
Eqs.~\eqref{PressureInPhysicalSpace_qq_fs} through 
\eqref{PressureInPhysicalSpace_qww_fs}.
It follows from 
 Eqs.~\eqref{DivergenceFreeInPhysicalSpace_qandw_fs} 
 and \eqref{HomogeneitySymmetryInversionMirrorfs} that
$\tWImag_{ijk}$ may be represented linearly in terms of 
 $\tWImag_{111}$, $\tWImag_{112}$, $\tWImag_{122}$, and $\tWImag_{222}$
    as the primary components 
    (details in Appendix~\ref{Appensec:DetailsThirdOrderModel}).
The equations of evolution for these primary components
are derived from 
Eqs.~\eqref{CLMInPhysicalSpace_www_fs},
\begin{align}
\label{vEvolutionOftWijkPrimary_SOM}
&
\frac{\exp[-H(\bk,\bl)]}{|\bk|^2 |\bl|^2 |\bk+\bl|^2}
\bigg(
     \frac{\partial}{\partial t}
    -k_1\frac{\partial}{\partial k_2}
    -l_1\frac{\partial}{\partial l_2}
\bigg)
\Big(|\bk|^2 |\bl|^2 |\bk+\bl|^2 \exp\big[H(\bk,\bl)\big]\tWImag_{222}(\bk,\bl)\Big)
\notag\\&\
=
 (k_l+l_l)\tW_{(2l)22}(\bk,\bl) 
-(k_2+l_2)\frac{(k_m+l_m)(k_l+l_l)}{|\bk+\bl|^2}\tW_{(lm)22}(\bk,\bl)
-\bigg( k_l\tW_{(2l)22}
       -k_2\frac{k_m k_l}{|\bk|^2}\tW_{(lm)22}\bigg)(-\bk-\bl,\bl)
\notag\\&\hskip4mm
-\bigg( l_l\tW_{(2l)22}     
       -l_2\frac{l_m l_l}{|\bl|^2}\tW_{(lm)22}\bigg)(-\bk-\bl,\bk), 
\notag\\[6pt]&
\frac{\exp[-H(\bk,\bl)]}{|\bk|^2|\bl|^2}
\bigg(
 \frac{\partial }{\partial t}          
-k_1\frac{\partial}{\partial k_2}
-l_1\frac{\partial}{\partial l_2}
\bigg)
\Big(|\bk|^2|\bl|^2\exp\big[H(\bk,\bl)\big]\tWImag_{122}(\bk,\bl)\Big)
+\bigg(1-\frac{2(k_1+l_1)^2}{|\bk+\bl|^2}\bigg)\tWImag_{222}(\bk,\bl)
\notag\\&\
=
 (k_l+l_l)\tW_{(1l)22}(\bk,\bl)  
-(k_1+l_1)\frac{(k_m+l_m)(k_l+l_l)}{|\bk+\bl|^2}\tW_{(lm)22}(\bk,\bl)
-\bigg( k_l\tW_{(2l)12}
       -k_2\frac{k_mk_l}{|\bk|^2}\tW_{(lm)12}\bigg)(-\bk-\bl,\bl)
\notag\\&\hskip4mm
-\bigg( l_l\tW_{(2l)12}     
       -l_2\frac{l_ml_l}{|\bl|^2}\tW_{(lm)12}\bigg)(-\bk-\bl,\bk),
\notag\\[-7pt]&
\\[-0pt]&\ \
\frac{\exp[-H(\bk,\bl)]}{|\bl|^2}
\bigg(
       \frac{\partial }{\partial t}           
      -k_1\frac{\partial}{\partial k_2}
      -l_1\frac{\partial}{\partial l_2}\bigg)
   \Big(|\bl|^2\exp\big[H(\bk,\bl)\big]\tWImag_{112}(\bk,\bl)\Big) 
+\bigg(1-\frac{2 (k_1+l_1)^2}{|\bk+\bl|^2}\bigg)\tWImag_{122}(-\bk-\bl,\bl)
\notag\\&\ 
+\bigg(1-\frac{2(k_1)^2}{|\bk|^2}\bigg)\tWImag_{122}(\bk,\bl)
=
 (k_l+l_l)\tW_{(1l)12}(\bk,\bl)  
-(k_1+l_1)\frac{(k_m+l_m)(k_l+l_l)}{|\bk+\bl|^2}\tW_{(lm)12}(\bk,\bl)
\notag\\&\hskip4mm
-\bigg( k_l\tW_{(1l)12}
       -k_1\frac{k_m k_l}{|\bk|^2}\tW_{(lm)12}\bigg)(-\bk-\bl,\bl)
-\bigg( l_l\tW_{(2l)11}      
       -l_2\frac{l_m l_l}{|\bl|^2}\tW_{(lm)11}\bigg)(-\bk-\bl,\bk),
\notag\\[6pt]&
\exp[-H(\bk,\bl)]
\bigg(\frac{\partial }{\partial t}           
      -k_1\frac{\partial}{\partial k_2}
      -l_1\frac{\partial}{\partial l_2}\bigg)
   \Big(\exp\big[H(\bk,\bl)\big]\tWImag_{111}(\bk,\bl)\Big)
+\bigg(1-\frac{2(k_1+l_1)^2}{|\bk+\bl|^2}\bigg)\tWImag_{112}(\bl,-\bk-\bl)
\notag\\&\
+\bigg(1-\frac{2(k_1)^2}{|\bk|^2}\bigg)\tWImag_{112}(\bl,\bk)
+\bigg(1-\frac{2(l_1)^2}{|\bl|^2}\bigg)\tWImag_{112}(\bk,\bl)
=
 (k_l+l_l)\tW_{(1l)11}(\bk,\bl) 
-(k_1+l_1)\frac{(k_m+l_m)(k_l+l_l)}{|\bk+\bl|^2}\tW_{(lm)11}(\bk,\bl)
\notag\\&\hskip4mm
-\bigg( k_l\tW_{(1l)11}
       -k_1\frac{k_m k_l}{|\bk|^2}\tW_{(lm)11}\bigg)(-\bk-\bl,\bl)
-\bigg( l_l\tW_{(1l)11}   
       -l_1\frac{l_m l_l}{|\bl|^2}\tW_{(lm)11}\bigg)(-\bk-\bl,\bk),
\notag
\end{align}
where
\begin{align*}
H(\bk,\bl)=H(\bk)+H(\bl)+H(\bk+\bl).
\end{align*}
The remaining  component equations of Eqs.~\eqref{CLMInPhysicalSpace_www_fs}
are redundant.
Equations~\eqref{vEvolutionOftWijkPrimary_SOM} may be solved  with
initial conditions and the boundary conditions of
\begin{align}
\label{BoundaryConditionOftWijk}
\lim_{k_2\rightarrow \pm\infty}\tWImag_{ijk}(\bk,\bl)=
\lim_{l_2\rightarrow \pm\infty}\tWImag_{ijk}(\bk,\bl)=0,
\end{align}
owing to the physical boundedness of $\W_{ijk}(\br,\bs)$.
All the state variables may be represented in terms of $\tW_{(ij)kl}$ 
linearly via the equations presented.
 
As control variables, $\tW_{(ij)kl}$ are simplified further 
with the solution of
 Eqs.~\eqref{DivergenceFreeInPhysicalSpace_qandw_fs},
\begin{align}
 \tW_{(ij)3l}(\bk,\bl)
=
-\frac{k_1}{k_3}\tW_{(ij)1l}(\bk,\bl)
-\frac{k_2}{k_3}\tW_{(ij)2l}(\bk,\bl),\ \
ij=11, 12, 13, 22, 23, 33,
\ l=1,2,3.
\label{DivergenceFreeInPhysicalSpace_wiwjwkwl_fs}
\end{align}
The specifically listed $\tW_{(ij)kl}$
 are the primary components of $\tW_{(ij)kl}$ and
  the primary control variables.
 The components not present are 
 found via Eqs.~\eqref{HomogeneitySymmetryInversionMirrorfs}.

The last of Eqs.~\eqref{DivergenceFreeInPhysicalSpace_qandw_fs} are
integrated  to produce 
\begin{align}
\int_{\mathbb{R}^3\times\mathbb{R}^3} d\bk d\bl\,
        \big[ k_i \big(\tW_{(ik)jl}(\bk,\bl)+\tW_{(ij)kl}(\bk,\bl)\big)
             +l_i \tW_{(ij)kl}(\bk,\bl)\big]  =0.
\end{align}
Further, $\tW_{(ij)kl}$ are constrained by 
Eqs.~\eqref{HomogeneitySymmetryInversionMirrorfs} and \eqref{twiwj_CSInWNS}
and those outlined
in Subsec.~\ref{Subsec:ConstraintsInvolvingFourthOrderCorrelations}.

The third-order model faces the issue of
whether the multi-point correlations
are experimentally measurable and how its predictions are evaluated,
this is briefly commented on here:
(a) Two-point correlations involving $\w_i(\bx)$
and $\w_j(\by)$, such as $\W_{ij}(\br)$,
$\W_{ijk}(\mathbf{0},\br)$,
and $\W_{(ij)kl}(\mathbf{0},\br)$,
may be measured experimentally for $\bx$ and $\by$ 
in certain subregions by two hot-wire probes
(restricted by  interferences with the flow and between the probes)
\cite{Bruun1995}.
(b) These two-point correlations may be computed from
$\tW_{ij}$, $\tW_{ijk}$, and $\tW_{(ij)kl}$ within the model,
the predicted values may be compared with the measured ones.

We use the Cauchy-Schwarz inequality
and the non-negativity of variance of products  to construct
constraints for the state and control variables.
  These constraints are classified into two groups, based on whether 
they involve only $\tW_{ij}$ or involve $\tW_{(ij)kl}$, and are
   discussed in the two succeeding subsections.

\subsection{\label{Subsec:ConstraintsInvolvingSecondOrderCorrelations}
  Constraints of inequality involving only $\tW_{ij}$}

This subsection focuses on the constraints
of inequality involving only $\tW_{ij}$ 
that may be enforced within either of the two models. 
The constraints are constructed on the basis of 
 the Cauchy-Schwarz inequality
 $\big|\overline{a b}\big|^2\leq \overline{a a}\,\overline{b b}$ applied
to $\overline{\wi(\bx) \wj(\by)}$ and
related structure functions.

It is straight-forward to apply 
$\big|\overline{a b}\big|^2\leq \overline{a a}\,\overline{b b}$ to 
\begin{align*}
 \overline{\wi(\bx) \wj(\by)},\
 \overline{\wi(\bx) \wj,_k(\by)},\
\overline{\wi,_k\s(\bx) \wj,_l\s(\by)},\, \ldots
\end{align*}
to generate constraints for $\tW_{ij}$.
The structure of these integral inequalities in the wave-number space 
motivate us to adopt
\begin{align}
0\leq \tW_{\underlinei \underlinei}(\bk),\ \
 \big|\tWij(\bk)\big|^2
\leq
\tW_{\underlinei \underlinei}(\bk)\, \tW_{\underlinej\underlinej}(\bk),\ \ i< j.
\label{twiwj_CSInWNS}
\end{align}
These six inequalities play a rather elementary or fundamental role,
as demonstrated below by examples.

Firstly,  as a special and physical case,
Eqs.~\eqref{twiwj_CSInWNS} guarantee the positive
 semi-definiteness of the Reynolds stress tensor 
 $\overline{\wi(\bx)\wj(\bx)}$ and
 the non-negativity of the viscous dissipation 
    $\overline{\wj,_k\s(\bx)\wj,_k\s(\bx)}$;
 These are the realizability conditions
 usually addressed
 in engineering turbulence modeling methodology \cite{Schumann1977}.

Secondly, it is  verified directly 
 (see  Appendix~\ref{Appensec:Twomoreexamplesinvolving})
that Eqs.~\eqref{twiwj_CSInWNS} result in
\begin{align}
\label{tviwj_twivj_CSInWNS}
0\leq \overline{\tv_{\underlinei} \tv_{\underlinei}}(\bk),\quad
\big|\overline{\tv_i \tv_j}(\bk)\big|^2
\leq
\overline{\tv_{\underlinei} \tv_{\underlinei}}(\bk)\,
\overline{\tv_{\underlinej} \tv_{\underlinej}}(\bk),
\quad
\big|\overline{\tw_i \tv_j}(\bk)\big|^2
\leq
\tW_{\underlinei\underlinei}(\bk)\,
\overline{\tv_{\underlinej} \tv_{\underlinej}}(\bk).
\end{align}
Here, $\Vorticity_i$ denotes the vorticity fluctuation field,
\begin{align}
\label{Vorticityetc}
\Vorticity_i(\bx)=\epsilon_{imn}\w_{m,n}(\bx),\ \
\overline{\tv_i\tv_j}(\bk)=\epsilon_{imn}\epsilon_{jkl}k_nk_l\tW_{mk}(\bk),
\quad
 \overline{\tilde\w_i\tilde\Vorticity_j}(\bk)
 =\imaginary \epsilon_{jpq} k_q\tW_{ip}(\bk),
\end{align}
$\epsilon_{ijk}$ is the alternating tensor. 
Together with Eqs.~\eqref{twiwj_CSInWNS},
Eqs.~\eqref{tviwj_twivj_CSInWNS} guarantee satisfaction of
the integral constraints
generated by the application of
$\big|\overline{a b}\big|^2\leq \overline{a a}\,\overline{b b}$ to 
\begin{align*}
 \overline{\wi(\bx)\, \Vorticity_j(\by)},\
 \overline{\Vorticity_i(\bx)\, \Vorticity_j(\by)},\
\overline{\Vorticity_i(\bx)\, \Vorticity_j,_k(\by)},\, \ldots
\end{align*}

Thirdly, the structure of the aforementioned integral inequalities,
together with Eqs.~\eqref{twiwj_CSInWNS} and \eqref{tviwj_twivj_CSInWNS}, 
suggests that the following group be considered,
\begin{align}
\label{SFConstraintsRedundancyTest}
 \l|\tc(\bk)\r|^2\leq \ta(\bk)\,\tb(\bk),\quad 0\leq \ta(\bk),\quad 0\leq \tb(\bk);
\quad
 \l|\fc\r|^2\leq \fa\,\fb,\quad 0\leq\fa,\quad 0\leq \fb,
\end{align}
where $\fa$, $\fb$, and $\fc$ are non-stochastic functions of $\bk$, spatial vectors,
and time,
associated with $\ta$, $\tb$, and $\tc$, respectively.
Inequalities~\eqref{SFConstraintsRedundancyTest} 
can be operated on to obtain (see Appendix~\ref{Appensec:Twomoreexamplesinvolving})
\begin{align}
\l|\int_{\mathbb{R}^3}d\bk\,\tc(\bk)\fc\r|^2
\leq 
\int_{\mathbb{R}^3}d\bk\,\ta(\bk)\fa \int_{\mathbb{R}^3}d\bk\,\tb(\bk)\fb.
\label{SFConstraintsRedundancyTest01&02}
\end{align}
Inequalities~\eqref{SFConstraintsRedundancyTest}
and \eqref{SFConstraintsRedundancyTest01&02} are applicable to
the constraints generated by
$\big|\overline{a b}\big|^2\leq \overline{a a}\,\overline{b b}$
like the combinatoric sets from
\begin{align}
a, b\in
\big\{&
         \w_i(\by)-\w_i(\bx)+\alpha [\w_i(\bz')-\w_i(\bz)],
\ \
     \Vorticity_i(\bz')-\Vorticity_i(\bz)-\beta[\Vorticity_i(\by)-\Vorticity_i(\bx)]
\big\},
\end{align}
where $\alpha$ and $\beta$ are
 non-stochastic functions of time and spatial vectors
 (details in Appendix~\ref{Appensec:Twomoreexamplesinvolving}).

A general argument for the role of  Eqs.~\eqref{twiwj_CSInWNS}
in cases other than Eqs.~\eqref{SFConstraintsRedundancyTest}
is not yet obtained. The advantage of Eqs.~\eqref{twiwj_CSInWNS} 
is that they are local
(relative to Eq.~\eqref{SFConstraintsRedundancyTest01&02})
and they eliminate the  great
computational size and complexity associated with 
 the global integral and space-dependent 
 constraint~\eqref{SFConstraintsRedundancyTest01&02}.
The resulting simplification allows us to solve the second-order
model numerically.

\subsection{\label{Subsec:ConstraintsInvolvingFourthOrderCorrelations}
Constraints of inequality involving $\tW_{(ij)kl}$}

The third-order model includes constraints involving $\tW_{(ij)kl}$.
 One way to construct such constraints is to apply the Cauchy-Schwarz inequality
$\big|\overline{a b}\big|^2\leq \overline{a a}\,\overline{b b}$
 to a  correlation among $\w_i(\bx)$, $\q(\by)$, 
   and their  spatial and temporal derivatives, such as
\begin{align*}
\overline{\w_i(\bx) \w_j(\by) \w_k(\bz)},\ \
\overline{\q(\bx)\, \w_i(\by)},\ \
\overline{\w_i(\bx) \w_j(\bx) \w_k(\by) \w_l(\bz)},
\ \
\overline{\q(\bx)\, \q(\by)},\ \
\overline{q(\bx) \w_i(\by) \w_{j}\s(\bz)},\ \
\overline{\w_i(\bx) \partial\w_j(\by)/\partial t},\ \
\ldots
\end{align*}
The condition is that the resulting inequality involves only
 the state and control variables within the  model.
 Similarly, we may apply $\big|\overline{a b}\big|^2\leq \overline{a a}\,\overline{b b}$
to structure functions involving the above correlations.
 These constraints are not formulated here, since the present study focuses
  on the second-order model.
To reduce computational size,
two closely related issues pertinent to our investigation
are yet to be studied:  
(a) how to generate selectively
    the structure function-based inequalities without much redundancy;
(b) whether there are elementary inequalities
 analogous to Eqs.~\eqref{twiwj_CSInWNS}.

Within the third-order model, we may also construct constraints 
involving $\tW_{(ij)kl}$
by applying the non-negativity of variance of products,
\begin{align}
0\leq \overline{(ab-\overline{ab})^2},\quad
 (\overline{ab})^2\leq \overline{aabb}.
\label{NonNegativeVariance}
\end{align}
For instance, in the simple cases of
\begin{align*}
\big\{a=w_{i}(\bx),\ b=w_{j}(\by)\big\}\ \text{and}\
 \big\{a=\w_{\underlinei}(\bx)-\w_{\underlinej}(\bx), \
          b=\w_{\underlinei}(\by)+\w_{\underlinej}(\by)\big\},
\end{align*}
Eq.~\eqref{NonNegativeVariance}, along with 
Eqs.~\eqref{HomogeneitySymmetryInversionMirrorfs}, produces
\begin{align}
\label{wiwj_Deviation}
\big(\W_{i j}(\br)\big)^2\leq 
\W_{(\underlinei\underlinei)\underlinej\underlinej}(\br,\br), 
\quad
\big|\W_{\underlinei\underlinei}(\br)-\W_{\underlinej\underlinej}(\br)\big|^2
\leq
   \W_{(\underlinei\underlinei)\underlinei\underlinei}(\br,\br)
  +\W_{(\underlinej\underlinej)\underlinej\underlinej}(\br,\br)
  +2\W_{(\underlinei\underlinei)\underlinej\underlinej}(\br,\br)
  -4\W_{(\underlinei\underlinej)\underlinei\underlinej}(\br,\br).
\end{align}
More such inequalities may be produced via
\begin{align*}
a, b\in\big\{&
 \w_{\underlinei}(\by)+\alpha \w_{\underlinej}(\bx),\
\Vorticity_{\underlinei}(\by)+\alpha\, \Vorticity_{\underlinej}(\bx),\,
\w_j(\by)-\w_j(\bx)+\alpha [\w_j(\bz)-\w_j(\bx)],
\Vorticity_j(\by)-\Vorticity_j(\bx)+\alpha[\Vorticity_j(\bz)-\Vorticity_j(\bx)],\, 
  \ldots\big\}.
\end{align*}
 The  condition on the choice of $a$ and $b$ is that 
$\overline{ab}$ and
$\overline{aabb}$ are resolvable within the third-order model.

Since Eqs.~\eqref{vEvolutionOftWPrimarySOM} and \eqref{vEvolutionOftWijkPrimary_SOM}
imply that $\tW_{ij}$ depend linearly on $\tW_{(ij)kl}$
(peculiar to homogeneous turbulence),
the left-hand sides of Eqs.~\eqref{wiwj_Deviation} and \eqref{NonNegativeVariance}
   depend on $\tWAsy_{(ij)kl}$ quadratically
 while the right-hand sides depend on $\tWAsy_{(ij)kl}$ linearly.
It then follows that, along with constraints from  the Cauchy-Schwarz inequality,
Eqs.~\eqref{wiwj_Deviation} and \eqref{NonNegativeVariance}
 impose an upper bound on $|\W_{(ij)kl}|$ and in turn an
upper bound on $|\W_{ij}|$ and an upper bound on  $|\W_{ijk}|$.

\subsection{\label{subsec:ClosureObjectiveFunction}Closure and objective functional}

To make the models closed, a conventional scheme, like QN \cite{Orsazg1970}, 
 adds another set of equality constraints, such as approximating
the fourth order correlations in terms of the lower ones within the third-order model.
 The added ones act  effectively as exact equality constraints in  simulation
 and tend to be incompatible with some of the constraints discussed above,
 as demonstrated specifically 
 in the case of isotropic turbulence \cite{OBrienFrancis1962, Ogura1963}.
This incompatibility is also hinted at by the inequalities~\eqref{wiwj_Deviation}.
In the light of the large number and varied origins of  constraints,
it is expected that such an equality closure approach
cannot satisfy all of them.
These observations motivate us to pursue a framework
where all  the constraints
  and the dynamical equations are satisfied
  through the adoption of an objective  to be optimized,
  inspired by theories of optimal control and optimization.
This strategy  transfers the task of closure to 
the construction of an adequate objective.

Since the present framework involves ensemble statistical average
of the fluctuating velocity and pressure fields,
a probability density functional is required,
 parallel to Eqs.~\eqref{NSEqns}. However,
from the perspective of numerical simulation,
the domain of motion is approximated by a computational box,
 the fluctuation velocity and pressure fields are approximated
 by the use of a truncated Fourier series,
 an effective probability density function $f$ 
   may then be used to describe
  the discretized fluctuations statistically.
Next,  $f$ is  used to  construct a Shannon entropy.
Motivated by the underlying  idea of unbiased guess
and many successful applications
 \cite{Shannon1948, Jaynes1957, Jaynes1983, CoverThomas2006},
the maximization of Shannon entropy is employed 
to determine the structure of $f$;
The optimal  $f$ has the broadest spread-out.
The analysis is straight-forward but lengthy, 
and is presented in Appendix~\ref{Appensec:ShannonEntropy}.

It is impractical to implement the Shannon entropy maximization 
because of its
 formidable computational size and numerical integration of high dimensions.
The analysis, however, reveals the necessity of $\W_{kk}(\br)$
 as an alternative objective to be maximized,
  on the following grounds.
  (a) The optimal $f$ is characterized 
 partially by large values of the second order moments, 
 this results in overall large  $\W_{\underlinei\underlinei}(\br)$. 
 (b) Since $\W_{\underlinei\underlinei}(\br)$ are the lowest order correlations in the 
 framework, they satisfy the need
 that an objective characterize directly and collectively the effects of 
 all the dynamical equations 
and all the constraints.
(c) A scalar is required, independent of the coordinate systems, 
transforming  $\W_{\underlinei\underlinei}(\br)$ into  $\W_{kk}(\br)$.

To have a single value to evaluate, 
the $\br$-dependence of  $\W_{kk}(\br)$ may be resolved in two  ways:
$\W_{kk}(\bo)$ (proportional to turbulent energy per unit volume) and 
$\int_{\mathbb{R}^3}\W_{kk}(\br)\,d\br$ (overall).
They are related to each other through
\begin{align*}
\W_{kk}(\mathbf{0})=\int_{\mathbb{R}^3}d\br\,\W_{kk}(\br)\,\delta(\br),
\end{align*}
where $\delta(\br)$ is the Dirac delta. The relation illustrates the 
 local character of $\W_{kk}(\mathbf{0})$ (localized by  $\delta(\br)$)
 against the overall global character of
 $\int_{\mathbb{R}^3} d\br\, \W_{kk}(\br)$:
The maximization of the latter
produces a more uniform distribution of $\W_{kk}(\br)$
throughout the space $\br$, in contrast to the former 
(whose $\W_{kk}(\br)$ has much larger values in a neighborhood of $\br=\bo$,
because of $\max \W_{kk}(\bo)$).
This greater uniformity reflects the essence of the optimal $f$ being the broadest,
without unduly larger $\W_{kk}(\br)$ in a subregion.
Further, as presented in Appendix~\ref{Appensec:Turbulentenergyasobjectivefunction}, 
 simulation of the second-order model
in the asymptotic steady state under 
$\max\W_{kk}(\bo)$ produces the  numerical order pattern,
$\WAsy_{11}(\bo)>\WAsy_{22}(\bo)>\WAsy_{33}(\bo)$,
 which is inconsistent with 
the experimentally obtained $\WAsy_{11}(\bo)>\WAsy_{33}(\bo)>\WAsy_{22}(\bo)$ 
\cite{Piquet1999}.
Thus, we take
\begin{align}
 \int_{\mathbb{R}^3}d\br\, \W_{kk}(\br)
=&\,\int_{\mathbb{R}^3}d\br \int_{\mathbb{R}^3} d\bk \,\tW_{kk}(\bk)\cos(\br\cdot\bk)
\ \ 
\text{(to be maximized)}
\label{ObjectiveFunctionMaxIntWkkbr}
\end{align}
as the objective functional in the framework,
which can be expressed in terms of the primary control variables
in the second-order model
   (through the solution of Eqs.~\eqref{vEvolutionOftWPrimarySOM})
and in the third-order model 
 (through the solutions of Eqs.~\eqref{vEvolutionOftWPrimarySOM} 
 and \eqref{vEvolutionOftWijkPrimary_SOM}).

The above argument for Eq.~\eqref{ObjectiveFunctionMaxIntWkkbr} is 
essentially a mathematical application of Shannon entropy maximization.
We now interpret and substantiate the content of Eq.~\eqref{ObjectiveFunctionMaxIntWkkbr}
from a physical perspective, especially relate the integral structure
to the nonlocal nature of turbulent flows.
 A turbulent flow may be viewed physically as containing transient eddies
of various sizes. These eddies  occur irregularly in space and time
and consist of irregular regions of fluctuating velocity and vorticity,
they are responsible for  
the outstanding characteristic of 
turbulence -- its much greater ability to transport and mix  momentum,
kinetic energy, and contaminants than 
a corresponding laminar state and  molecular diffusion
\cite{DurbinPetterssonReif2011, TennekesLumley1972}.
Next, 
we apply this qualitative description
to homogeneous shear turbulence quantitatively, 
with the help of the two-point correlation 
$\overline{\w_k(\bx)\w_k(\bx+\br)}$
 which may be used to define integral length
scale in turbulence study.
 (i) Statistically, $\overline{\w_k(\bx)\w_k(\bx+\br)}$ is supposed to 
 characterize the turbulent eddies centered at $\bx$.
 The range of size of the eddies may be estimated  by the range of $\br$ where 
$\overline{\w_k(\bx)\w_k(\bx+\br)}$ is nontrivial.
 (ii) These $\bx$-centered eddies are directly responsible
 for the large-scale turbulent transport, 
 the collective spatial intensity  of these
 eddies may be roughly described by $\overline{\w_k(\bx)\w_k(\bx+\br)}$
 as a function of $\br$,
 reflecting that the larger the correlation value at $\br$, the stronger 
 the spreading and mixing.
 The contribution of these  $\bx$-centered eddies to the mixing 
  in $\mathbb{R}^3$ may be approximately quantified via
$\int_{\mathbb{R}^3}d\br\, \overline{\w_k(\mathbf{x})\w_k(\bx+\br)}$.
(iii) The collective contribution of all the  eddies in $\mathbb{R}^3$ to 
the turbulent mixing may be approximated 
formally through the unbounded
$\int_{\mathbb{R}^3} d\bx\int_{\mathbb{R}^3}d\br\, \overline{\w_k(\bx)\w_k(\bx+\br)}$
or effectively through
the finite $\int_{\mathbb{R}^3}d\br\, \overline{\w_k(\mathbf{0})\w_k(\br)}$,
because of homogeneity. 
(iv) Strong turbulent mixing corresponds to large value of 
$\int_{\mathbb{R}^3}d\br\, \overline{\w_k(\mathbf{0})\w_k(\br)}$,
compatible with and favoring the maximization operation
in  Eq.~\eqref{ObjectiveFunctionMaxIntWkkbr}.
Since $\W_{kk}(\mathbf{0})\geq |\W_{kk}(\mathbf{\br})|$ and $\W_{kk}(\mathbf{\br})$ may be
negative for some $\br\not=\mathbf{0}$, Eq.~\eqref{ObjectiveFunctionMaxIntWkkbr}
promotes  the positiveness of  $\W_{kk}(\mathbf{\br})$ as a whole.
This property is further examined in 
Subsec.~\ref{subsec:NumericalResultsandDiscussion}, in
the specific simulation of the second-order model,
with regard to Figs.~\ref{CCr1} through \ref{CCr3}.

\subsection{Basic features and SOCP}

Because this is an attempt to model  turbulence
via the interdisciplinary approach of optimal correlations,
we recap what has been presented so far
to help better explain the idea.
It is a hierarchical formulation of statistical turbulence modeling, composed of
multi-point spatial correlations. The issue of closure is resolved through
 maximization of the objective functional~\eqref{ObjectiveFunctionMaxIntWkkbr},
subject to the constraints of equality and inequality outlined previously.
Two models are constructed:
The second-order model takes the second order 
correlations as the state variables and the contracted and degenerate
third order correlations $\Contractedtoc_{ij}$ as the control variables;
The third-order model contains the correlations up to
the degenerate fourth order, the degenerate correlations  $\tW_{(ij)kl}$
 act as the control variables
and all the others as the state variables.
Therefore, statistical turbulence modeling 
is explored hierarchically via optimal control and convex optimization.

All the constraints of equality for the correlations
are intrinsic, either derived from
the fundamental equations~\eqref{NSEqns} or from the self-consistency
of correlation definitions or from the symmetries of inversion and mirror.
The constraints of inequality are generated
systematically from the Cauchy-Schwarz inequality
and the non-negativity of variance of products, they hold generally and
 the vast majority of them do not have simple physical interpretations.
 The mathematical nature of these constraints raises the question of
 why they are required within the framework.
We address this issue from several angles:
Firstly, they are built on the basis that the fluctuating velocity
and pressure fields are continuous and smooth, as assumed for 
the solutions of the Navier-Stokes equations,
and the averaging is meaningful. 
These are feasible and reasonable conditions.
Secondly, the generality of the constraints
 implies that the inequalities are compatible with
the dynamical equations~\eqref{vEvolutionOftWPrimarySOM} 
 and \eqref{vEvolutionOftWijkPrimary_SOM}, as demanded by
 the the framework's feasibility.
Thirdly, each inequality is loose individually owing to its generality,
the very many such inequalities may provide  collectively a rather tight
restriction to the structures of the correlations, together with the 
dynamical equations and other constraints.
Fourthly, the inequalities are not built from  specific experimental 
data, this indicates the potential to extend and apply the framework
to other turbulent flows.

The adoption of an objective functional
is to satisfy all the dynamical equations and 
constraints, which resolves the issue of non-realizability in the analytical
theories of turbulence.
The adequate objective functional to adopt is a major issue addressed in the work.
Inspired by the many successful applications of Shannon entropy maximization
in other fields,
we apply it here and  infer  Eq.~\eqref{ObjectiveFunctionMaxIntWkkbr} as the 
objective functional. We then proceed to substantiate the inferred form
on the basis that  physically turbulent eddies strongly promote
large-scale spreading and mixing of momentum 
and that statistically the correlation $\overline{\w_k(\bx)\w_k(\bx+\br)}$ 
and $\max\int_{\mathbb{R}^3}d\br\, \overline{\w_k(\mathbf{0})\w_k(\br)}$ may
 quantify this mechanism approximately.

In our modeling of homogeneous shear turbulence, 
the state of  motion is represented by  multi-point spatial correlations. 
The closure strategy of optimal control and the mathematical structures of
Eqs.~\eqref{vEvolutionOftWPrimarySOM}  and \eqref{vEvolutionOftWijkPrimary_SOM}
suggest that the highest order correlations are taken as the control variables
and the lower order ones as  state variables.
Further, the divergence-free condition offers grounds to introduce 
primary state and control variables.
All the state variables, the constraints, and the objective functional
can be represented in terms of the primary control variables,
 $\Contractedtoc_{ij}$ in the second-order model and
  $\tW_{(ij)kl}$ in the third-order model.
 That is, the second-order model is an optimization problem in the 
 space of primary  $\Contractedtoc_{ij}$ and 
 the third-order model an optimization problem in the 
 space of primary  $\tW_{(ij)kl}$.
 Further, since  $\Contractedtoc_{ij}$ can be 
 represented linearly in terms of  $\tW_{(ij)kl}$, we view 
 the third-order model as an optimization problem in the 
 enlarged control variable space. This viewpoint may also serve
to explain  mathematically why the third-order model is an improvement
 over the second-order model.
The determination of $\Contractedtoc_{ij}$ in the second-order model
or $\tW_{(ij)kl}$  in the third-order model
is achieved through maximizing the objective, 
subject to all the constraints relevant.
A specific implementation of the second-order model
is presented in Sec.~\ref{Sec:AsymptoticStatesofSOM}.
Because of statistical averaging, all the correlations and  relations 
in the framework are mathematically well-behaved and regular,
in contrast to the irregularly fluctuating velocity and pressure fields. 
Consequently, a discretization scheme of the correlations
and relations faces less demand 
than DNS regarding mesh size in space and time, among other advantages.

It is noticed that the dynamical equations~\eqref{vEvolutionOftWPrimarySOM} 
and \eqref{vEvolutionOftWijkPrimary_SOM} are linear,
 the constraints are either linear or quadratically convex,
 and the objective~\eqref{ObjectiveFunctionMaxIntWkkbr} is linear
 in terms of the primary control variables. Consequently,
 the second-order and the third-order models, in their discretized forms,
are second-order cone programs (SOCP) (see  Appendix~\ref{Appensec:SOCP}), 
according to the 
criteria set in \cite{Loboetal1998, AlizadehGoldfarb2003}.

Homogeneous shear turbulence is approximated mathematically as SOCP.
As a  well-formulated mathematical problem, its solutions, the multi-point
 correlations in the wave-number and physical spaces,
   are amenable numerically.
For example, the asymptotic distributions of
the second order correlations in the second-order model
are obtained via optimization (see  Figs.~\ref{CCr1} through \ref{CCr3}
in Subsec.~\ref{subsec:NumericalResultsandDiscussion}).
 Since the models are truncated,
    containing and resolving only lower order correlations,
 the adequacy of the solutions
needs to be evaluated against  DNS data
of homogeneous shear turbulence.
Also, the solved  correlations  in the wave-number and physical spaces
may be further analyzed to answer questions like
the separation between the length scales of production and 
dissipation of turbulent energy.

\section{\label{sec:AsymptoticStates}Asymptotic states}

To test the framework against experimental and DNS data,
it is applied to asymptotic states at large time that
  are characterized by the form,
\begin{align}
\psi=\psiAsy\exp(\sigma t).
\label{AsymptoticForm}
\end{align}
Here, $\psi$ represents any of 
\begin{align*}
 \Big\{&\tW_{ij},  \Contractedtoc_{ij}, \tWImag_{ijk}, \tW_{(ij)kl}, 
\W_{ij}, \W_{ijk}, \W_{(ij)kl}, \Q, \Q_i, \Q_{ij}, \ldots\Big\},
\end{align*}
$\psiAsy$ is the time-independent part of $\psi$, 
 and $\sigma$ a  constant 
 denoting the exponential growth rate of the correlations. 
 The structure is allowable by the
 mean flow~\eqref{AverageVelocityField},
  the dynamical equations, and the  constraints discussed.

 For the third-order model,
we obtain an upper bound for $\sigma$ by
 applying  Eqs.~\eqref{wiwj_Deviation} and \eqref{AsymptoticForm},
\begin{align}
\sigma\leq \max\sigma=0.
\label{MaxsigmaFromDeviationConstraint}
\end{align}
Therefore, the turbulent energy per unit volume, along with all the correlations,
  cannot grow indefinitely,
it evolves toward an asymptotic steady state of $\sigma=0$,
 if not decaying. 
 This result is more restrictive than the growth rate values suggested
 by some experimental and DNS data, e.g. $\sigma=0.10$, $0.12$, $0.14$, $0.20$,
 as summarized in Section 4.5 of \cite{Piquet1999} and Table 1 of 
 \cite{IsazaCollins2009}.
 
Since the second-order model cannot include constraints 
generated by the non-negativity of variance of products,
it allows $\sigma\in(0,1)$ to occur 
(see  Appendix~\ref{Appendsec:UpperBoundsImplications} for details).

Next, substitution of Eq.~\eqref{AsymptoticForm} into
Eqs.~\eqref{vEvolutionOftWPrimarySOM} and \eqref{vEvolutionOftWijkPrimary_SOM} yields
\begin{align}
\label{EvolutionOftWij_Asy}   
&
\frac{k_1\exp[-2 H(\sigma,\bk)]}{2|\bk|^4}
\frac{\partial}{\partial k_2}\Big(|\bk|^4\exp\big[2 H(\sigma,\bk)\big] \tWAsy_{22}(\bk)\Big)
=
-\frac{k_2k_l}{|\bk|^2}\ContractedtocAsy_{l2}(\bk)
+\ContractedtocAsy_{22}(\bk),
\notag\\[8pt]&
\frac{k_1\exp[-2 H(\sigma,\bk)]}{|\bk|^2}
 \frac{\partial}{\partial k_2}\Big(|\bk|^2\exp\big[2 H(\sigma,\bk)\big]\tWAsy_{12}(\bk) \Big)
\notag\\[-7pt]&
\\[-7pt]&\ \
=
 \bigg(1-\frac{2(k_1)^2}{|\bk|^2}\bigg)\tWAsy_{22}(\bk)
+\ContractedtocAsy_{12}(\bk)
+\ContractedtocAsy_{21}(\bk)
-\frac{k_l}{|\bk|^2}\Big( k_1\ContractedtocAsy_{l2}(\bk)
                         +k_2\ContractedtocAsy_{l1}(\bk)\Big),
\notag\\[8pt]&
\frac{k_1\exp[-2 H(\sigma,\bk)]}{2}
 \frac{\partial}{\partial k_2}\Big(\exp\big[2 H(\sigma,\bk)\big]\tWAsy_{11}(\bk)\Big) 
=
\bigg(1-\frac{2(k_1)^2}{|\bk|^2}\bigg)\tWAsy_{12}(\bk)
      -\frac{k_1k_l}{|\bk|^2}\ContractedtocAsy_{l1}(\bk)  
      +\ContractedtocAsy_{11}(\bk),
      \notag
\end{align}
and
\begin{align}
\label{EvolutionOftWijk_Asy}  
&
\frac{\exp[-H(\sigma,\bk,\bl)]}{|\bk|^2|\bl|^2|\bk+\bl|^2}
\bigg(k_1\frac{\partial}{\partial k_2}+l_1\frac{\partial}{\partial l_2}\bigg)
            \Big(|\bk|^2|\bl|^2|\bk+\bl|^2\exp\big[H(\sigma,\bk,\bl)\big]
            \tWImagAsy_{222}(\bk,\bl)\Big)
\notag\\&\
=
-(k_l+l_l)\tWAsy_{(2l)22}(\bk,\bl)
+(k_2+l_2)\frac{(k_m+l_m)(k_l+l_l)}{|\bk+\bl|^2}\,\tWAsy_{(lm)22}(\bk,\bl)
+\bigg( k_l\tWAsy_{(2l)22}
       -k_2\frac{k_mk_l}{|\bk|^2}\,\tWAsy_{(lm)22}\bigg)(-\bk-\bl,\bl)
\notag\\&\hskip5mm
+\bigg( l_l\tWAsy_{(2l)22}     
       -l_2\frac{l_m\,l_l}{|\bl|^2}\tWAsy_{(lm)22}\bigg)(-\bk-\bl,\bk),
\notag\\[8pt]&
\frac{\exp[-H(\sigma,\bk,\bl)]}{|\bk|^2|\bl|^2}
\bigg(k_1\frac{\partial}{\partial k_2}+l_1\frac{\partial}{\partial l_2}\bigg)
       \Big(|\bk|^2|\bl|^2\exp\big[H(\sigma,\bk,\bl)\big]\tWImagAsy_{122}(\bk,\bl)\Big)
\notag\\&\
=
 \bigg(1-\frac{2(k_1+l_1)^2}{|\bk+\bl|^2}\bigg)\tWImagAsy_{222}(\bk,\bl)
-(k_l+l_l)\tWAsy_{(1l)22}(\bk,\bl) 
+(k_1+l_1)\frac{(k_m+l_m)(k_l+l_l)}{|\bk+\bl|^2}\,\tWAsy_{(lm)22}(\bk,\bl)
\notag\\&\hskip5mm
+\bigg( k_l\,\tWAsy_{(2l)12}
       -k_2\frac{k_mk_l}{|\bk|^2}\tWAsy_{(lm)12}\bigg)(-\bk-\bl,\bl)
+\bigg( l_l\tWAsy_{(2l)12}     
       -l_2\frac{l_ml_l}{|\bl|^2}\tWAsy_{(lm)12}\bigg)(-\bk-\bl,\bk), 
\notag\\[8pt]&
\frac{\exp[-H(\sigma,\bk,\bl)]}{|\bl|^2}
 \bigg(k_1\frac{\partial}{\partial k_2}+l_1\frac{\partial}{\partial l_2}\bigg)
            \Big(|\bl|^2\exp\big[H(\sigma,\bk,\bl)\big]\tWImagAsy_{112}(\bk,\bl)\Big)
\notag\\[-6pt]&
\\[-6pt]&\ \
=
 \bigg(1-\frac{2 (k_1+l_1)^2}{|\bk+\bl|^2}\bigg)\tWImagAsy_{122}(-\bk-\bl,\bl)
+\bigg(1-\frac{2(k_1)^2}{|\bk|^2}\bigg)\tWImagAsy_{122}(\bk,\bl)
-(k_l+l_l)\tWAsy_{(1l)12}(\bk,\bl) 
\notag\\&\hskip5mm
+(k_1+l_1)\frac{(k_m+l_m)(k_l+l_l)}{|\bk+\bl|^2}\,\tWAsy_{(lm)12}(\bk,\bl)
+\bigg( k_l\tWAsy_{(1l)12}
       -k_1\frac{k_mk_l}{|\bk|^2}\tWAsy_{(lm)12}\bigg)(-\bk-\bl,\bl)
\notag\\&\hskip5mm
+\bigg( l_l\tWAsy_{(2l)11}    
       -l_2\frac{l_ml_l}{|\bl|^2}\tWAsy_{(lm)11}\bigg)(-\bk-\bl,\bk),
\notag\\[8pt]&
\exp\big[-H(\sigma,\bk,\bl)\big]
\bigg(k_1\frac{\partial}{\partial k_2}+l_1\frac{\partial}{\partial l_2}\bigg)
          \Big(\exp\big[H(\sigma,\bk,\bl)\big]\tWImagAsy_{111}(\bk,\bl)\Big)
\notag\\&\
=
 \bigg(1-\frac{2(k_1+l_1)^2}{|\bk+\bl|^2}\bigg)\,\tWImagAsy_{112}(\bl,-\bk-\bl)
+\bigg(1-\frac{2(k_1)^2}{|\bk|^2}\bigg)\,\tWImagAsy_{112}(\bl,\bk)
\notag\\&\hskip5mm
+\bigg(1-\frac{2(l_1)^2}{|\bl|^2}\bigg)\,\tWImagAsy_{112}(\bk,\bl)
-(k_l+l_l)\tWAsy_{(1l)11}(\bk,\bl)
+(k_1+l_1)\frac{(k_m+l_m)(k_l+l_l)}{|\bk+\bl|^2}\,\tWAsy_{(lm)11}(\bk,\bl)
\notag\\&\hskip4mm
+\bigg( k_l\tWAsy_{(1l)11}
       -k_1\frac{k_mk_l}{|\bk|^2}\,\tWAsy_{(lm)11}\bigg)(-\bk-\bl,\bl)
+\bigg( l_l\tWAsy_{(1l)11}     
       -l_1\frac{l_ml_l}{|\bl|^2}\,\tWAsy_{(lm)11}\bigg)(-\bk-\bl,\bk).
       \notag
\end{align}
Here,
\begin{align*}
 H(\sigma,\bk,\bl)
=
 H(2\sigma/3,\bk)
+H(2\sigma/3,\bl)
+H(2\sigma/3,\bk+\bl),
\quad
 H(\sigma,\bk)
=
-\frac{k_2}{k_1}\bigg(\frac{1}{2}\sigma+(k_1)^2+(k_3)^2+\frac{1}{3}\,(k_2)^2\bigg).
\end{align*}
These equations may be solved
under the boundary conditions from Eqs.~\eqref{BoundaryConditionOftWij}
and \eqref{BoundaryConditionOftWijk},
\begin{subequations}
\begin{align}
\label{BoundaryConditionAInftyFortWij}
\lim_{k_2\rightarrow \pm\infty}\tWAsy_{ij}(\bk)=0,
\end{align}
\vspace{-5mm}
\begin{align}
\label{BoundaryConditionAInftyFortWijk}
\lim_{k_2\rightarrow \pm\infty}\tWImagAsy_{ijk}(\bk,\bl)=
\lim_{l_2\rightarrow \pm\infty}\tWImagAsy_{ijk}(\bk,\bl)=0.
\end{align}
\end{subequations}
The constraints corresponding to the asymptotic states can be derived from 
the aforementioned ones
with the help of Eq.~\eqref{AsymptoticForm}, the objective functional
\eqref{ObjectiveFunctionMaxIntWkkbr} is replaced with
\begin{align}
\label{ObjectiveFunctionMaxIntWkkbrAsymp}
 \int_{\mathbb{R}^3}d\br\, \WAsy_{kk}(\br)&
=\int_{\mathbb{R}^3}d\br \int_{\mathbb{R}^3} d\bk\, \tWAsy_{kk}(\bk)\cos(\br\cdot\bk)
\ \ 
\text{(to be maximized)},
\end{align}
owing to the linear nature of the functional and fixed $\sigma$.

One great challenge to the third-order model is its computational size,
even restricted to the asymptotic state of $\sigma=0$.
There are  thirty primary control variables,
 $\tWAsy_{(ij)kl}(\bk,\bl)$, listed in
 Eqs.~\eqref{DivergenceFreeInPhysicalSpace_wiwjwkwl_fs}
 and defined in  6-dimensional  space;
 the number of corresponding discrete control variables 
 under a moderate mesh size
 is of the order of $10^9$ or higher.
Next, there are $\WAsy_{ijk}(\br,\bs)$,
 $\WAsy_{(ij)kl}(\br,\bs),$ and 
$\QAsy_{ij}(\br,\bs)$  to be computed 
at collocation points of similar order of magnitude,
for the purpose of enforcing the constraints outlined in
Subsec.~\ref{Subsec:ConstraintsInvolvingFourthOrderCorrelations}.
 This computational size is of huge-scale
  according to the criteria listed in \cite{Nesterov2014}
 and poses challenges to both 
   hardware (computer memory, number of computers, etc.)
 and  software (a solver, especially the algorithm
  for such a huge-scale problem).
 These issues need further study
 in the context of randomized  algorithms, distributed computing, etc.
  There is a close connection between the third-order model and
 big data  \cite{Cevheretal2014}, and research 
 on big data is expected to help explore the model.
 Accordingly the third-order model  may be 
 classified as a big model.

 This issue of huge-scale and its computational challenge raises a serious 
question about  the prospect and merit of the framework proposed,
especially compared with DNS, which has been employed to 
study homogeneous shear turbulence
 (see the references in \cite{Sekimotoetal2016}). It is observed that,
as it resolves all structure scales of a turbulent flow,
DNS is a fundamental tool in turbulence research, 
 but it faces its own challenges.
 The problem most relevant to the present study is discussed below.

 The  Navier-Stokes equations are recognized as a chaotic dynamical system
 \cite{Bohretal1998, HolmesLumleyBerkooz1996},
  they govern the continuous turbulent fields of velocity and pressure.
  As an approximation to the continuous fields,
  DNS produces   essentially discrete distributions,
  its truncation and computational errors  act as a
disturbing source to the continuous fields.
This disturbing source affects quantitatively the individual
solutions or realizations under given initial and boundary conditions
solved by DNS.
The question is whether this disturbance qualitatively 
  affects the average fields predicted by DNS,
  because of the chaotic behavior of the Navier-Stokes equations.
  
 In the case of homogeneous shear turbulence of concern here,
 DNS studies suggest that
 $\sigma=0.10, 0.18, 0.20$ (see Table 1 of \cite{IsazaCollins2009})
 and ``Ideal HST in unbounded domains grows indefinitely, both in intensity 
  and length scale. During the initial stages of shearing
  an isotropic turbulent flow, linear effects result in algebraic
growth of the turbulent kinetic energy, which is later transformed to exponential 
due to nonlinearity.'' (p.2 of \cite{Sekimotoetal2016}).
These results are inconsistent with
the stringent upper bound requirement~\eqref{MaxsigmaFromDeviationConstraint}.
Also, the structures of $\tWAsy_{ij}(\bk)$
and the oscillation of $\cos(\bk\cdot\br)$
in the Fourier transforms may yield that 
$\WAsy_{ij}(\br)$ have an effectively finite support;
 the length scales of the support
 are implicitly determined by the structure of the models,
  unlike the explicit timescale $S^{-1}$ in Eqs.~\eqref{AverageVelocityField}.
To demonstrate this spatial boundedness,
   a simulation and  analysis is presented 
in Sec.~\ref{Sec:AsymptoticStatesofSOM}.
The above inconsistencies highlight the difficulty faced by DNS 
in simulating adequately asymptotic states of
homogeneous shear turbulence. In comparison, the present framework has
   a potential advantage: 
  its spatial correlations are regular mathematical functions 
  without chaotic behavior, its mathematical structure can be specialized
  to exploit homogeneity,
  and its average turbulent flow structure is controlled directly by
  generic constraints from the Cauchy-Schwarz inequality
  and the non-negativity of variance of products, the latter
  guarantees Eq.~\eqref{MaxsigmaFromDeviationConstraint}. 
  The trade-off for this advantage is the huge-scale computational
  size of the third-order model.
  
Furthermore, as discussed in Sec.~\ref{sec:Conclusion},
there is a possibility to generalize the framework to
 inhomogeneous turbulence,
 which may provide one more tool to tackle turbulent flows.
 Such a possibility may also serve to justify   exploration
 of this huge-scale model.

\section{\label{Sec:AsymptoticStatesofSOM}Asymptotic steady state of the second-order model}

The asymptotic steady state solution of the second-order model
is studied numerically,
owing  to its computational feasibility at present and its ability to provide
  valuable information concerning
  the adequacy of the objective functional,
  the role of the non-negativity of variance of products,
 and the potential of the framework.

As a special case of Eq.~\eqref{AsymptoticForm},
the asymptotic states of the second-order model are characterized by
\begin{align}
\big\{\tW_{ij}, \Contractedtoc_{ij}\big\}(\bk,t)
=\Big\{\tWAsy_{ij}, \ContractedtocAsy_{ij}\Big\}(\bk)\exp(\sigma t).
\label{AsymptoticFormSOM}
\end{align}
Further, the equations governing $\tWAsy_{ij}$ and $\ContractedtocAsy_{ij}$ 
can be derived from the relevant ones discussed previously
and are collected below.

\paragraph{Consequence of symmetry}
Equations~\eqref{HomogeneitySymmetryInversionMirrorfs}
   and \eqref{ContractedSymmetryInversionMirrorfs} 
   imply that it is sufficient
to solve $\tWAsy_{ij}$ and $\ContractedtocAsy_{ij}$ in
the subdomain $\{k_1\leq 0, k_3\geq 0, k_2\in\mathbb{R}\}$,
which is adopted.
As intended, Eqs.~\eqref{HomogeneitySymmetryInversionMirrorfs} produce
the experimentally observed \cite{Piquet1999, SagautCambon2008},
\begin{align}
 \WAsy_{13}(\bo)=\WAsy_{23}(\bo)=0.
\end{align}

\paragraph{Primary variables}
Equations~\eqref{vHST_DivergenceFreeInPhysicalSpace_wiwjwk_fs_SOM}
 and \eqref{DivergenceFreeInPhysicalSpace_wiwj_fs} imply that 
 $\ContractedtocAsy_{ij}$, $i=1,2,3$, $j=1,2$, are taken as the 
primary variables of $\ContractedtocAsy_{mn}$ and
  $\tWAsy_{11}$, $\tWAsy_{12}$, and $\tWAsy_{22}$ as the primary components
of $\tWAsy_{ij}$.
The boundedness of $\tWAsy_{ij}$ and $\ContractedtocAsy_{ij}$ and
the structures of Eqs.~\eqref{vHST_DivergenceFreeInPhysicalSpace_wiwjwk_fs_SOM}
 through \eqref{vEvolutionOftWPrimarySOM}
suggest the transformations,
\begin{subequations}
\label{GammijtWiiTransformed}
\begin{align}
\label{vContrcatedtoc_Transformed}
\ContractedtocAsy_{ij}(\bk)=(k_3)^2\,\dContractedtocAsy_{ij}(\bk),\ \
 i=1, 2, 3,\ j=1, 2;
\end{align}
\vspace{-6mm}
\begin{align}
\label{tWij_Transformed}
\tWAsy_{ij}(\bk)=(k_3)^2\,\dWAsy_{ij}(\bk),\ \
ij=11, 12, 22.
\end{align}
\end{subequations}
That is,  $\dContractedtocAsy_{ij}$, $i=1,2,3$, $j=1,2$, act as the
primary control variables, which may be observed clearly in
Eqs.~\eqref{dWij_Asy_Sol}.

\paragraph{Formal solution of dynamical equations}
With the aid of Eqs.~\eqref{GammijtWiiTransformed} and \eqref{BoundaryConditionAInftyFortWij},
 we solve Eqs.~\eqref{EvolutionOftWij_Asy} to get
\begin{align}
\label{dWij_Asy_Sol}
\dWAsy_{22}(\bk)
=\,&
 \frac{2}{|k_1||\bk|^4}
 \int_{-\infty}^{k_2} dk'_2
   \Expal(\bk;k'_2;\sigma) |\bk'|^4
\bigg( \frac{k'_2 k'_l}{|\bk'|^2}\dContractedtocAsy_{l2}(\bk')-\dContractedtocAsy_{22}(\bk')\bigg),
\notag\\[6pt]
 \dWAsy_{12}(\bk)
=\,&
-\frac{2}{|k_1|^2[|k_1|^2+(k_3)^2]|\bk|^2}
 \int_{-\infty}^{k_2}dk'_2 
                  \Expal(\bk;k'_2;\sigma)
                  [F(\bk)-F(\bk')]\, |\bk'|^4
  \bigg(\frac{k'_2 k'_l}{|\bk'|^2}\dContractedtocAsy_{l2}(\bk')-\dContractedtocAsy_{22}(\bk')\bigg)
\notag\\&
+\frac{1}{|k_1| |\bk|^2}\int_{-\infty}^{k_2}dk'_2 
                      \Expal(\bk;k'_2;\sigma)
\bigg[
      k'_l\Big(k_1 \dContractedtocAsy_{l2}(\bk')+k'_2 \dContractedtocAsy_{l1}(\bk')\Big)
    -|\bk'|^2\Big(\dContractedtocAsy_{12}(\bk')+\dContractedtocAsy_{21}(\bk')\Big)
\bigg],
\notag\\[6pt]
       \dWAsy_{11}(\bk)
=\,&
 \frac{2}{|k_1|^3 [|k_1|^2+(k_3)^2]^2}
 \int_{-\infty}^{k_2}dk'_2
           \Expal(\bk;k'_2;\sigma)
           [F(\bk)-F(\bk')]^2 |\bk'|^4
\bigg( \frac{k'_2\,k'_l}
         {|\bk'|^2}\dContractedtocAsy_{l2}(\bk')-\dContractedtocAsy_{22}(\bk')\bigg)
\\&
-\frac{2}{|k_1|^2 [|k_1|^2+(k_3)^2]}
 \int_{-\infty}^{k_2}dk'_2
                      \Expal(\bk;k'_2;\sigma)
                      [F(\bk)-F(\bk')]\,
\notag\\&\hskip40mm\times 
\bigg[
  k'_l \Big(k_1\dContractedtocAsy_{l2}(\bk')
              +k'_2\dContractedtocAsy_{l1}(\bk')\Big)
-|\bk'|^2\Big(\dContractedtocAsy_{12}(\bk')+\dContractedtocAsy_{21}(\bk')\Big)
\bigg] 
\notag\\&
+\frac{2}{|k_1|}
  \int_{-\infty}^{k_2}dk'_2
          \Expal(\bk;k'_2;\sigma)
\bigg( \frac{k_1 k'_l}{|\bk'|^2} \dContractedtocAsy_{l1}(\bk')  
      -\dContractedtocAsy_{11}(\bk')
\bigg).
\notag
\end{align}
Here, $\bk'=(k_1,k'_2,k_3)$, $k_1<0$, and
\begin{subequations}
\begin{align}
\label{ExponentialFunctionPart}
\Expal(\bk;k'_2;\sigma)
=
\exp\bigg[
\frac{2 (k_2-k'_2)}{k_1}\bigg(\frac{\sigma}{2}+(k_1)^2+(k_3)^2
+\frac{(k_2)^2+(k'_2)^2+k_2k'_2}{3}\bigg)
\bigg],
\end{align}
\vspace{-1em}
\begin{align}
 F(\bk)=
 \frac{(k_3)^2}{\big[|k_1|^2+(k_3)^2\big]^{1/2}}
               \arctan\frac{k_2}{\big[|k_1|^2+(k_3)^2]^{1/2}}
          -\frac{|k_1|^2 k_2}{|\bk|^2}.
\end{align}
\end{subequations}

\paragraph{Constraints}
Equations~\eqref{twiwj_CSInWNS} become
\begin{align}
0\leq \tWAsy_{\underlinei \underlinei}(\bk),\ \
 \Big|\tWAsy_{ij}(\bk)\Big|^2
\leq
\tWAsy_{\underlinei \underlinei}(\bk)\,\tWAsy_{\underlinej\underlinej}(\bk),
\label{ElementaryInequalitiesFortwiwj_Asy}
\end{align}
they constrain $\dContractedtocAsy_{ij}$
via Eqs.~\eqref{DivergenceFreeInPhysicalSpace_wiwj_fs},
\eqref{tWij_Transformed}, and \eqref{dWij_Asy_Sol}.
Symmetries~\eqref{ContractedSymmetryInversionMirrorfs} require that
\begin{align}
\label{CVgamma_TransformedAsy_SymmetryAboutk3}
\frac{\partial \dContractedtocAsy_{ij}(k_1,k_2,k_3)}{\partial k_3}\bigg|_{k_3=0}=0,\ \
ij=11,21,12,22;\quad
\dContractedtocAsy_{ij}(k_1,k_2,0)=0,\ \ ij=31,32.
\end{align}
The relevant ones in Eqs.~\eqref{HomogeneitySymmetryInversionMirrorfs} 
are satisfied automatically.
The global constraints~\eqref{vIntrinsicEquality_kk_jjk_Half_SOM} reduce to
\begin{align}
\label{dCVgamma_Constraint_Integral_Asy}
&
\int_{-\infty}^0 dk_1 \int_{0}^{+\infty} dk_3
    \int_{\mathbb{R}} dk_2(k_3)^2\dContractedtocAsy_{11}(\bk)=0,
\quad
\int_{-\infty}^0 dk_1 \int_{0}^{+\infty} dk_3
    \int_{\mathbb{R}} dk_2(k_3)^2\dContractedtocAsy_{22}(\bk)=0,
\notag\\&
\int_{-\infty}^0 dk_1 \int_{0}^{+\infty} dk_3 
      \int_{\mathbb{R}} dk_2
      (k_3)^2
   \Big(\dContractedtocAsy_{12}(\bk)+\dContractedtocAsy_{21}(\bk)\Big)
=0,
\\&
\int_{-\infty}^0 dk_1 \int_{0}^{+\infty} dk_3 
      \int_{\mathbb{R}} dk_2
      k_3
\Big(
 k_1 \dContractedtocAsy_{31}(\bk)
+k_2 \dContractedtocAsy_{32}(\bk)
\Big)
=0.
\notag
\end{align}
The  model possesses scaling invariance under
\begin{align}
\big\{\tWAsy_{ij}, \ContractedtocAsy_{ij}\big\}
  \rightarrow
\lambda\big\{\tWAsy_{ij}, \ContractedtocAsy_{ij}\big\},
\quad \forall \lambda>0.
\label{ScalingSOMAsy}
\end{align}
It implies that, to obtain definite solutions, bounds 
need to be imposed  on the control variables explicitly like
\begin{align}
\big|\dContractedtocAsy_{ij}(\bk)\big|\leq C=1,\ \ i=1,2,3,\ j=1,2.
 \label{tocBounds}
\end{align}
 Though the resulting
   solutions do not yield absolute distributions of $\tWAsy_{ij}(\bk)$,
 they provide definite normalized quantities represented
 by the anisotropy tensor,
\begin{align}
b^{(\infty)}_{ij}
=\frac{\WAsy_{ij}(\bo)}{\WAsy_{kk}(\bo)}-\frac{1}{3}\,\delta_{ij},
\label{AnisotropicTensor}
\end{align}
which is used to compare against experimental data.

Equality constraint~\eqref{vIntrinsicRelationForUkk(0)Evolution}
reduces to
\begin{align}
\frac{\sigma}{2}\WAsy_{jj}(\bo)
+\WAsy_{12}(\bo)
+\overline{\w_j,_k\!(\bx)  \w_j,_k\!(\bx)}^{(\infty)}=0.
\label{IntrinsicRelationForUkk(0)Asy}
\end{align}
Its satisfaction  is used to
check partially the adequacy of numerical solutions.

\paragraph{Removal of singularity}
Under the supposedly bounded and continuous
distributions of $\dContractedtocAsy_{ij}$ and $\dWAsy_{ij}$,
the singularity of Eqs.~\eqref{dWij_Asy_Sol}
  at $k_1=0$ is apparent. It may be shown that
$\dContractedtocAsy_{ij}(\bk)=\dWAsy_{ij}(\bk)=0$  at $k_1=0$ 
(see Appendix~\ref{Appendsec:LimitingBehavior} for details).
Considering computational feasibility,
we restrict the supports of $\dContractedtocAsy_{ij}$ and $\dWAsy_{ij}$
to $k_1\leq \max k_{1}<0$ with $|\max k_{1}|$ being small.
This simple treatment is adequate for the state $\sigma=0$.

\paragraph{Support estimates}
It is essential to the simulation that
the domains of the control variables 
$\dContractedtocAsy_{ij}$  and the state variables $\dWAsy_{ij}$
 be approximated by bounded supports.
The existence of such supports may be inferred from
the physical requirement that $\WAsy_{kk}(\mathbf{0})$ be finite
as follows.
With the use of
\begin{align*}
\Big|\WAsy_{12}(\bo)\Big| 
 \leq \frac{1}{2}\Big(\WAsy_{11}(\bo)+\WAsy_{22}(\bo)\Big),
\end{align*}
Eq.~\eqref{IntrinsicRelationForUkk(0)Asy} is cast in the form
\begin{align}
\label{WAsykk_Primary}
 \int_{-\infty}^0 dk_1 \int_{\mathbb{R}} dk_2 \int_0^{+\infty} dk_3
 \big(1-\sigma-2|\bk|^2\big)\, \tWAsy_{kk}(\bk)
\geq \WAsy_{33}(\bo)/4,
\end{align}
which, along with  $\WAsy_{33}(\bo)> 0$ and $\tWAsy_{kk}(\bk)\geq 0$, suggests that
\begin{align}
 \tWAsy_{kk}(\bk)\ &
\text{is large predominantly inside a neighborhood of}\  |\bk|^2\leq (1-\sigma)/2.
\label{tWAsykk_PredominantRegion_A}
\end{align}
It is then inferred from 
\begin{align}
\tWAsy_{kk}(\bk) 
=
 \big[(k_1)^2+(k_3)^2\big]\dWAsy_{11}(\bk)
+2 k_1 k_2 \dWAsy_{12}(\bk)
+\big[(k_2)^2+(k_3)^2\big]\dWAsy_{22}(\bk) 
\label{tWkk_tW112212}
\end{align}
and Eqs.~\eqref{ElementaryInequalitiesFortwiwj_Asy}
and \eqref{tWAsykk_PredominantRegion_A} that
\begin{align}
 |\dWAsy_{ij}(\bk)|\ &
 \text{are large predominantly inside a neighborhood of}\  |\bk|^2\leq (1-\sigma)/2.
\label{tWAsyij_Distribution_Primary}
\end{align}
The above relation may be further justified as follows.
(i) The presence of $|\bk|^2$ results from the viscous dissipation
term of Eq.~\eqref{IntrinsicRelationForUkk(0)Asy}, its upper bound
may reflect the effect of the  viscous dissipation.
(ii) The more negative the value of $\sigma$, the greater the support
of $\dWAsy_{ij}(\bk)$. This dependence is physically sound, 
since larger wave-numbers tend to carry less fluctuation turbulent energy
but produce more viscous dissipation and lead to greater 
exponential rate of decay.

The integral structures of solutions~\eqref{dWij_Asy_Sol}
and the requirement~\eqref{tWAsyij_Distribution_Primary} suggest
 $|\bk|\leq [(1-\sigma)/2]^{1/2}$ as a preliminary estimate
for the support of $\dContractedtocAsy_{ij}$.
Considering the rough nature of this estimate,
the restriction on computational size from the aspect of computational feasibility,
and the convenience of a structured mesh for coding, we take
 the hexahedral estimate for the effective support of
 $\dContractedtocAsy_{ij}$,
\begin{align*}
&
 |k_1|\leq \lambda_1 [(1-\sigma)/2]^{1/2},\quad
 |k_3|\leq \lambda_3 [(1-\sigma)/2]^{1/2},
\quad
k_2\in \big[\LowerSupportBoundForgammaij(\sigma),
            \UpperSupportBoundForgammaij(\sigma)\big];
\ \ 
\LowerSupportBoundForgammaij(\sigma)
 \equiv-\lambda_{2L} [(1-\sigma)/2]^{1/2},
\notag\\&
 \UpperSupportBoundForgammaij(\sigma)
 \equiv \lambda_{2U} [(1-\sigma)/2]^{1/2},
\quad
\lambda_i \in [1,3].
\end{align*}
The incorporation of $k_1\leq\max k_1$ 
and the restriction to $\sigma=0$
lead to the finite support estimate of $\dContractedtocAsy_{ij}$,
\begin{align}
\label{tCVgammaAsyij_SupportEstimate}
\EstimatedSupportForgammaij(0)
=\,&
\bigg\{\bk:\ 
k_1\in \bigg[-\frac{\lambda_1}{2^{1/2}}, \max k_1\bigg],\
k_3\in \bigg[0, \frac{\lambda_3}{2^{1/2}}\bigg], \ 
k_2\in \Big[\LowerSupportBoundForgammaij(0),
               \UpperSupportBoundForgammaij(0)\Big]
\bigg\},\ \ 
    \lambda_i \in [1,3].
\end{align}
The specific values of $\lambda_i$ are fixed from consideration of
  adequacy of numerical solutions via trial-and-error.

For an estimate of the support of $\dWAsy_{ij}$, 
denoted as $\EstimatedSupportForWij(0)$,
 solutions~\eqref{dWij_Asy_Sol} indicate
 that the two supports coincide with each other in the directions of $k_1$ and $k_3$,
 and  $\LowerSupportBoundForgammaij(0)$ 
 is the lower bound  to  $\EstimatedSupportForWij(0)$ along $k_2$. 
 Regarding the upper bound of $\EstimatedSupportForWij(0)$ along $k_2$,
   denoted as $\max k^{U}_2$, 
 the integral nature of the solutions implies that 
 $\max k^{U}_2>\UpperSupportBoundForgammaij(0)$. In  simulation,
  $\max k^{U}_2$ is treated as a constant whose value is fixed
 via numerical trials. Thus,
\begin{align}
\label{tWAsyij_SupportEstimate}
\EstimatedSupportForWij(0)=
\bigg\{\bk:\ 
k_1\in \bigg[-\frac{\lambda_1}{2^{1/2}}, \max k_1\bigg],\ 
k_3\in \bigg[0, \frac{\lambda_3}{2^{1/2}}\bigg],\
k_2\in \Big[\LowerSupportBoundForgammaij(0),
               \max k^{U}_2\Big]
\bigg\}.
\end{align}

\paragraph{Objective}
The distributions of $\tWAsy_{ij}(\bk)$
 and the oscillation of $\cos(\bk\cdot\br)$
in the Fourier transforms may result in $\WAsy_{ij}(\br)$
being negligibly small at large $|\br|$.
Thus, Eq.~\eqref{ObjectiveFunctionMaxIntWkkbrAsymp} becomes
\begin{align}
\label{ObjectiveFunctionMaxIntWkkbrAsy}
 \int_{\mathbb{R}^3}d\br\,\WAsy_{kk}(\br)
=
 32\int_{-\infty}^{0} dk_1 
  \int_{-\infty}^{+\infty} dk_2\int_{0}^{+\infty} dk_3\,
    \tWAsy_{kk}(\bk)\,
    \frac{\sin(L_1k_1) \sin(L_2k_2) \sin(L_3k_3)}{k_1 k_2 k_3}.
\end{align}
Here, $\prod_{k=1}^{3}[-L_k, L_k]$
denotes the effective support for $\WAsy_{kk}(\br)$.
The values of $L_k$ indicate the macro length scales
beyond which the spatial correlations are negligible,
which is tested and fixed below numerically.

\paragraph{SOCP}
Constrained by Eqs.~\eqref{dWij_Asy_Sol},
\eqref{ElementaryInequalitiesFortwiwj_Asy} through
\eqref{dCVgamma_Constraint_Integral_Asy},
\eqref{tocBounds}, \eqref{tCVgammaAsyij_SupportEstimate},
and \eqref{tWAsyij_SupportEstimate},
the maximization problem~\eqref{ObjectiveFunctionMaxIntWkkbrAsy}
is a SOCP in its discretized form.

\subsection{\label{subsec:DiscretizationAndComputation}Discretization and computation}

The basic ideas involved in the numerical simulation are outlined here,
 the lengthy but straight-forward 
  details are presented in Appendices~\ref{Appensec:NumericalDiscretization}
 through \ref{Appensec:Algorithm}.

\paragraph{Discretization and integrations}
The bounded supports for $\dContractedtocAsy_{ij}$  and  $\dWAsy_{ij}$
are specified, respectively, as
\begin{align}
\label{CVSVSupport_OS}
 \EstimatedSupportForgammaij(0)
=
[-0.71, -0.01]\times
[-1.35, 1.25]
\times
[0.0,\,0.9],\quad
 \EstimatedSupportForWij(0)
=
[-0.71, -0.01]\times
[-1.35, 1.65]
\times
[0.0, 0.9],
\end{align}
with the help of support estimates~\eqref{tCVgammaAsyij_SupportEstimate}
and \eqref{tWAsyij_SupportEstimate}.
In  physical space,
 $L_1=250$, $L_2=60$, and $L_3=150$ are
  selected for Eq.~\eqref{ObjectiveFunctionMaxIntWkkbrAsy}.
 We need to specify these values through numerical trial tests 
  because  complicated nonlinear relations
 are involved in the optimization problem. 
The supports are discretized, respectively, with  structured hexahedral meshes,
\begin{align*}
&
\EstimatedSupportForgammaij(0)
=
\bigcup_{n_1=1}^{N_1-1} 
\bigcup_{n_2=1}^{N_2-1} 
\bigcup_{n_3=1}^{N_3-1} 
{\cal H}(n_1,n_2,n_3),
\quad
\EstimatedSupportForWij(0)
=
\bigcup_{n_1=1}^{N_1-1} 
\bigcup_{n_2=1}^{M_2-1} 
\bigcup_{n_3=1}^{N_3-1} 
{\cal H}(n_1,n_2,n_3),
\\[4pt]& 
{\cal H}(n_1,n_2,n_3)=
[k_{1,n_1},k_{1,n_1+1}]\times
      [k_{2,n_2},k_{2,n_2+1}]
      \times
      [k_{3,n_3},k_{3,n_3+1}].
\notag
\end{align*}
In  each hexahedral element ${\cal H}(n_1,n_2,n_3)$
 of $\EstimatedSupportForgammaij(0)$,
the distribution of $\dContractedtocAsy_{ij}(\bk)$
 is approximated trilinearly in $\bk$, together with the nodal values
 of $\dContractedtocAsy_{ij}$,
\begin{align*}
&
\dContractedtocAsy_{ij}(\bk)
=
 \frac{ \dContracted_{ij}(n_1,n_2,n_3)(k_{2,n_2+1}-k_2)
       +\dContracted_{ij}(n_1,n_2+1,n_3)(k_2-k_{2,n_2})}{k_{2,n_2+1}-k_{2,n_2}}
 \frac{k_{1,n_1+1}-k_1}{k_{1,n_1+1}-k_{1,n_1}}
 \frac{k_{3,n_3+1}-k_3}{k_{3,n_3+1}-k_{3,n_3}}+\ldots,
\notag\\&
Z=\Big\{\dContracted_{ij}(n_1,n_2,n_3):\ 
            n_1=1,\ldots, N_1,\
  n_2=1,\ldots, N_2, \
     n_3=1,\ldots, N_3;\ i, j\Big\}\subset\bbR^{6 \NCV},
     \ \ \NCV=N_1 N_2  N_3.
\end{align*}
The piecewise linear approximation in $\EstimatedSupportForgammaij(0)$
is adequate at small mesh element sizes,
 since integration operations are mainly involved for
  $\dContractedtocAsy_{ij}$.
Accordingly, Eqs.~\eqref{dWij_Asy_Sol},
\eqref{dCVgamma_Constraint_Integral_Asy},
and \eqref{ObjectiveFunctionMaxIntWkkbrAsy}
 are discretized and represented in terms of nodal values $Z$.
 
Function `{\small NIntegrate}' of {\small MATHEMATICA}
is used to compute the 1-dimensional integrals
in Eqs.~\eqref{dWij_Asy_Sol} over the
 mesh elements.
Algorithms `{\small Cuhre}' and  `{\small Divonne}'
 of the open source software {\small CUBA} library
 \cite{Hahn2005, Hahn2006, httpcuba} are used to compute
4-dimensional integrals like 
Eq.~\eqref{ObjectiveFunctionMaxIntWkkbrAsy}.

\paragraph{Constraints and optimization}
Constraints~\eqref{CVgamma_TransformedAsy_SymmetryAboutk3}
and \eqref{tocBounds} are enforced at the mesh nodes.
Constraints~\eqref{ElementaryInequalitiesFortwiwj_Asy}
are imposed at collocation points defined below:
(i) the points located in the middle of the
mesh element edges parallel to the $k_2$-axis,
denoted as {1C};
(ii) the points located at the center of the mesh 
element facets parallel to the $k_2$-axis,
denoted as {2C};
(iii) the points located at the center of the mesh elements, denoted as {4C}.
We enforce Eqs.~\eqref{ElementaryInequalitiesFortwiwj_Asy} at
{1C} or the combination of {1C} and {4C}, etc.
to evaluate the impact of the collocation points.

The discretized second-order model is then cast in the form of
\begin{align}
\label{SOMasSOCP}
&
\minimize\ \ \  a^T Z
\notag\\[-6.5pt]&
\\[-6.5pt]&
\subjectto\ \ Z\in\cC.
\notag
\end{align}
Here, $-a$ is fixed from 
 the discretized  objective~\eqref{ObjectiveFunctionMaxIntWkkbrAsy},
 and $\cC$ is composed of the discretized 
 constraints~\eqref{ElementaryInequalitiesFortwiwj_Asy} through 
    \eqref{dCVgamma_Constraint_Integral_Asy} and  \eqref{tocBounds}.
 Problem~\eqref{SOMasSOCP} is a SOCP.

\paragraph{Parallel computing}
Since the discretized model is a large-scale SOCP,
the  widely used and tested splitting conic solver
{\small SCS} \cite{ODonoghueetal2016, ODonoghueetalPackage2016} is employed,
which is a first-order method.
With regard to  Eqs.~\eqref{ElementaryInequalitiesFortwiwj_Asy},
their imposition at \{1C, 2C, 4C\} under a coarser mesh 
or the imposition only at {1C} under a finer mesh
   demands large RAM (well above 64GB available in a workstation),
  specifically in the step of generating the
standard form of a SOCP necessary for the solver.
Hence, it is essential to formulate and solve
the discretized model~\eqref{SOMasSOCP}
 as a parallel  computing problem, with the help of the 
 global consensus algorithm
    of the alternating direction method of multipliers (ADMM)
    \cite{Boydetal2010, ParikhBoyd2013}.
 The ADMM consensus algorithm consists of the following iterations $k$,
\begin{subequations}
\label{SOM_ADMMAlgorithmProjectionmain}
\begin{align}
 Z^{k}=\overX^{k}:=\frac{1}{N}\sum_{i=1}^N X_i^k,
\label{SOM_ADMMAlgorithmProjectionZmain}
\end{align}
\vspace{-0.9em}
\begin{align}
 X_i^{k+1}:=\underset{X_i\in\cC_i}{\argmin}\,
             \big|\big|X_i-\Lik\big|\big|_2^2
           =\Projection_{\cC_i}\big(\Lik\big),
\label{SOM_ADMMAlgorithmProjection2main}
\end{align}
\vspace{-1.5em}
\begin{align}
\tilde{Y}_i^{k+1}:=\tilde{Y}_i^k+X_i^{k+1}-\overX^{k+1},\quad
\frac{1}{N}\sum_{i=1}^N \tilde{Y}_i^0=0.
\label{SOM_ADMMAlgorithmProjectionYmain}
\end{align}
\end{subequations}
Here, there are $N$ computing processing elements used in parallel,
  $\{\cC_i: i=1,\ldots, N\}$ is a partition of $\cC$.
 In the $i$-th element,
$X_i\in\bbR^{6 \NCV}$ is the local variable,
 $Y_i=\rho\,\tilde{Y}_i$ 
   the dual variable associated with the consensus constraint
$X_i-Z=0$,
 $\rho \,(>0)$ the penalty parameter in the augmented Lagrangian,
 and $Z$ acts as the common global variable.
Next, $||\cdot||_2$ is the Euclidean norm,
$\Lik=\overX^k-(a/\rho+\tilde{Y}_i^k)$,
and $\Projection_{\cC_i}\big(\Lik\big)$ denotes Euclidean projection of vector $\Lik$
onto $\cC_i$, $i=1,\ldots,N$.
The  $X_i$-update~\eqref{SOM_ADMMAlgorithmProjection2main}
 is computed by {\small SCS}.
 (See Appendix~\ref{Appensec:Algorithm} for details.)

The convergence of the iterative process~\eqref{SOM_ADMMAlgorithmProjectionmain}
is guaranteed analytically
 \cite{Boydetal2010, EcksteinBertsekas1992}.
To terminate the iterative process,
 the conventional stopping criteria \cite{Boydetal2010} are adopted,
\begin{align}
\label{StoppingCriteriamain}
&
 \big|\big| R^k\big|\big|_2=\bigg(\sum_{i=1}^N \big|\big|X_i^k-\overX^k\big|\big|_2^2\bigg)^{1/2}
 \leq \sqrt{N}\big(\epsilon^{\text{abs}}
         +\big|\big|\overX^k\big|\big|_2\epsilon^{\text{rel}}\big),
\notag\\[-9pt]&
\\[-8pt]&
\big|\big| S^k\big|\big|_2=\Big(N \rho^2 \big|\big|\overX^k-\overX^{k-1}\big|\big|_2^2\Big)^{1/2}
\leq \rho \sqrt{N} \big( \epsilon^{\text{abs}}
                        +\big|\big|\overX^k\big|\big|_2\epsilon^{\text{rel}}\big),
\notag
\end{align}
where $\big|\big| R^k\big|\big|_2$
and $\big|\big| S^k\big|\big|_2$ are, respectively,
 the squared norms of the primal and dual residuals, $R^k$ and $S^k$,
at the $k$-th iteration,
$\epsilon^{\text{abs}}$ and $\epsilon^{\text{rel}}$ denote, respectively, 
the absolute and relative tolerances.
The implementation is based mainly on the works of  
\cite{Boydetal2010, ParikhBoyd2013, EcksteinBertsekas1992, DiamondBoyd2016,
GrantBoyd2014, ODonoghueetalPackage2016, Boydetal2011, AmerMPICH2015}.
The values of $\rho=1$, $\epsilon^{\text{abs}}=10^{-4}$, and 
$\epsilon^{\text{rel}}=10^{-3}$ are employed and are adequate for the present
exploration.

\paragraph{Specifics}
In the numerical simulation, 
a single mesh size along the directions of $k_1$, $k_2$, and $k_3$ is used
to discretize the supports uniformly; 
two specific mesh sizes $0.1$ and $0.05$ are employed and denoted
by (I) and (II), respectively. No finer meshes are employed because of 
the resulting large computational size.

Different versions of the discretized model result from various combinations of
mesh size and constraints in order to test their effects.
Regarding the constraints, we select  {L$\,\&\,$1C},
 L$\,\&\,\text{1C}\,\&\,\text{4C}$,
 and {L$\,\&\,\text{1C}\,\&\,\text{2C}\,\&\,\text{4C}$},
 where  L$\,\&\,\text{1C}\,\&\,\text{4C}$, say,
 stands for the imposition of all the linear constraints
   and Eqs.~\eqref{ElementaryInequalitiesFortwiwj_Asy} at $\{\text{1C},\text{4C}\}$.
 The numbers of parallel computing processing elements
 are $N=3$, $6$, $12$, respectively, for {L$\,\&\,$1C\,(I)},
 {L$\,\&\,\text{1C}\,\&\,\text{4C}$\,(I)}, 
          and {L$\,\&\,\text{1C}\,\&\,\text{2C}\,\&\,\text{4C}$\,(I)},
 and $N=12$ for {L$\,\&\,\text{1C}$\,(II)}.
 The adequacy of the mathematical formulation and numerical solutions 
is indicated partially by the fact that separate calculations of 
$\WAsy_{12}(\bo)$ and $\overline{w_{j,k}\,w_{j,k}}^{(\infty)}(\bo)$
 yield the ratio
\begin{align*}
 \WAsy_{12}(\bo)\big/\overline{w_{j,k}\,w_{j,k}}^{(\infty)}(\bo)=-1.000,
\end{align*}
consistent with the exact value of $-1$ dictated 
by the global balance equation~\eqref{IntrinsicRelationForUkk(0)Asy} 
under $\sigma=0$.
Also, the choice of $L_k$ is adequate, 
as indicated by the distribution trends
 of the correlation functions
 $\RAsy_{\underlinei\underlinei}(r_1,0,0)$,
 $\RAsy_{\underlinei\underlinei}(0,r_2,0)$,
and  $\RAsy_{\underlinei\underlinei}(0,0,r_3)$
and their negligible values at $r_k=L_k$
 displayed in Figs.~\ref{CCr1} through \ref{CCr3},
  that are required by the self-consistency of being a support. Here,
the correlation functions are defined via
\begin{align*}
 \RAsy_{\underlinei\underlinei}(\br)
 =\WAsy_{\underlinei\underlinei}(\br)/\WAsy_{\underlinei\underlinei}(\bo).
\end{align*}

\subsection{\label{subsec:NumericalResultsandDiscussion}Numerical results and discussion}

To evaluate the proposed framework,
 experimental data summarized in p.81 of \cite{Piquet1999}
and  DNS data from \cite{RogersMoin1987, Sekimotoetal2016} are 
taken as the basis of comparison.
To this end,
the optimal solutions are used to compute the following dimensionless quantities: 
(a) the nontrivial components of the anisotropy tensor $b^{(\infty)}_{ij}$
   of Eq.~\eqref{AnisotropicTensor};
(b) the ratio of turbulent energy $K$
   to viscous dissipation $\varepsilon$ \cite{Piquet1999},
\begin{align*}
 (S K/\varepsilon)^{(\infty)}
         =\WAsy_{kk}(\bo)\big/\overline{w_{j,k}\,w_{j,k}}^{(\infty)}(\bo),
\end{align*}
where $S$ is from Eq.~\eqref{AverageVelocityField};
(c) the ratio of
 $G=-\overline{\w_i \w_j}\,V_{i,j}$ (a dimensional quantity
 defined in \cite{Piquet1999} without the use of Eqs.~\eqref{DimensionlessQuantityDefn}) 
 to dissipation,
\begin{align*}
  (G/\varepsilon)^{(\infty)}
=-2\,\WAsy_{12}(\bo)\big/\overline{w_{j,k}\,w_{j,k}}^{(\infty)}(\bo),
\end{align*}
which characterizes the extent of turbulent mean shear \cite{Piquet1999}.
It is noted that the experimental values quoted in Table~\ref{OSDataVsExperimental}
are ``plausible target of asymptotic data'' 
because of ``the strong discrepancy between data'' \cite{Piquet1999},
but they are taken as the basis for comparison, since they are 
  representative.

Table~\ref{OSDataVsExperimental} lists  the predicted values of 
\begin{align*}
 \big(b^{(\infty)}_{11},\ b^{(\infty)}_{22},\ b^{(\infty)}_{33},\ b^{(\infty)}_{12},
                 \ (S K/\varepsilon)^{(\infty)},\ (G/\varepsilon)^{(\infty)}\big)
\end{align*}
for  various versions of the discretized model under $\sigma=0$.
 A few observations may be made about 
 the pattern of the predicted and the comparison with  
  experimental values.

\begin{table*}
\caption{\label{OSDataVsExperimental}
   Data from optimal solutions and experiments.}
\begin{tabular}{ll}
\hline\hline
 Constraints  & $\big(b^{(\infty)}_{11},\ b^{(\infty)}_{22},\ b^{(\infty)}_{33},\ b^{(\infty)}_{12},
                 \ (S K/\varepsilon)^{(\infty)},\ (G/\varepsilon)^{(\infty)}\big)$ \\[2pt]
\hline
L$\,\&\,$1C\,(I)   & ($0.3908, -0.2674, -0.1234, -0.1329, 7.521, 2.000$)  \\
L$\,\&\,\text{1C}\,\&\,\text{4C}$\,(I)   & ($0.3944, -0.2642, -0.1302, -0.1388, 7.205, 2.000$)  \\
L$\,\&\,\text{1C}\,\&\,\text{2C}\,\&\,\text{4C}$\,(I) \ \ &
                                          ($0.4153, -0.2676, -0.1478, -0.1351, 7.400, 2.000$)  \\
\hline
L$\,\&\,\text{1C}$\,(II) & ($0.5031, -0.2951,  -0.2080,  -0.1161, 8.613, 2.000$)\\
\hline
Experimental \cite{Piquet1999} & ($0.203, \ \, -0.143,\ \, -0.06,\ \ \ -0.156,\ \, 5.54, \ \, 1.73$)  \\
\hline\hline
\end{tabular}
\end{table*}

\paragraph{}
Under mesh (I), 
the three combinations of constraints produce nearly 
the same quantitative results, 
though the corresponding distributions of $\tWAsy_{ij}$
in the wave-number space differ in details.
 The same relative numerical order 
 as that of the experimental data is produced as follows,
\begin{align}
\label{ExperimentalPattern}
  b^{(\infty)}_{11}>0>b^{(\infty)}_{33}>b^{(\infty)}_{22}.
\end{align}
Under mesh (II), 
 {L$\,\&\,$1C} produces the results quantitatively similar to
 those of (I), the order pattern~\eqref{ExperimentalPattern}
 holds, though the values of $b^{(\infty)}_{11}$, 
  $b^{(\infty)}_{11}-b^{(\infty)}_{33}$,
 and $b^{(\infty)}_{11}-b^{(\infty)}_{22}$ are higher.
 No imposition of 2C or 4C is carried out under (II),  because of
 the following considerations: 
First, as displayed in Table~\ref{OSDataVsExperimental},
 the same relative order pattern~\eqref{ExperimentalPattern} is obtained,
 independent of 
 the constraints imposed under (I) and independent of mesh size reduction
  from (I) to (II) under  {L$\,\&\,\text{1C}$}.
 We infer from this independence
 that such a relative order pattern may not change under 
 {L$\,\&\,\text{1C}\,\&\,\text{4C}$\,(II)} and 
 {L$\,\&\,\text{1C}\,\&\,\text{2C}\,\&\,\text{4C}$\,(II)}
within the model.
Second, the required parallel processing elements are about $50$ and $100$,
    not available to this study.

\paragraph{}
 For the numerical solutions, 
 separate calculations of $\overline{w_{j,k} w_{j,k}}^{(\infty)}(\bo)$
 and  $\WAsy_{12}(\bo)$ give
 $(G/\varepsilon)^{(\infty)}=2.000$, consistent with the value of 2
 dictated by \eqref{IntrinsicRelationForUkk(0)Asy} under $\sigma=0$.
 Unequal to $2$,
the experimental value of $(G/\varepsilon)^{(\infty)}=1.73$
 may result from  the deviation from 
 the asymptotic state of $\sigma=0$ in experiments and experimental errors.

Regarding the values of 
$b^{(\infty)}_{11}$,  $b^{(\infty)}_{22}$, and  $b^{(\infty)}_{33}$,
the predictions differ significantly from the experiments.
This discrepancy may be attributed to the non-enforceability
of the non-negativity of variance of products~\eqref{NonNegativeVariance}
within the second-order model, as explained below.
The results of {L$\,\&\,\text{1C}$\,(II)} are taken as the basis
for comparison.

As mentioned in Subsec.~\ref{subsec:PrimaryEquationsEvolutionInSOM},
the global equality 
constraints~\eqref{vIntrinsicEquality_kk_jjk_Half_SOM}/\eqref{dCVgamma_Constraint_Integral_Asy}
provide a distributive mechanism of distributing $\tWAsy_{kk}(\bk)$
among its components $\tWAsy_{\underlinei\underlinei}(\bk)$
 and   $\WAsy_{kk}(\bzero)$
among its components $\WAsy_{\underlinei\underlinei}(\bzero)$:
 a larger $\WAsy_{11}(\bzero)$ is accompanied by a smaller  $\WAsy_{22}(\bzero)$
and/or   $\WAsy_{33}(\bzero)$.
 Equivalently,
 a positive larger $b^{(\infty)}_{11}$ is accompanied by a negative larger
 $b^{(\infty)}_{22}$ and/or  $b^{(\infty)}_{33}$,
 as required by  $b^{(\infty)}_{kk}=0$.
This is the pattern displayed by 
the prediction data of {L$\,\&\,\text{1C}$ (II)}: the predicted
$b^{(\infty)}_{\underlinei\underlinei}(\bzero)$ are about twice to thrice
of the experimental values.

 The predicted values show that
 maximization of the objective~\eqref{ObjectiveFunctionMaxIntWkkbrAsy}
makes $\WAsy_{11}(\bzero)$ unduly large relative to
 $\WAsy_{22}(\bzero)$ and  $\WAsy_{33}(\bzero)$ within the second-order model.
 As indicated by Eqs.~\eqref{ScalingSOMAsy} and \eqref{tocBounds},
 the model does not have a natural mechanism to bound from above
    $\WAsy_{11}(\bo)$, $\WAsy_{11}(\bo)-\WAsy_{22}(\bo)$,
  and  $\WAsy_{11}(\bo)-\WAsy_{33}(\bo)$.
Hence, a combination of the distributive mechanism and the 
 maximization   causes 
 such a large difference between the predicted and the experimental values.
 
To remedy this problem,
we need a  mechanism  
to  bound from above $\WAsy_{11}(\bo)$, $\WAsy_{11}(\bo)-\WAsy_{22}(\bo)$,
  and  $\WAsy_{11}(\bo)-\WAsy_{33}(\bo)$.
 The non-negativity requirement~\eqref{NonNegativeVariance}
 may play such a role, as evidenced by the specific 
 examples~\eqref{wiwj_Deviation},
\begin{align}
\label{NNVPIsotropization}
\Big(\WAsy_{i j}(\br)\Big)^2\leq 
\WAsy_{(\underlinei\underlinei)\underlinej\underlinej}(\br,\br),
\quad
\Big|\WAsy_{\underlinei\underlinei}(\br)-\WAsy_{\underlinej\underlinej}(\br)\Big|^2
\leq
   \WAsy_{(\underlinei\underlinei)\underlinei\underlinei}(\br,\br)
  +\WAsy_{(\underlinej\underlinej)\underlinej\underlinej}(\br,\br)
  +2\WAsy_{(\underlinei\underlinei)\underlinej\underlinej}(\br,\br)
  -4\WAsy_{(\underlinei\,\underlinej)\underlinei\underlinej}(\br,\br).
\end{align}
These inequalities,
together with  others from Eq.~\eqref{NonNegativeVariance},
expectedly produce certain upper bounds on
$|\WAsy_{i j}(\bzero)|$ and
 $\big|\WAsy_{\underlinei\underlinei}(\bzero)-\WAsy_{\underlinej\underlinej}(\bzero)\big|$ 
 with $\tWAsy_{(ij)kl}$ as the control variables,
 because the left-hand sides of these inequalities
   depend on $\tWAsy_{(ij)kl}$ quadratically
 while the right-hand sides depend on $\tWAsy_{(ij)kl}$ linearly.
This observation provides a ground for the need of the third-order model. 
 Within the third-order model,
the coupling interaction among the two mechanisms
of distribution and bounding and the maximization of 
the objective~\eqref{ObjectiveFunctionMaxIntWkkbrAsy} 
is expected to make 
  $b^{(\infty)}_{\underlinei\underlinei}$
  closer to the experimental values
  than those produced by the second-order model.

\paragraph{}  
Both the distributive and
the bounding mechanisms seem related to the issue of
isotropization \cite{Davidson2004, Rota1951}, which needs to be explored 
within the third-order model.

\begin{figure*}[ht]
\centering
\subfloat{\includegraphics[width = 3.5in]{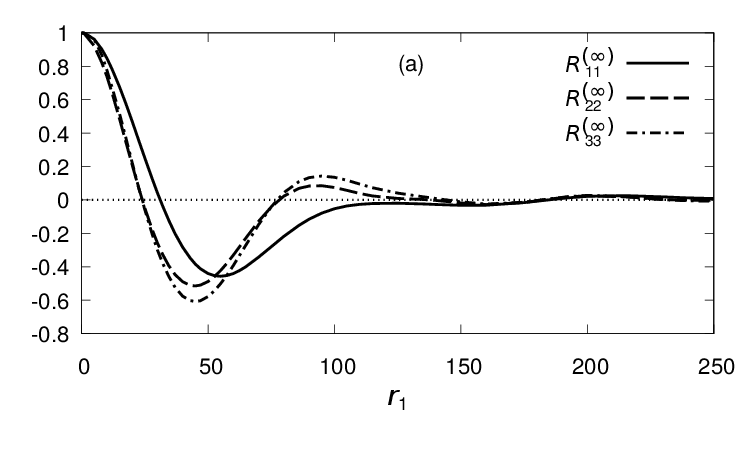}}
\psfrag{rxxx}{$r_x/L_z$}
\subfloat{\includegraphics[width = 3.5in]{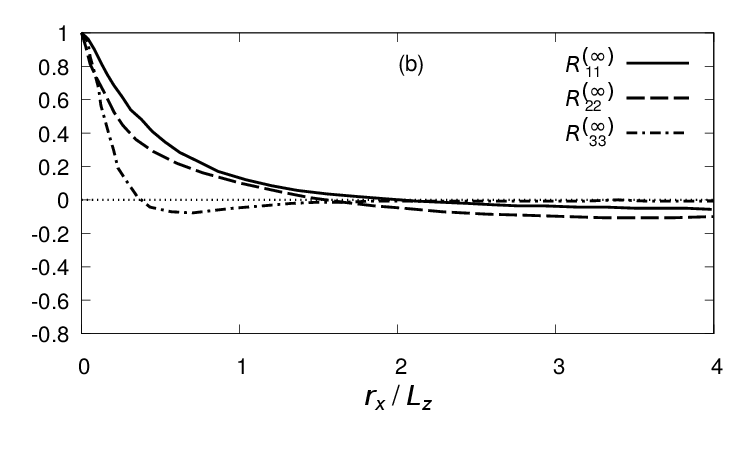}}
\vspace{-1em}
\caption{Two-point correlation functions $\RAsy_{\underlinei\underlinei}(\cdot,0,0)$.
         (a) Simulation of the second-order model.
             The streamwise length scale $L_{\text{strm}}\sim 250$.
         (b) DNS data. 
             $\RAsy_{11}$,  $\RAsy_{22}$, and  $\RAsy_{33}$ are extracted, respectively, from
             Figs.~5(a), 5(b),
              and 5(c) with $A_{xz}=8$ of \cite{Sekimotoetal2016}. $\hskip127mm $}
\label{CCr1}
\end{figure*}

\begin{figure*}[ht]
\centering
\subfloat{\includegraphics[width = 3.5in]{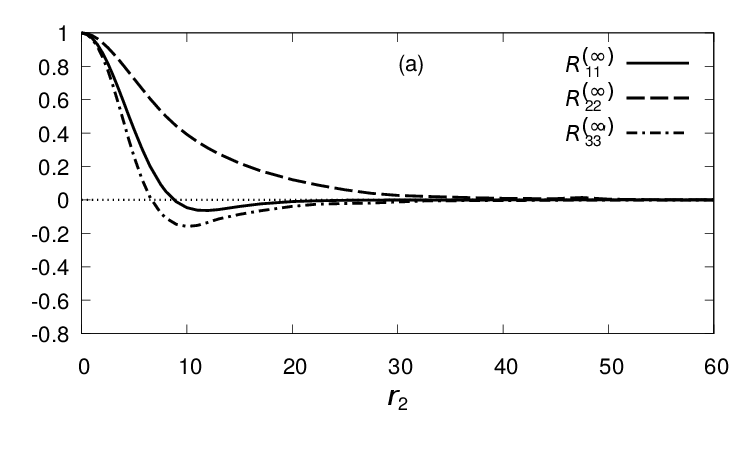}}
\subfloat{\includegraphics[width = 3.5in]{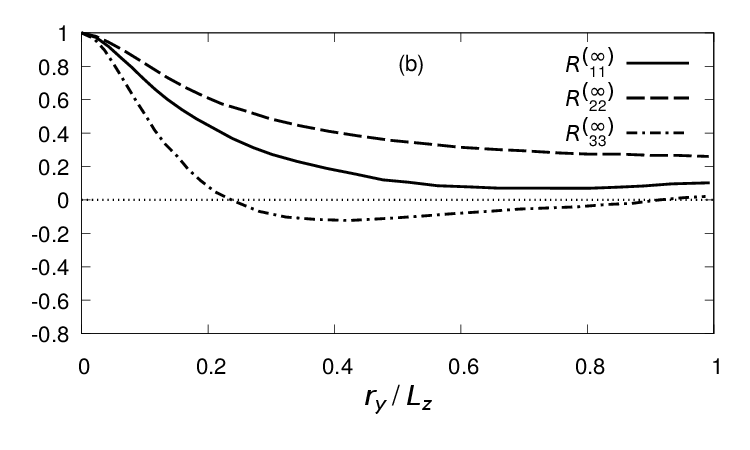}}
\vspace{-1em}
\caption{Two-point correlation functions $\RAsy_{\underlinei\underlinei}(0,\cdot,0)$.
(a) Simulation of the second-order model.
    The vertical length scale $L_{\text{vert}}\sim 50$.
(b) DNS data. 
    $\RAsy_{11}$,  $\RAsy_{22}$, and  $\RAsy_{33}$ are extracted, respectively, from
    Figs.~5(d), 5(e), and 5(f) with $A_{xz}=8$ of \cite{Sekimotoetal2016}.
     $\hskip143mm $}
\label{CCr2}
\end{figure*}

\begin{figure*}[ht]
\centering
\subfloat{\includegraphics[width = 3.5in]{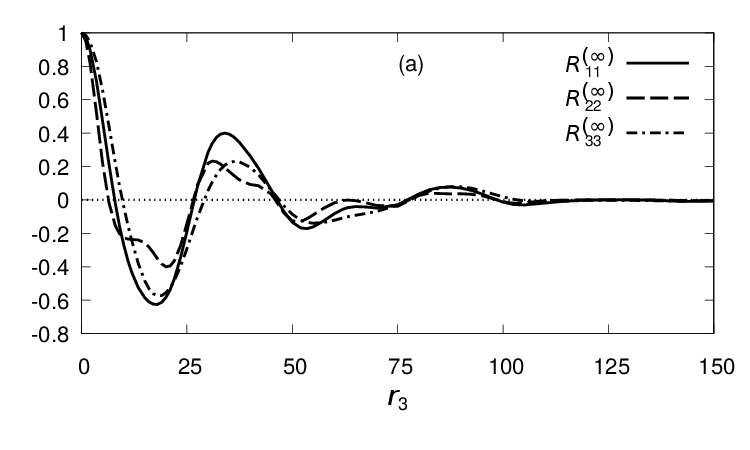}}
\subfloat{\includegraphics[width = 3.5in]{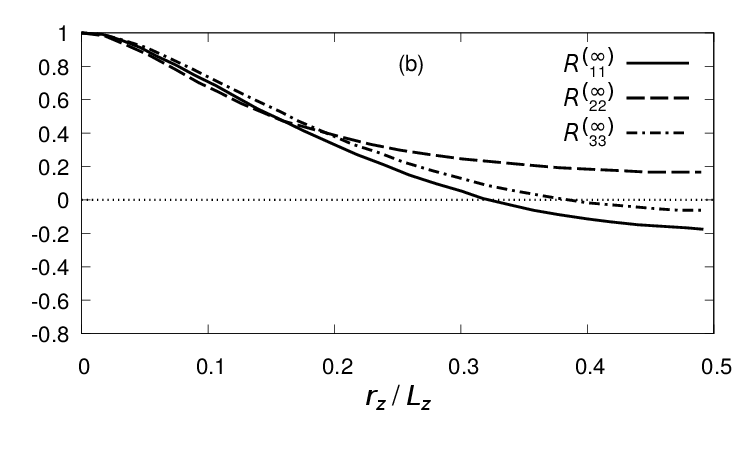}}
\vspace{-1em}
\caption{Two-point correlation functions $\RAsy_{\underlinei\underlinei}(0,0,\cdot)$.
(a) Simulation of the second-order model.
    The spanwise length scale $L_{\text{span}}\sim 120$.
(b) DNS data. 
    $\RAsy_{11}$,  $\RAsy_{22}$, and  $\RAsy_{33}$ are extracted, respectively, from
    Figs.~5(g), 5(h), and 5(i) with $A_{xz}=8$ of \cite{Sekimotoetal2016}.
     $\hskip127.5mm$}
\label{CCr3}
\end{figure*}

 We evaluate below the correlation function
 predictions of the second-order model
under L$\,\&\,\text{1C}$\,(II) against 
the DNS studies of \cite{RogersMoin1987, Sekimotoetal2016}.
Figures~\ref{CCr1}(a), \ref{CCr2}(a), and \ref{CCr3}(a) show the predicted distributions of 
the correlation functions $\RAsy_{\underlinei\underlinei}$
  along $r_1$, $r_2$, and $r_3$, respectively.
  Also, for the convenience of comparison, we have extracted  DNS data from Figs.~5
  of \cite{Sekimotoetal2016} and present them 
  in Figs.~\ref{CCr1}(b), \ref{CCr2}(b), and \ref{CCr3}(b). 

Considering the fundamental defect of the second-order model ---
its lack of a bounding mechanism discussed above, 
we analyze and compare these predictions with  DNS data in a qualitative fashion.
Clearly the ranges of $r_k$ in Figs.~\ref{CCr1}(a), \ref{CCr2}(a), and \ref{CCr3}(a)
indicate the existence of 
 three distinct characteristic macro length scales 
beyond which the two-point correlations are effectively negligible:
$L_{\text{strm}}\sim 250$ (streamwise along the $r_1$-direction), 
$L_{\text{vert}}\sim 50$ (vertical along the $r_2$-direction), 
$L_{\text{span}}\sim 120$ (spanwise along the $r_3$-direction), 
and they are in the numerical order of 
$L_{\text{strm}}>L_{\text{span}}>L_{\text{vert}}$.
This order may reflect  underlying physics: 
(i) The only nontrivial mean velocity component 
in the sheared flow is the streamwise velocity
$V'_1=x'_2$   (dimensionless from Eqs.~\eqref{AverageVelocityField} 
                 and \eqref{DimensionlessQuantityDefn}),
  it produces streaks \cite{Sekimotoetal2016}, 
  and thus,  the corresponding correlations have the longest range and
  $L_{\text{strm}}$ is the largest  (Fig.~\ref{CCr1}(a)). 
 This appears to be supported partially by Fig.~6 of  \cite{RogersMoin1987} 
 (which is obtained in transient state and not reproduced here).
(ii) In the vertical $r_2$-direction, the mean shearing $\partial V'_1/\partial x'_2=1$
 tends to suppress the velocity fluctuations, 
which results in the smallest $b^{(\infty)}_{22}$ ($\WAsy_{22}(\bzero)$) among 
    $b^{(\infty)}_{\underlinei\underlinei}$ ($\WAsy_{\underlinei\underlinei}(\bzero)$)
    \cite{Piquet1999, KidaTanaka1994}.
    This suppression may also make the range of the vertical correlations the shortest
    and $L_{\text{vert}}$ the smallest   (Fig.~\ref{CCr2}(a)).
(iii) Along the spanwise $r_3$-direction, $V'_3=0$
 and no suppression is present, their combined effect may make  
 $L_{\text{span}}$ the middle scale between $L_{\text{strm}}$
 and $L_{\text{vert}}$  (Fig.~\ref{CCr3}(a)).
 We cannot check this numerical order of the characteristic length scales 
 by  using the DNS data in  Figs.~\ref{CCr1}(b), \ref{CCr2}(b), and \ref{CCr3}(b)
 because the latter
 do not have a sufficiently large computational box/flow domain
 to yield negligible values of $\RAsy_{\underlinei\underlinei}$
 at the respective boundaries.
 The boundedness of these length scales is incompatible with
   that length scales in (ideal) homogeneous shear turbulence
   are accepted as growing indefinitely large in DNS \cite{Sekimotoetal2016}.
 
 It is interesting to notice that 
the presence of such bounded macro length scales
may be essential for the experimental measurement of the flow properties of
 homogeneous shear turbulence. Otherwise, unbounded macro correlation scales
  imply that measurement of  $\W_{ij}(\br)$
 would be affected significantly by a laboratory setup of finite size
 and even measurement of Reynolds stress  $\W_{ij}(\bzero)$ would become
 unreliable, considering the close interaction among
  $\W_{ij}(\bzero)$ and  $\W_{ij}(\br)$. This result of three bounded macro
 length scales may be viewed as  a merit contributed by the second-order model.

 Regarding the distribution patterns of $\RAsy_{\underlinei\underlinei}(\br)$,
 the predictions are quite different from those of DNS reproduced in 
  Figs.~\ref{CCr1}(b), \ref{CCr2}(b), and \ref{CCr3}(b).
(i) Compared with Fig.~\ref{CCr1}(b),
 the negative values of the streamwise distributions in Fig.~\ref{CCr1}(a)
  are much larger in magnitude.
  The predicted local maxima of $\RAsy_{\underlinei\underlinei}$
  and  minima of $\RAsy_{11}$ and  $\RAsy_{22}$
   do not exist in the DNS data either.
(ii) About the vertical distributions, 
the trends of $\RAsy_{22}$ and  $\RAsy_{33}$ in Fig.~\ref{CCr2}(a)
  are compatible with those in
      Fig.~\ref{CCr2}(b), but the predicted
        $\RAsy_{11}$ has a locally negative minimum which is absent in 
        Fig.~\ref{CCr2}(b). 
(iii) Figure~\ref{CCr3}(a) shows
   the complex wave-like spanwise distributions with local minima and maxima,
   incompatible with the monotonically decreasing pattern displayed in
Fig.~\ref{CCr3}(b).

The above comparisons reveal a major flawed feature of 
the prediction --- the local negative minima of  $\RAsy_{\underlinei\underlinei}(\br)$ are
either too large in magnitude or present  spuriously.
The cause can be attributed to the mathematical structure of the second-order model
  in the space of  control variables $\dContractedtocAsy_{ij}$,
  especially the structures of $\WAsy_{ij}$ and the constraints in terms of
  $\dContractedtocAsy_{ij}$: 
  the maximization of objective~\eqref{ObjectiveFunctionMaxIntWkkbrAsy} 
  still produces this feature, even though it has the tendency  to
  promote overall large and positive $\WAsy_{\underlinei\underlinei}(\br)$.
It is expected that
the third-order model has the potential to modify this feature
and improve the prediction of $\RAsy_{\underlinei\underlinei}(\br)$,
based on the following considerations:
(a) Physically, the third-order model is an improved model which includes
     the third order correlations (their evolution and associated constraints),
      the nonnegativity of variance of products, etc.
(b) Mathematically,
  $\tWAsy_{(ij)kl}$ act as the control variables, the optimization problem is 
    formulated in the enlarged control variable
    space of $\tWAsy_{(ij)kl}$,
     the mathematical structures of $\WAsy_{ij}$ and the constraints 
     in term of  $\tWAsy_{(ij)kl}$
     are different  from those of the second-order model. 
These new structures may correct the flawed feature,
  together with the maximization of
   objective~\eqref{ObjectiveFunctionMaxIntWkkbrAsy} 
    which tends to promote overall large and positive $\WAsy_{\underlinei\underlinei}(\br)$.
With this improvement, the correlation patterns may be closer to those
produced by DNS.

\section{\label{sec:Conclusion}Conclusion}

The ideas from  theories of optimal control and convex optimization 
are used to model
homogeneous shear turbulence of an incompressible Newtonian fluid.
The intent is to explore an approach of optimal correlations
to help resolve the issues of non-realizability 
and restriction to homogeneity encountered by 
analytical theories of turbulence.
The multi-point spatial correlations of velocity and pressure fluctuations
up to the degenerate fourth order are included in the framework.
Two models are formulated:
The second-order model takes the second order correlations as  state variables and
 the contracted and degenerate third order correlations as the control variables;
The third-order model takes both the second and the third order correlations
as  state variables and
 the degenerate fourth order correlations as the control variables.

 For both models,
the primary dynamical equations of evolution  are presented.
The sources of constraints are discussed, namely
 the correlation definitions, the inversion and mirror symmetries, 
the Cauchy-Schwarz inequality, and the non-negativity of variance of products.
 The constraints of the non-negativity of variance of products
 provide a  mechanism to bound all the correlations and to
 dictate that the maximum
 exponential growth rate of the correlations in the asymptotic state be zero,
 contradicting  DNS data.
 Shannon entropy  is used to argue for
 $\max\int_{\mathbb{R}^3}d\br\, \W_{kk}(\br)$ 
as the alternative objective functional, the latter is then substantiated
from the perspective of strong mixing in physical turbulence. 
 The models are second-order cone programs in their discretized forms.

Restricted by computational feasibility at present, 
  only the asymptotic steady state of
the second-order model is  solved numerically. 
As part of the basis for evaluation,
the non-trivial values of the anisotropy tensor $b^{(\infty)}_{ij}$ are compared
between the prediction and the experiments \cite{Piquet1999},
and they are given here:
(a) $b^{(\infty)}_{11}=0.5031$ $> b^{(\infty)}_{33}=-0.2080$ 
       $> b^{(\infty)}_{22}=-0.2951$ (Prediction) versus
$b^{(\infty)}_{11}=0.203$ $> b^{(\infty)}_{33}=-0.06$ $> b^{(\infty)}_{22}=-0.143$
 (Experiment);
(b) $b^{(\infty)}_{12}=-0.1161$ (Prediction) versus 
$b^{(\infty)}_{12}=-0.156$ (Experiment).
 The qualitative agreement with regard to the relative
 numerical order pattern among  $b^{(\infty)}_{11}$,
  $b^{(\infty)}_{22}$, and  $b^{(\infty)}_{33}$ is encouraging,
indicating the potential of the framework.
However, there are significant quantitative differences between the predicated
 and the experimental values.
  The non-enforceability of the non-negativity of variance of products 
 within the second-order model
 is identified as the cause.
 Regarding the prediction of the second order correlation functions
      $\RAsy_{\underlinei\underlinei}(\br)$,
 the results are mixed too: 
(a)  The streamwise, vertical, and spanwise characteristic macro length scales
are estimated as finite for the two-point correlations, 
 in contrast to  DNS  where they are accepted to be indefinitely large.
The  relative numerical order of the scales appear  
consistent with the physical flow.
(b) The $\br$-dependent distribution patterns display a major flawed feature 
that local negative minima are either too large in magnitude or present spuriously.
 The mathematical structures of $\WAsy_{ij}$ and the constraints in  the space of
 control variables  $\dContractedtocAsy_{ij}$ are recognized as the cause for
 this erroneous prediction.
 It is then inferred  that 
  the third-order model is necessary and has  potential to  improve predictions,
 on the basis of the following considerations: 
 (i) It is an improvement over the second-order model.
   It contains the third order and degenerate fourth order correlations,
   the non-negativity of variance of products, etc.
      The resulting mathematical structure of the optimization problem
      exists in the enlarged
      space of control variables $\tWAsy_{(ij)kl}$ and may  predict better.
 (ii) The model contains both the distributive mechanism of distributing turbulent energy
 among its three components
 and the bounding mechanism to bound from above these components
 and  their differences.
 However, implementation of the  model needs further study
 as it is a huge-scale problem. 
 Also, algorithms specific to  multi-dimensional
 integrals (up to 10-dimensions)
 that arise in the model need to be developed in order to improve 
 convergence rates and compute efficiently the huge number of integrals
 necessary for the implementation of the constraints.
 Alternatively, to avoid this issue of integration,
 we may explore how to solve
 the third-order model in physical space.

 It is noted that the dynamical equations of evolution 
for the mean flow fields and
for the spatial correlations 
may be easily derived for inhomogeneous turbulent flows
of an incompressible Newtonian fluid.
 The above listed sources of constraints, such as 
 the correlation definitions, 
the Cauchy-Schwarz inequality, and the non-negativity of variance of products,
are also the sources of constraints for such flows
and may be formulated accordingly.
Further, the issue of Shannon entropy, its substitute,
and the physical interpretation may be discussed similarly.
Therefore, the present exploration of homogeneous shear turbulence
as a second-order cone program
may shed light on modeling inhomogeneous turbulent flows.
A similar comment may be made regarding some other motion systems too.
Further, it will be interesting to explore the possibility of
a second-order model suitable for LES.

\begin{acknowledgments}
I thank Professor M. Ramakrishna,  Aerospace Engineering, IIT Madras,
 for  meaningful discussions, and
I thank Dr. Parag Ravindran,  Mechanical Engineering, IIT Madras,
 for  help rewording the manuscript.
  I thank the developers of the open-source packages {\small CUBA} Library,
{\small CVXPY}, {\small SCS}, and 
{\small MPICH} whose codes made the simulation easier,
 and I thank CVX Research, Inc. for granting me a free academic license for 
 {\small CVX} Professional
  to use {\small MOSEK}.
\end{acknowledgments}

The computational codes and their output data files
 may be accessed via
 \\ ${\, }$ \qquad
  \verb|https://drive.google.com|
  \verb|/drive/folders/1s4e6__AkdQQTMCCXKVmoCr3iPyrh0b|
  \verb|6T?usp=sharing|

\appendix

 \vspace{-3.5mm}

\section{\label{Appensec:InversionAndMirror}Symmetries of inversion and mirror}

The derivation of the inversion and mirror symmetries~\eqref{InversionMirror} 
is outlined in this appendix.

The geometric and kinematic symmetries underlying the mean flow
field~\eqref{AverageVelocityField} suggest that,
under the spatial inversion $\{\bx,\by,\bz\}\rightarrow \{-\bx,-\by,-\bz\}$,
the statistical correlations  transform according to
\begin{align}
\label{uHomogeneity_Inversion_Statistical}
&
\overline{w_i(\bx) w_j(\by)}=\overline{(-w_i(-\bx))(-w_j(-\by))},
\quad
\overline{w_i(\bx) w_j(\by) w_k(\bz)}=\overline{(-w_i(-\bx)) (-w_j(-\by)) (-w_k(-\bz))},
\notag\\[-6pt]&
\\[-6pt]&
\overline{\q(\bx) \wj(\by)}=\overline{\q(-\bx) (-\wj(-\by))},\ldots
\notag
\end{align}
which results in the inversion symmetry equalities of Eqs.~\eqref{InversionMirror}
under homogeneity.

Next, the geometric and kinematic symmetries of the mean flow field suggest
 the plane inversion symmetry under the plane transformation
$\{\bx,\by,\bz\}$
       $\rightarrow$ $\{(-x_1,-x_2,x_3)$, $(-y_1,-y_2,y_3)$, $(-z_1,-z_2,z_3)\}$:
 $w_1$ and $w_2$ change their signs and $w_3$ and $q$ do not change  signs
 in the correlations, such as
\begin{align}
& 
\overline{w_1(\bx) w_1(\by)}=\overline{(-w_1(-x_1,-x_2,x_3))(-w_1(-y_1,-y_2,y_3))},
\quad
\overline{w_1(\bx) w_3(\by)}=\overline{(-w_1(-x_1,-x_2,x_3)) w_3(-y_1,-y_2,y_3)},
\notag\\&
\overline{\q(\bx) \w_2(\by)}=\overline{\q(-x_1,-x_2,x_3) (-\w_2(-y_1,-y_2,y_3))},\ldots
\end{align}
Combining this plane inversion symmetry with the spatial inversion symmetry 
 \eqref{uHomogeneity_Inversion_Statistical}
leads to the statistical symmetry of mirror
under $\{\bx,\by,\bz\}$
       $\rightarrow$ $\{\bx'=(x_1,x_2,-x_3)$,
       $\by'=(y_1,y_2,-y_3)$, $\bz'=(z_1,z_2,-z_3)\}$,
\begin{align}
\overline{w_i(\bx)\,w_j(\by)}
=(-1)^{\delta_{\underlinei 3}+\delta_{\underlinej 3}}\,\,
          \overline{w_{\underlinei}(\bx')\,w_{\underlinej}(\by')},
\quad
\overline{q(\bx)\,w_i(\by)}
=(-1)^{\delta_{\underlinei 3}}\,\,
         \overline{q(\bx')\,w_{\underlinei}(\by')}, \ldots
\end{align}
Further, imposition of homogeneity produces
the mirror symmetry constraints of Eqs.~\eqref{InversionMirror}. 

\vspace{-1.5mm}

\section{\label{Appensec:DetailsSecondOrderModel}Detail for the second-order model}

The derivation of constraints~\eqref{vIntrinsicEquality_kk_jjk_Half_SOM}
is outlined here.  Integration of
 $(k_k+l_k)\tWImag_{kij}(\bk,\bl)=0$ of
 Eqs.~\eqref{DivergenceFreeInPhysicalSpace_qandw_fs} gives
\begin{align*}
 \int_{\mathbb{R}^3}  \int_{\mathbb{R}^3} d\bk\, d\bl\,
 k_k \Big(\tWImag_{kji}(\bl,\bk)+\tWImag_{kij}(\bl,\bk)\Big)
=0.
\end{align*}
This equality is operated upon as follows,
\begin{align*}
\int_{-\infty}^{+\infty} dk_1=\int_{-\infty}^0 dk_1+\int_{0}^{+\infty} dk_1,\quad
(\bk, \bl)\rightarrow-(\bk, \bl),
\quad
\tWImag_{kij}(-\bk,-\bl)=-\tWImag_{kij}(\bk,\bl).
\end{align*}
The second part and the third part are applied 
to the term $\int_{0}^{+\infty} dk_1$ of the first part;
together with the use of  Eqs.~\eqref{ContractedTOC},
the operation results in the mentioned constraints.

\section{\label{Appensec:DetailsThirdOrderModel}Detail for the third-order model}

The third order correlations
$\tW_{ijk}$ are  state variables.
Along with Eqs.~\eqref{HomogeneitySymmetryInversionMirrorfs},
 the divergence-free constraints for $\tW_{ijk}$
 in Eqs.~\eqref{DivergenceFreeInPhysicalSpace_qandw_fs} 
 are solved to yield
\begin{align}
\label{DivergenceFreeInPhysicalSpace_wiwjwk_fs}
\tWImag_{113}(\bk,\bl)
=&
-\frac{l_1}{l_3}\tWImag_{111}(\bk,\bl)
-\frac{l_2}{l_3}\tWImag_{112}(\bk,\bl),
\quad
\tWImag_{123}(\bk,\bl)
=
-\frac{l_1}{l_3}\tWImag_{112}(\bl,\bk)
-\frac{l_2}{l_3}\tWImag_{122}(\bk,\bl),
\notag\\ 
\tWImag_{223}(\bk,\bl)
=&
-\frac{l_1}{l_3}\tWImag_{122}(-\bl-\bk,\bk)
-\frac{l_2}{l_3}\tWImag_{222}(\bk,\bl),
\notag\\ 
\tWImag_{133}(\bk,\bl)
=&\,
 \frac{k_1 l_1}{k_3 l_3}\tWImag_{111}(\bk,\bl)
+\frac{l_1 k_2}{k_3 l_3}\tWImag_{112}(\bl,\bk)
+\frac{k_1 l_2}{k_3 l_3}\tWImag_{112}(\bk,\bl)
+\frac{k_2 l_2}{k_3 l_3}\tWImag_{122}(\bk,\bl),
\notag\\ 
\tWImag_{233}(\bk,\bl)
=&\,
 \frac{k_1 l_1}{k_3 l_3}\tWImag_{112}(\bk,-\bl-\bk)
+\frac{k_2 l_1}{k_3 l_3}\tWImag_{122}(-\bl-\bk,\bk)
+\frac{k_1 l_2}{k_3 l_3}\tWImag_{122}(-\bl-\bk,\bl)
+\frac{k_2 l_2}{k_3 l_3}\,\tWImag_{222}(\bk,\bl),
\\
\tWImag_{333}(\bk,\bl)
=&
-\frac{l_1 k_1 (k_1+l_1)}{k_3 l_3 (k_3+l_3)}\tWImag_{111}(\bk,\bl)
-\frac{l_1 (k_1+l_1)k_2 }{k_3 l_3 (k_3+l_3)}\tWImag_{112}(\bl,\bk)
-\frac{k_1 (k_1+l_1) l_2}{k_3 l_3 (k_3+l_3)}\tWImag_{112}(\bk,\bl)
\notag\\ &
-\frac{k_1 l_1 (k_2+l_2)}{k_3 l_3 (k_3+l_3)}\tWImag_{112}(\bk,-\bl-\bk)
-\frac{l_1 k_2 (k_2+l_2)}{k_3 l_3 (k_3+l_3)}\tWImag_{122}(\bk,-\bl-\bk)
-\frac{k_1 l_2 (k_2+l_2)}{k_3 l_3 (k_3+l_3)}\tWImag_{122}(\bl,-\bl-\bk)
\notag\\  &
-\frac{(k_1+l_1)k_2 l_2 }{k_3 l_3 (k_3+l_3)}\tWImag_{122}(\bk,\bl)
-\frac{k_2 l_2 (k_2+l_2)}{k_3 l_3 (k_3+l_3)}\tWImag_{222}(\bk,\bl).
\notag
\end{align}
  The components not present above can be obtained
 with the aid of Eqs.~\eqref{HomogeneitySymmetryInversionMirrorfs}.

\section{\label{Appensec:Twomoreexamplesinvolving}
Examples of inequalities involving only the second order correlations}

With the help of Eqs.~\eqref{Vorticityetc},
Eqs.~\eqref{tviwj_twivj_CSInWNS} can be verified directly,
\begin{align}
&
 \overline{\tilde\Vorticity_1\tilde\Vorticity_1}(\bk)
\geq
 \bigg(|k_3|\sqrt{\tW_{22}(\bk)}-|k_2|\sqrt{\tW_{33}(\bk)}\bigg)^2
+2|k_2 k_3|\bigg(\sqrt{\tW_{22}(\bk) \tW_{33}(\bk)}-\big|\tW_{23}(\bk)\big|\bigg),
\notag\\&
 \overline{\tilde\Vorticity_1\tilde\Vorticity_1}(\bk)\,
 \overline{\tilde\Vorticity_3\tilde\Vorticity_3}(\bk)
-\Big(\overline{\tilde\Vorticity_1\tilde\Vorticity_3}(\bk)\Big)^2
=
 (k_2)^2\big[(k_3)^2+2(k_1)^2\big]
    \Big(\tW_{11}(\bk)\tW_{22}(\bk)-\big(\tW_{12}(\bk)\big)^2\Big)
    \notag\\&\hskip3.5mm
+(k_1 k_2)^2\Big(\tW_{22}(\bk)\tW_{33}(\bk)-\big(\tW_{23}(\bk)\big)^2\Big)
+(k_2)^2\big[(k_2)^2+2(k_1)^2+2(k_3)^2\big]
      \Big(\tW_{11}(\bk)\tW_{33}(\bk)-\big(\tW_{13}(\bk)\big)^2\Big),
\\[-10pt]&
\notag\\[-4pt]&
\tW_{11}(\bk)\,\overline{\tilde\Vorticity_1\tilde\Vorticity_1}(\bk)
-\Big|\overline{\tilde\w_1\tilde\Vorticity_1}(\bk)\Big|^2
=
 \big[(k_3)^2+2\,(k_2)^2\big]
          \Big(\tW_{11}(\bk)\tW_{22}(\bk)-\big(\tW_{12}(\bk)\big)^2\Big)
+(k_2)^2\Big(\tW_{11}(\bk)\tW_{33}(\bk)-\big(\tW_{13}(\bk)\big)^2\Big),
\notag\\&
\tW_{11}(\bk)\,\overline{\tilde\Vorticity_2\tilde\Vorticity_2}(\bk)
-\l|\overline{\tilde\w_1\tilde\Vorticity_2}(\bk)\r|^2
=
 (k_1)^2\Big(\tW_{11}(\bk)\tW_{33}(\bk)-\big(\tW_{13}(\bk)\big)^2\Big).
\notag
\end{align}
The remaining ones are obtained by symmetry.

For the derivation of
Eq.~\eqref{SFConstraintsRedundancyTest01&02},
inequalities~\eqref{SFConstraintsRedundancyTest} are combined,
operated on  with the help of the following inequality,
\begin{align*}
[(\lambda\, \ta f_a)^{1/2}-(\tb f_b/\lambda)^{1/2}]^2\geq 0,
\quad
 \lambda=\bigg(\int_{\mathbb{R}^3}d\bk\,\tb(\bk)\fb
                  \Big\slash \s \int_{\mathbb{R}^3}d\bk\,\ta(\bk)\fa \bigg)^{1/2},
\end{align*}
and then integrated to obtain
\begin{align*}
\l|\int_{\mathbb{R}^3}d\bk\,\tc(\bk)\fc\r|^2
\leq 
\int_{\mathbb{R}^3}d\bk\,\ta(\bk)\fa \int_{\mathbb{R}^3}d\bk\,\tb(\bk)\fb.
\end{align*}

 As an example,  consider
\begin{align}
\label{StructureFunctions_SOC}
a=\w_i(\by)-\w_i(\bx)+\alpha [\w_i(\bz')-\w_i(\bz)],
\quad
b=\Vorticity_j(\bz')-\Vorticity_j(\bz)-\beta[\Vorticity_j(\by)-\Vorticity_j(\bx)].
\end{align}
The Cauchy-Schwarz inequality 
$\big|\overline{a b}\big|^2\leq \overline{a a}\,\overline{b b}$
is  applied to obtain
\begin{align}
\label{StructureFunction_wvIneq_SOC}
\bigg|\int_{\mathbb{R}^3} d\bk\, \overline{\tw_i \,\tv_j}(\bk) \fc\bigg|^2
\leq
\int_{\mathbb{R}^3} d\bk\, \tW_{\underlinei\underlinei}(\bk) \fa
\int_{\mathbb{R}^3} d\bk\, \overline{\tv_{\underlinej}\tv_{\underlinej}}(\bk) \fb,
\end{align}
where
\begin{align}
\label{StructureFunction_wvIneqCoeff_SOC}
&
 \fa
=
4\bigg(\Big|\sin\frac{\br\cdot\bk}{2}\Big|^2
+\Big|\alpha\sin\frac{(\bs'-\bs)\cdot\bk}{2}\Big|^2
+2\sin\frac{\br\cdot\bk}{2}\,
  \alpha\sin\frac{(\bs'-\bs)\cdot\bk}{2}\cos\frac{(\bs'+\bs-\br)\cdot\bk}{2}
\bigg)
\notag\\&\hskip4mm
\geq
4\bigg( 
 \Big|\sin\frac{\br\cdot\bk}{2}\Big|
-\Big|\alpha\sin\frac{(\bs'-\bs)\cdot\bk}{2}\Big|
\bigg)^2,
\notag\\&
\fb
=
4\bigg(
 \Big|\sin\frac{(\bs'-\bs)\cdot\bk}{2}\Big|^2
+\Big|\beta\,\sin\frac{\br\cdot\bk}{2}\Big|^2
-2\sin\frac{(\bs'-\bs)\cdot\bk}{2}\,\beta \sin\frac{\br\cdot\bk}{2}
      \cos\frac{(\bs'+\bs-\br)\cdot\bk}{2}
\bigg)
\notag\\&\hskip4mm
\geq
4\bigg(
 \Big|\sin\frac{(\bs'-\bs)\cdot\bk}{2}\Big|
-\Big|\beta\,\sin\frac{\br\cdot\bk}{2}\Big|
\bigg)^2,
\\&
\fc=
\imaginary\,
(1+\alpha\beta) 
\Big[ 
 \sin[\bk\cdot(\bs'-\br)]
-\sin[\bk\cdot(\bs-\br)]
-\sin(\bk\cdot\bs')+\sin(\bk\cdot\bs)
\Big],
\notag\\&
\fa\,\fb
-\l|\fc\r|^2
=
16
\bigg(
 \alpha\bigg|\sin\frac{(\bs'-\bs)\cdot\bk}{2}\bigg|^2
  -\beta\bigg|\sin\frac{\br\cdot\bk}{2}\bigg|^2
+(1-\alpha\beta)
  \sin\frac{\br\cdot\bk}{2}
  \sin\frac{(\bs'-\bs)\cdot\bk}{2}
  \cos\frac{(\bs'+\bs-\br)\cdot\bk}{2}
\bigg)^2.
\notag
\end{align}
Inequalities~\eqref{StructureFunction_wvIneqCoeff_SOC}
imply automatic satisfaction of Eq.~\eqref{StructureFunction_wvIneq_SOC},
by the use of Eqs.~\eqref{twiwj_CSInWNS},
\eqref{tviwj_twivj_CSInWNS}, \eqref{SFConstraintsRedundancyTest}, and
\eqref{SFConstraintsRedundancyTest01&02}.

Next, $\big|\overline{a b}\big|^2\leq \overline{a a}\,\overline{b b}$
is applied to two more examples,
\begin{subequations}
\label{MoreStructureFunctions_SOC}
\begin{align}
a=\w_i(\by)-\w_i(\bx)+\alpha [\w_i(\bz')-\w_i(\bz)],
\quad
b=\w_j(\bz')-\w_j(\bz)-\beta [\w_j(\by)-\w_j(\bx))];
\end{align}
\vspace{-2em}
\begin{align}
a=\Vorticity_i(\by)-\Vorticity_i(\bx)+\alpha [\Vorticity_i(\bz')-\Vorticity_i(\bz))],
\quad
b=\Vorticity_j(\bz')-\Vorticity_j(\bz)-\beta[\Vorticity_j(\by)-\Vorticity_j(\bx)].
\end{align}
\end{subequations}
A straight-forward operation results in the following constraints of inequality,
\begin{subequations}
\label{StructureFunction_wwvvIneq_SOCShortened}
\begin{align}
\bigg|\int_{\mathbb{R}^3} d\bk\, \tW_{ij}(\bk) \fc\bigg|^2
\leq
\int_{\mathbb{R}^3} d\bk\, \tW_{\underlinei\underlinei}(\bk) \fa
\int_{\mathbb{R}^3} d\bk\, \tW_{\underlinej\underlinej}(\bk) \fb,
\end{align}
\vspace{-1.5em}
\begin{align}
\bigg|\int_{\mathbb{R}^3} d\bk\, \overline{\tv_i \,\tv_j}(\bk) f_c\bigg|^2
\leq
\int_{\mathbb{R}^3} d\bk\, \overline{\tv_{\underlinei}\tv_{\underlinei}}(\bk)\fa
 \int_{\mathbb{R}^3} d\bk\, \overline{\tv_{\underlinej}\tv_{\underlinej}}(\bk) \fb.
\end{align}
\end{subequations}
Here, $\fa$ and $\fb$ are given
by Eqs.~\eqref{StructureFunction_wvIneqCoeff_SOC}, and 
\begin{align}
\label{StructureFunction_wwvvIneq_SOCShortenedCoeff}
&
\fc=
 \cos[\bk\cdot(\bs'-\br)]
-\cos[\bk\cdot(\bs-\br)]
-\cos(\bk\cdot\bs')
+\cos(\bk\cdot\bs)
+2\alpha \big(1-\cos[\bk\cdot(\bs'-\bs)]\big)
 \notag\\&\hskip6.5mm
-\alpha\beta \big( \cos[\bk\cdot(\bs'-\br)]
-\cos[\bk\cdot(\bs-\br)]-\cos(\bk\cdot\bs')+\cos(\bk\cdot\bs) \big)
-2\,\beta \big(1-\cos(\bk\cdot\br)\big),
\\&
\fa\,\fb-|\fc|^2
=
16(1+\alpha \beta)^2
 \bigg|\sin\frac{\br\cdot\bk}{2}\,\sin\frac{(\bs'-\bs)\cdot\bk}{2}\bigg|^2
\bigg(
 1-\Big|\cos\frac{(\bs'+\bs-\br)\cdot\bk}{2}\Big|^2
\bigg).
\notag
\end{align}
Therefore, inequalities~\eqref{StructureFunction_wwvvIneq_SOCShortened}
are satisfied automatically because of  elementary
constraints~\eqref{twiwj_CSInWNS}.

\section{\label{Appensec:ShannonEntropy}Application of Shannon entropy}

This appendix discusses the application of Shannon entropy 
to the framework.
The argument intends to be conceptual and illustrative.
It serves the purpose of obtaining an objective functional.

\paragraph{Discretized representation}
Consider the equations of motion~\eqref{NSEqns}
for $\w_i$  and $q$.
In numerical simulation,
 the  domain of motion $\mathbb{R}^3$ is replaced 
with a cube ${\cal D}=[-L,L]^3$  with $L$  chosen 
appropriately.

The fluctuation fields $\w_i$ and $q$ are represented in terms of a truncated 
function basis $\{\psi_m(\bx):\ m=1,\ldots,M\}$,
\begin{align}
 \w_i(\bx,t)=c_{i,m}(t) \psi_m(\bx),
 \quad
 q=c_{0,m}(t) \psi_m(\bx),
\label{VelocityFluctuationRepresentation}
\end{align}
where  $\bc=\{c_{i,j}\}$ characterize the turbulent fluctuations
and contain  $M_c=4M$ elements.
The dynamical equations of $\bc(t)$ may be derived 
from Eqs.~\eqref{NSEqns} and \eqref{VelocityFluctuationRepresentation},
that are not required in a formal argument below.

On the basis of Eqs.~\eqref{VelocityFluctuationRepresentation},
 we introduce a probability density function $f(\bc)$
to  describe $\w_i$ and $q$ statistically,
instead of a complicated probability density functional 
 $\hat{f}\bm{(}\w_i(\bx,t),q(\bx,t)\bm{)}$.
 Though $\bc$ depends on $t$,  $f(\bc)$ may be viewed
   effectively as a function, 
      since $\bc$ is to be discretized.
 $|\w_i(\bx,t)|$ are  supposedly bounded.
Hence,
 $f$ has a support $[-c_0,c_0]^{M_c}$ for some positive constant $c_0$
 which is yet to be fixed as part of the treatment.
Further, the structure of $f$ is constrained by the definition of $f$
($f(\bc)\geq 0$, $\int f(\bc)\, d\bc=1$),
 $\int \bc f(\bc)\, d\bc=\mathbf{0}$, 
 the dynamical equations (to be discretized in time), and the
 constraints for the spatial correlations developed
 (without the imposition of homogeneity throughout ${\cal D}$,
 especially in the boundary subregion).
All these constraints may be expressed in terms of $\bc$
via intermediate representations, such as
\begin{align*}
 \overline{\w_k(\bx,t)}
= \psi_m(\bx) \int_{\{|c_{i,j}|\leq c_0\}} 
                            c_{k,m}\, f(\bc)\, d\bc,\ \
 \overline{\w_k(\bx,t)\w_l(\by,t)}
= \psi_m(\bx) \psi_n(\by)
       \int_{\{|c_{i,j}|\leq c_0\}} c_{k,m}\, c_{l,n}\, f(\bc) \, d\bc,\
       \ldots
\end{align*}

\paragraph{Differential and Shannon entropies}
We define the differential entropy 
 \cite{Shannon1948, Jaynes1983, CoverThomas2006},
\begin{align}
H(f)=
-\int_{\{|c_{i,j}|\leq c_0\}} f(\bc) \log f(\bc) \, d\bc.
\label{DifferentialEntropy}
\end{align}
For convenience,
we do not adopt the concept of
relative entropy  \cite{Jaynes1983, CoverThomas2006},
since it does not affect the alternative objective
 obtained (see  Eq.~\eqref{ObjectiveFunctionMaxIntWkkbr}).
If necessary,  a relative entropy may be defined
by the use of the uniform probability density function over
the support of $f$.

To evaluate $H(f)$ numerically, a uniform mesh (of size $\Delta$ in each direction)
 is adopted to discretize the support of $f$.
The centers of mesh elements are 
\begin{align}
 {\cal M}=\prod_{k,l} \big\{c_{k,l}=&-c_0+(m_{k,l}-1/2)\Delta:
\ 
          m_{k,l}=1,\ldots,2c_0/\Delta\big\}.
\end{align}
We approximate $f$ simply via the nodal values
$\{f(\theta)\}=\{f(\theta): \theta\in{\cal M}\}$
(piecewise constant).
We then resort to
the basic  composite midpoint rule for equal mesh elements
to discretize and approximate  $H(f)$,
 \begin{align}
H(f)
 \approx
-\Delta^{M_c}\sum_{\theta\in {\cal M}}
      f(\theta) \log f(\theta) 
=
 H_{\Delta}(f)
+\log \big(\Delta^{M_c}\big)
      \sum_{\theta\in {\cal M}} \big[f(\theta)\Delta^{M_c}\big].
\label{DifferentialEntropyApprox}
\end{align}
$H_{\Delta}(f)$ is the Shannon entropy associated with the discretized
$f$ \cite{CoverThomas2006},
\begin{align}
\label{ShannonEntropy}
H_{\Delta}(f)=
-\sum_{\theta\in {\cal M}}
      \big[f(\theta)\Delta^{M_c}\big] \log\big[f(\theta)\Delta^{M_c}\big],
\end{align}
with $f(\theta)\Delta^{M_c}$ representing the $\theta$-element probability.

Similarly, the correlations
and  all the aforementioned constraints may be represented
in terms of the nodal values $\{f(\theta)\}$,
   these resulting constraints are to be enforced
 at adequately selected collocation points  in ${\cal D}$ at time instant $t$.
 The unknown $\{f(\theta)\}$ 
are constrained by all these discretized constraints.
According to information theory \cite{Shannon1948, Jaynes1957, Jaynes1983, CoverThomas2006}, 
 $H_{\Delta}(f)$ is a measure of  uncertainty
of the discretized fluctuations (for $\{c_{i,j}\}$ to take the discrete values
in ${\cal M}$),  $\{f(\theta)\}$ may be found through 
 maximization of $H_{\Delta}(f)$ subject to the  discretized constraints.
This idea is adopted here.

In numerical simulation, an adequately small and finite
value of $\Delta$ is fixed 
to carry out the above procedure of maximization.
In this procedure, by the definition of $f$,
\begin{align}
\sum_{\theta\in {\cal M}} \big[f(\theta)\Delta^{M_c}\big]=1
\label{pdfApproxNormalized}
\end{align}
 is enforced. Combination of
    Eqs.~\eqref{DifferentialEntropyApprox}  and \eqref{pdfApproxNormalized} yields
 \begin{align}
\max  H_{\Delta}(f) \  \Leftrightarrow\  \max H(f),
\end{align}
in the sense that the optimal discrete values $\{f(\theta)\}$ 
 from   $\max H_{\Delta}(f)$ approximate
  the optimal continuous density function $f$ from  $\max H(f)$.
Therefore, from a numerical simulation point of view,  the maximization of Shannon entropy 
effectively implies the maximization of the differential entropy $\max H(f)$.
Considering  the debate regarding  $\max H(f)$ as a principle,
 we take $\max H_{\Delta}(f)$ as the starting point and 
$\max H(f)$ as the approximate and effective consequence,
because $f$ needs to be solved numerically, if ever possible.

\paragraph{Alternative objective}
There are tremendous difficulties
 to solve  $\max H(f)$, such as
 the computationally formidable size of a discretized $f$, 
   numerical integration in very high $M_c$-dimensions, etc.
 An alternative objective needs to be sought
 which is computationally feasible or explorable
 and is closely related to  $\max H(f)$.

 The solution of $\max H(f)$
is  the broadest distribution $f$ with a large spread-out, 
characterized partially by large values of the second order moments, 
according to probability theory,
\begin{align*}
 \int_{\{|c_{i,j}|\leq c_0\}}c_{\underlinek,m}\, c_{\underlinek,n}
           f(\bc) \, d\bc,
\end{align*}
which affects the second order correlations through
\begin{align}
\W_{\underlinek \underlinek }(\br)=
 \psi_m(\bo) \psi_n(\br)
   \int_{\{|c_{i,j}|\leq c_0\}}c_{\underlinek,m}\, c_{\underlinek,n}
        f(\bc)\, d\bc,
\label{SOCviapdf}
\end{align}
which holds expectedly for $\br=\by-\bx$ with $\bx$ and $\by$
away from the boundary subregion.
This selection of the second order moments
and the second order correlations is needed,
because an objective should  reflect
the direct and collective effect of all the 
dynamical equations and the constraints for the correlations.
To construct a single objective  from Eq.~\eqref{SOCviapdf},
 a scalar quantity is required which is
 independent of the Cartesian coordinate systems chosen
 to describe the turbulent motion,
  in order to satisfy  objectivity.
  The natural choice is the trace of the 
   second order correlations, $\W_{kk}(\br)$.

\section{\label{Appensec:SOCP}Second-order cone programming}

Both models, in their discretized forms,
are SOCP, according to the 
criteria set in \cite{Loboetal1998, AlizadehGoldfarb2003}.

Within the third-order model,
$\tW_{(ij)kl}$ act as the control variables,
$\tW_{ij}$, $\tWImag_{ijk}$, $\tQ$, $\tQImag_j$, $\tQ_{ij}$,
and their corresponding quantities in physical space are
 state variables.
With the help of the method of characteristics, Eqs.~\eqref{vEvolutionOftWPrimarySOM}
and \eqref{vEvolutionOftWijkPrimary_SOM} may be formally solved,
the state variables are related to the control variables linearly, 
and thus,
all the state variables, the constraints,
and the dynamical equations of evolution are expressed explicitly
in terms of $\tW_{(ij)kl}$, 
  as is the  objective functional~\eqref{ObjectiveFunctionMaxIntWkkbr}.

Computationally, $\tW_{(ij)kl}(\bk,\bl,t)$ need to be discretized
inside an adequate bounded support in the $(\bk,\bl)$-space and in $t$.
To establish the discretized third-order model as a SOCP,
it is sufficient to consider the following:
(a) The support is discretized into a finite element mesh.
(b) At instant $t=t^m$,
  inside each element the spatial distributions of 
  $\tW_{(ij)kl}(\bk,\bl,t^m)$ are approximated
  linearly in terms of the $\tW_{(ij)kl}$ at the nodes of the element;
  the collection of all these nodal control variables is denoted by $\{\delta_i^{m}\}$.
(c) At fixed $\bk$ and $\bl$,
    the temporal distributions of $\tW_{(ij)kl}(\bk,\bl,t)$
   in $t\in[t^{m-1},t^{m}]$ are approximated
  linearly in terms of  $\tW_{(ij)kl}(\bk,\bl,t^{m-1})$ and  $\tW_{(ij)kl}(\bk,\bl,t^m)$.
(d) At $t^{m}$, all the state variables depend linearly on $\{\delta_i^{m}\}$
    (and on $\{\delta_i^{m-1}\}$ too, which is known).
(e) All the linear constraints and the  objective have linear
forms in terms of  $\{\delta_i^{m}\}$.
(f) The quadratic constraints from the Cauchy-Schwarz inequality take the form, 
\begin{align}
\label{QuadraticFormPeculiar}
\l|\l|\l[\s
\begin{array}{cc}
 2(c_i\,\delta^{m}_i+c_0)\\
 (a_i-b_i)\,\delta^{m}_i+a_0-b_0
\end{array}
\s\r]\r|\r|_2
\leq (a_i+b_i)\,\delta^{m}_i+a_0+b_0,
\end{align}
where $a_k$, $b_k$, and $c_k$ are independent of $\{\delta_i^{m}\}$,
 $a_0$, $b_0$, and $c_0$ depend on $\{\delta_i^{m-1}\}$.
(g) The quadratic constraints from Eq.~\eqref{NonNegativeVariance}
are of the form,
\begin{align}
&
 \l|\l|\l[\s
\begin{array}{c}
 \frac{1}{2}(1-a'_i\,\delta_i^{m}-a'_0)\\
c'_i\,\delta_i^{m}+c'_0
\end{array}
\s\r]\r|\r|_{2}
\leq \frac{1}{2}(1+a'_i\delta_i^{m}+a'_0),
\quad
0\leq a'_i\delta_i^{m}+a'_0,
\label{QuadraticFormVariance_SOCP}
\end{align}
where $a'_k$ and $c'_k$ are independent of $\{\delta_i^{m}\}$,
 $a'_0$ and $c'_0$ depend on $\{\delta_i^{m-1}\}$.
It then follows
that the discretized third-order model is a SOCP,  according to the 
criteria in 
\cite{Loboetal1998, AlizadehGoldfarb2003}. 
Similarly, in a discretized form,
the second-order model is shown to be a SOCP with $\Contractedtoc_{ij}$
as the control variables;
The relevant details of discretization
in the asymptotic states are given in 
Appendix~\ref{Appensec:NumericalDiscretization}.

\section{\label{Appendsec:UpperBoundsImplications}Other upper bound for growth rate}

Combination of
  Eqs.~\eqref{vIntrinsicRelationForUkk(0)Evolution}, \eqref{twiwj_CSInWNS},
  and \eqref{AsymptoticForm} results in
\begin{subequations}
\begin{align}
\frac{\sigma}{2}\WAsy_{jj}(\bo)
=
-\WAsy_{12}(\bo)
-\overline{\w_j,_k\!(\bx) \w_j,_k\!(\bx)}^{(\infty)},
\label{IntrinsicRelationAsyForsigma}
\end{align}
\vspace{-2em}
\begin{align}
\Big|\WAsy_{12}(\bo)\Big| 
\leq \sqrt{\WAsy_{11}(\bo) \WAsy_{22}(\bo)}
 \leq \frac{1}{2}\Big(\WAsy_{11}(\bo)+\WAsy_{22}(\bo)\Big)
 =\frac{1}{2}\Big(\WAsy_{jj}(\bo)-\WAsy_{33}(\bo)\Big),
\label{InequalityFromWijPSD}
\end{align}
\end{subequations}
which, along with $\WAsy_{\underlinek\underlinek}(\bo)>0$
   and the positive viscous dissipation, leads to
\begin{align}
\sigma
\leq  1 
-\frac{2\,\overline{\w_j,_k\!(\bx) \w_j,_k\!(\bx)}^{(\infty)}+\WAsy_{33}(\bo)}{\WAsy_{jj}(\bo)}
   <1.
\label{IntrinsicUpperBoundForsigmaFromReducedModel}
\end{align}
This inequality implies that the second-order  
model  may allow the asymptotic solution with $\sigma\in(0,1)$
 to emerge out of a transient state,
violating  $\max\sigma=0$ of Eq.~\eqref{MaxsigmaFromDeviationConstraint}.
To obtain an affirmative answer to this possibility,
 we resort to the mathematical structure that the second-order model 
 and specifically $\tWAsy_{ij}$
    depend on $\sigma$
 only via $\Expal(\bk;k'_2;\sigma)$ defined by Eq.~\eqref{ExponentialFunctionPart}.
Function  $\Expal(\bk;k'_2;\sigma)$ indicates that, 
 only in  certain subregions,
 $\sigma$ affects $\tWAsy_{ij}$  significantly, 
 as explained next.
 
 Suppose that in the case of $\sigma=0$
 there exists a feasible solution
 $\big\{\tWAsy_{ij}$, $\dContractedtocAsy_{ij}\big\}_{\sigma=0}$
 which takes the value of zero in the subregion,
\begin{align}
 {\cal R}_0=\big\{\bk:\ k_1\in \big[-(K_0)^{1/2},0\big]\big\},
\label{SubRegionForBoundOfsigma}
\end{align}
where $K_0$ is a positive constant.
We argue that this feasible solution is
also a feasible solution for certain positive values of $\sigma$.
Under the condition of 
\begin{align}
|\sigma|< K_0/50,\ \ \text{say},
\label{BoundForPerturbationOfsigma}
\end{align}
there is,
\begin{align}
&
(k_1)^2+(k_3)^2+\frac{(k_2)^2+(k'_2)^2+k_2k'_2}{3}
\geq K_0
>>\frac{|\sigma|}{2},\ \ 
\forall k_1 \leq -(K_0)^{1/2}.
\end{align}
That is, within the range~\eqref{BoundForPerturbationOfsigma}, 
$\sigma$ has a negligible effect on $\Expal(\bk;k'_2;\sigma)$ and $\tWAsy_{ij}(\bk)$,
$\forall \bk\not\in {\cal R}_0$.
 For $\bk\in {\cal R}_0$, $\sigma$ has no effect on $\tWAsy_{ij}$,
 owing to the condition underlying ${\cal R}_0$.
 Hence, this $\big\{\tWAsy_{ij}$,  $\dContractedtocAsy_{ij}\big\}_{\sigma=0}$ 
 is also a feasible solution for
 the values of $\sigma$ 
 satisfying  the restriction~\eqref{BoundForPerturbationOfsigma}.
 This feasible solution guarantees the existence of  optimal solutions
 to the second-order model for the values of $\sigma$
  satisfying \eqref{BoundForPerturbationOfsigma}.
 Such a feasible solution of $\sigma=0$ may be obtained  numerically 
 (say, with an appropriate choice of a rather small support for $\dContractedtocAsy_{ij}$ 
 to reduce the computational size, allowable by the boundary conditions and  constraints
 of the model).
 As an example, we consider the asymptotic steady state simulation of
 the second-order model, specifically
 the supports in Eqs.~\eqref{CVSVSupport_OS} and 
 the associated  solutions. Equations~\eqref{CVSVSupport_OS} and
 \eqref{SubRegionForBoundOfsigma} yield  $(K_0)^{1/2}=0.01$,
 the corresponding values of
    $|\sigma|$ in \eqref{BoundForPerturbationOfsigma} can be obtained,
 though they are quite small.

\section{\label{Appendsec:LimitingBehavior}Limiting behavior under $k_1\rightarrow 0^{-}$}

Since the primary component solutions~\eqref{dWij_Asy_Sol} appear 
to be singular at $k_1=0$, this issue
and the associated limiting behavior under $k_1\rightarrow 0^{-}$
are discussed in this appendix.
Both $\dWAsy_{ij}$ and $\dContractedtocAsy_{ij}$
are supposedly bounded and smooth in the wave-number space
on the basis of $\WAsy_{ij}(\br)$ and $\WAsy_{(ij)k}(\br)$
being bounded and smooth. 
For simplicity, the discussion is restricted to $\sigma=0$
and the original formal solutions~\eqref{EvolutionOfdWij_Asy_Sol}
(leading to Eqs.~\eqref{dWij_Asy_Sol})
are analyzed,
\begin{subequations}
\label{EvolutionOfdWij_Asy_Sol}
\begin{align}
\label{EvolutionOfdW22_Asy_Sol}   
\dWAsy_{22}(\bk)
=
 \int_{-\infty}^{k_2} & dk'_2
   \frac{2 |\bk'|^4 \Expal(\bk;k'_2;\sigma)}{|k_1| |\bk|^4} 
  \bigg( \frac{k'_2 k'_l}{|\bk'|^2}\dContractedtocAsy_{l2}(\bk')
          -\dContractedtocAsy_{22}(\bk')\bigg),
\end{align}
\vspace{-1.2em}
\begin{align}
\label{EvolutionOfdW12_Asy_Sol}   
             \dWAsy_{12}(\bk)
=&
 \int_{-\infty}^{k_2} dk'_2 
  \frac{|\bk'|^2 \Expal(\bk;k'_2;\sigma)}{|k_1| |\bk|^2}
\bigg[
     -\bigg(1-\frac{2(k_1)^2}{|\bk'|^2}\bigg)\dWAsy_{22}(\bk')
+\frac{k'_l}{|\bk'|^2}\Big(k_1\dContractedtocAsy_{l2}(\bk')+k'_2\dContractedtocAsy_{l1}(\bk')\Big)
\notag\\&\hskip39.5mm
-\Big(\dContractedtocAsy_{12}(\bk')+\dContractedtocAsy_{21}(\bk')\Big)
\bigg],
\end{align}
\vspace{-1.5em}
\begin{align}
\label{EvolutionOfdW11_Asy_Sol}   
       \dWAsy_{11}(\bk)
=
  \int_{-\infty}^{k_2} dk'_2
      \frac{2 \Expal(\bk;k'_2;\sigma)}{|k_1|}
\bigg[-\bigg(1-\frac{2 (k_1)^2}{|\bk'|^2}\bigg)\dWAsy_{12}(\bk') 
       +\frac{k_1 k'_l}{|\bk'|^2}\dContractedtocAsy_{l1}(\bk')  
      -\dContractedtocAsy_{11}(\bk')
\bigg],
\end{align}
\end{subequations}
where  $\bk'=(k_1,k'_2,k_3)$, $k_1<0$.
They are obtained with the use of Eqs.~\eqref{EvolutionOftWij_Asy},
\eqref{BoundaryConditionAInftyFortWij},
 and \eqref{GammijtWiiTransformed}.

First, consider Eq.~\eqref{EvolutionOfdW11_Asy_Sol} in the form of
\begin{subequations}
\label{EvolutionOfdW11_Asy_SolRecast}
\begin{align}
\label{EvolutionOfdW11_Asy_SolRecasta}
\dWAsy_{11}(\bk)
=
 \int_{-\infty}^{k_2} dk'_2\,
            \DensitydWAsyaa(\bk;k'_2),
\end{align}
\vspace{-1.5em}
\begin{align}
\DensitydWAsyaa(\bk;k'_2)
=\frac{2 \Expal(\bk;k'_2;\sigma)}{|k_1|}
\bigg[-\bigg(1-\frac{2 (k_1)^2}{|\bk'|^2}\bigg)\dWAsy_{12}(\bk') 
       +\frac{k_1 k'_l \dContractedtocAsy_{l1}(\bk')}{|\bk'|^2}  
      -\dContractedtocAsy_{11}(\bk')
\bigg].
\end{align}
\end{subequations}
To make $\dWAsy_{11}$  bounded and  smooth, 
the control variables $\dContractedtocAsy_{ij}$ need to behave such that
the density function $\rho^{(\infty)}_{11}(\bk;k'_2)$ with $k_2 k_3\not=0$
has a definite limit
under $k_1\rightarrow 0^{-}$ and $k'_2\rightarrow k_2^{-}$, 
\begin{align}
\label{dW11LimitExistenceunderk1}
\lim_{k_1\rightarrow 0^{-}} \lim_{k'_2\rightarrow k_2^-}\DensitydWAsyaa(\bk;k'_2)
=
\lim_{k'_2\rightarrow k_2^-} \lim_{k_1\rightarrow 0^{-}}\DensitydWAsyaa(\bk;k'_2),
\end{align}
which results in
\begin{subequations}
\begin{align}
\label{dW11Limitunderk1}
\lim_{k_1\rightarrow 0^{-}} 
\bigg(
 \frac{|\bk|^2 \dWAsy_{12}(\bk)+|\bk|^2 \dContractedtocAsy_{11}(\bk)}{k_1} 
       -k_2\dContractedtocAsy_{21}(\bk)
       -k_3\dContractedtocAsy_{31}(\bk)
\bigg)=0,
\ 
\forall k_2 k_3\not=0,
\end{align}
\vspace{-1.5em}
\begin{align}
\label{dW11Limitunderk1A}
 \lim_{k_1\rightarrow 0^{-}} \DensitydWAsyaa(\bk;k'_2)=0.
\end{align}
\end{subequations}

Second,
Eq.~\eqref{dW11Limitunderk1} implies
  that $\dWAsy_{12}(\bk)/k_1$ needs to be well-defined,
   providing the ground to analyze Eq.~\eqref{EvolutionOfdW12_Asy_Sol}
 in the form,
\begin{subequations}
\begin{align}
\frac{\dWAsy_{12}(\bk)}{k_1}
=
 \int_{-\infty}^{k_2} dk'_2\,
            \DensitydWAsyab(\bk;k'_2),
\end{align}
\vspace{-1.5em}
\begin{align}
\DensitydWAsyab(\bk;k'_2)
=
 \frac{|\bk'|^2 \Expal(\bk;k'_2;\sigma)}{|k_1|^2 |\bk|^2}
 \bigg[
      \bigg(1-\frac{2(k_1)^2}{|\bk'|^2}\bigg)\dWAsy_{22}(\bk')
-\frac{k'_l \big(k_1\dContractedtocAsy_{l2}(\bk')+k'_2\dContractedtocAsy_{l1}(\bk')\big)}{|\bk'|^2}
+\dContractedtocAsy_{12}(\bk')+\dContractedtocAsy_{21}(\bk')
\bigg]. 
\end{align}
\end{subequations}
Applying a similar argument for the existence of limit
under $k_1\rightarrow 0^{-}$ and $k'_2\rightarrow k_2^{-}$
to $\DensitydWAsyab(\bk;k'_2)$ leads to  
\begin{align}
\label{dW12Limitunderk1}
&
\lim_{k_1\rightarrow 0^{-}}
 \bigg(
       \frac{|\bk|^2 \dWAsy_{22}(\bk)}{(k_1)^2}
-\frac{k_2[\dContractedtocAsy_{11}(\bk)+\dContractedtocAsy_{22}(\bk)]}{k_1}
+\frac{[(k_2)^2+(k_3)^2] \dContractedtocAsy_{12}(\bk)+(k_3)^2 \dContractedtocAsy_{21}(\bk)}{(k_1)^2}
\notag\\&\hskip13mm
-\frac{k_3 [k_2\dContractedtocAsy_{31}(\bk)+k_1\dContractedtocAsy_{32}(\bk)]}{(k_1)^2}
\bigg)=0,\ \ 
 \forall k_2 k_3\not=0.
\end{align}

Third, the structure of Eq.~\eqref{dW12Limitunderk1} motivates the
consideration of $\dWAsy_{22}(\bk)/(k_1)^2$, and
Eq.~\eqref{EvolutionOfdW22_Asy_Sol} is written as
\begin{subequations}
\begin{align}
\frac{\dWAsy_{22}(\bk)}{(k_1)^2}
=
 \int_{-\infty}^{k_2} dk'_2\,
            \DensitydWAsybb(\bk;k'_2),
\end{align}
\vspace{-1.5em}
\begin{align}
\DensitydWAsybb(\bk;k'_2)
=\frac{2|\bk'|^4 \Expal(\bk;k'_2;\sigma)}{|k_1|^3|\bk|^4}
 \bigg(\frac{k'_2 k'_l \dContractedtocAsy_{l2}(\bk')}{|\bk'|^2}
       -\dContractedtocAsy_{22}(\bk')\bigg).
\end{align}
\end{subequations}
Applying a similar argument for the existence of limit
under $k_1\rightarrow 0^{-}$ and $k'_2\rightarrow k_2^{-}$
to $\DensitydWAsybb(\bk;k'_2)$, we obtain
\begin{align}
\label{dW22Limitunderk1}
\lim_{k_1\rightarrow 0^{-}}
\frac{ k_2 k_1 \dContractedtocAsy_{12}(\bk)
 -(k_3)^2 \dContractedtocAsy_{22}(\bk)
 +k_2 k_3 \dContractedtocAsy_{32}(\bk)}{|k_1|^3}
 =0,\ \ 
 \forall k_2 k_3\not=0.
\end{align}

Since $\dContractedtocAsy_{ij}$, $ij=11,21,31,12,22,32$, 
are the primary control variables,
we may infer from Eqs.~\eqref{dW11Limitunderk1},
\eqref{dW12Limitunderk1}, and \eqref{dW22Limitunderk1} that
\begin{align}
\label{dWijLimitunderk1Inferred}
&
 \dContractedtocAsy_{11}(\bk)\propto k_1,\ \
 \dContractedtocAsy_{21}(\bk)\propto (k_1)^2,\ \
 \dContractedtocAsy_{31}(\bk)\propto (k_1)^2,\ \ 
 \dContractedtocAsy_{12}(\bk)\propto (k_1)^2,\ \ 
 \dContractedtocAsy_{22}(\bk)\propto (k_1)^3,\ \
\dContractedtocAsy_{32}(\bk)\propto (k_1)^3,
 \notag\\[4pt]&\hskip10mm
 \forall k_2 k_3\not=0 \ \text{and small $|k_1|$}.
\end{align}
Then, under the supposed continuity of $\dContractedtocAsy_{ij}$ in the 
wave-number space, 
Eqs.~\eqref{dWijLimitunderk1Inferred} imply that
\begin{align}
 \dContractedtocAsy_{ij}(0,k_2,k_3)=0.
\end{align}

Regarding the values of $\dWAsy_{ij}(0,k_2,k_3)$,
Eqs.~\eqref{dW11Limitunderk1} and \eqref{dW12Limitunderk1} imply
\begin{align}
 \dWAsy_{12}(0,k_2,k_3)=\dWAsy_{22}(0,k_2,k_3)=0.
\end{align}
A combination of Eqs.~\eqref{dW11Limitunderk1A} and \eqref{EvolutionOfdW11_Asy_SolRecasta}
implies
\begin{align}
 \dWAsy_{11}(0,k_2,k_3)=0.
\end{align}
A rigorous but lengthy proof for the above result
may also be offered  with the help of bounded $\dWAsy_{11}$,
Eq.~\eqref{dW11Limitunderk1}, and the property of $\Expal(\bk;k'_2;0)$.

\section{\label{Appensec:NumericalDiscretization}Discretization}

The supports $\EstimatedSupportForgammaij(0)$ and $\EstimatedSupportForWij(0)$
are estimated in Eqs.~\eqref{tCVgammaAsyij_SupportEstimate}
and \eqref{tWAsyij_SupportEstimate}.
They  are discretized, respectively, with  structured hexahedral meshes,
\begin{align}
\label{SupportsMesh}
&
\EstimatedSupportForgammaij(0)
=
\bigcup_{n_1=1}^{N_1-1} 
\bigcup_{n_2=1}^{N_2-1} 
\bigcup_{n_3=1}^{N_3-1} 
{\cal H}(n_1,n_2,n_3),
\quad
\EstimatedSupportForWij(0)
=
\bigcup_{n_1=1}^{N_1-1} 
\bigcup_{n_2=1}^{M_2-1} 
\bigcup_{n_3=1}^{N_3-1} 
{\cal H}(n_1,n_2,n_3),
\notag\\[-15pt]&
\\[2pt]&
{\cal H}(n_1,n_2,n_3)=
[k_{1,n_1},k_{1,n_1+1}]\times
      [k_{2,n_2},k_{2,n_2+1}]
      \times
      [k_{3,n_3},k_{3,n_3+1}].
\notag
\end{align}
The distribution of $\dContractedtocAsy_{ij}$
 in each element
    ${\cal H}(n_1,n_2,n_3)$ of $\EstimatedSupportForgammaij(0)$
 is approximated by the trilinear distribution, 
\begin{align}
\label{TriLinearApproximation}
&
\dContractedtocAsy_{ij}(\bk')
=
\chi_{[k_{1,n_1},\, k_{1,n_1+1}]}(k_1)
\chi_{[k_{2,n_2},\, k_{2,n_2+1}]}(k'_2)
\chi_{[k_{3,n_3},\, k_{3,n_3+1}]}(k_3)
      \pdContractedtocAsy_{ij}(\bk',n_1,n_2,n_3),
\notag\\[6pt]&
\pdContractedtocAsy_{ij}(\bk',n_1,n_2,n_3)
\notag\\
=
\,&
 \frac{ \dContracted_{ij}(n_1,n_2,n_3)(k_{2,n_2+1}-k'_2)
       +\dContracted_{ij}(n_1,n_2+1,n_3)(k'_2-k_{2,n_2})}{k_{2,n_2+1}-k_{2,n_2}}
 \frac{k_{1,n_1+1}-k_1}{k_{1,n_1+1}-k_{1,n_1}}
 \frac{k_{3,n_3+1}-k_3}{k_{3,n_3+1}-k_{3,n_3}}
\notag\\&
+\frac{ \dContracted_{ij}(n_1+1,n_2,n_3)(k_{2,n_2+1}-k'_2)
       +\dContracted_{ij}(n_1+1,n_2+1,n_3)(k'_2-k_{2,n_2})}{k_{2,n_2+1}-k_{2,n_2}}
 \frac{k_1-k_{1,n_1}}{k_{1,n_1+1}-k_{1,n_1}}
 \frac{k_{3,n_3+1}-k_3}{k_{3,n_3+1}-k_{3,n_3}}
\\&
+ \frac{ \dContracted_{ij}(n_1,n_2,n_3+1)(k_{2,n_2+1}-k'_2)
        +\dContracted_{ij}(n_1,n_2+1,n_3+1)(k'_2-k_{2,n_2})}{k_{2,n_2+1}-k_{2,n_2}}
 \frac{k_{1,n_1+1}-k_1}{k_{1,n_1+1}-k_{1,n_1}}
 \frac{k_3-k_{3,n_3}}{k_{3,n_3+1}-k_{3,n_3}}
\notag\\&
+\frac{ \dContracted_{ij}(n_1+1,n_2,n_3+1)(k_{2,n_2+1}-k'_2)
       +\dContracted_{ij}(n_1+1,n_2+1,n_3+1)(k'_2-k_{2,n_2})}{k_{2,n_2+1}-k_{2,n_2}}
 \frac{k_1-k_{1,n_1}}{k_{1,n_1+1}-k_{1,n_1}}
 \frac{k_3-k_{3,n_3}}{k_{3,n_3+1}-k_{3,n_3}}.
 \notag
\end{align}
Here, $\chi_{[k_{1,n_1},\, k_{1,n_1+1}]}$ and the like denote
the characteristic functions and
 $\dContracted_{ij}(n_1,n_2,n_3)$ denote the nodal values.
The piecewise trilinear approximations~\eqref{TriLinearApproximation}
may be adequate at small mesh element sizes
 since integration operations are mainly involved in
 the distribution of the control variables.

Substitution of Eqs.~\eqref{TriLinearApproximation}
into Eqs.~\eqref{dWij_Asy_Sol} gives
\begin{subequations}
\label{dWAsyij_Discrete}
\vspace{-0.5em}
\begin{align}
\label{dWAsy22_Discrete}
&
\dWAsy_{22}(\bk)
=
 \frac{2\chi_{[k_{1,n_1}, k_{1,n_1+1}]}(k_1)\chi_{[k_{3,n_3}, k_{3,n_3+1}]}(k_3)}{|k_1||\bk|^4}
\notag\\&\hskip15mm\times
    \sum_{n_2=1}^{N_2-1} 
 \int_{k_{2,1}}^{k_2} dk'_2
   \Expal(\bk;k'_2;0)
   |\bk'|^4\chi_{[k_{2,n_2},k_{2,n_2+1}]}(k'_2)
\bigg(
 \frac{k'_2 k'_l}{|\bk'|^2}\,\pdContractedtocAsy_{l2}(\bk',n_1,n_2,n_3)
 -\pdContractedtocAsy_{22}(\bk',n_1,n_2,n_3)
\bigg),
\end{align}
\vspace{-1em}
\begin{align}
\label{dWAsy12_Discrete}
&
 \dWAsy_{12}(\bk)
=
-\frac{2\chi_{[k_{1,n_1}, k_{1,n_1+1}]}(k_1)\chi_{[k_{3,n_3}, k_{3,n_3+1}]}(k_3)}
    {|k_1|^2[|k_1|^2+(k_3)^2]|\bk|^2}   
\notag\\&\hskip20mm\times
     \sum_{n_2=1}^{N_2-1} 
 \int_{k_{2,1}}^{k_2} dk'_2
                  \Expal(\bk;k'_2;0)
                  \big(F(\bk)-F(\bk')\big)
                  |\bk'|^4\chi_{[k_{2,n_2}, k_{2,n_2+1}]}(k'_2)
\notag\\&\hskip38mm\times 
\bigg(
 \frac{k'_2 k'_l}{|\bk'|^2}\,\pdContractedtocAsy_{l2}(\bk',n_1,n_2,n_3)
 -\pdContractedtocAsy_{22}(\bk',n_1,n_2,n_3)
\bigg)
\notag\\&\hskip17mm
+\frac{1}{|k_1||\bk|^2}
     \chi_{[k_{1,n_1}, k_{1,n_1+1}]}(k_1)\chi_{[k_{3,n_3}, k_{3,n_3+1}]}(k_3)
\notag\\&\hskip20mm\times
    \sum_{n_2=1}^{N_2-1} 
 \int_{k_{2,1}}^{k_2} dk'_2
                      \Expal(\bk;k'_2;0)
                      \chi_{[k_{2,n_2}, k_{2,n_2+1}]}(k'_2)
\notag\\&\hskip38mm\times
\bigg[
      k'_l\Big(
                k_1 \pdContractedtocAsy_{l2}(\bk',n_1,n_2,n_3)
               +k'_2 \pdContractedtocAsy_{l1}(\bk',n_1,n_2,n_3)
          \Big)
\notag\\&\hskip44mm
    -|\bk'|^2\Big(\pdContractedtocAsy_{12}(\bk',n_1,n_2,n_3)
                  +\pdContractedtocAsy_{21}(\bk',n_1,n_2,n_3)\Big)
\bigg],
\end{align}
\vspace{-1em}
\begin{align}
&
\label{dWAsy11_Discrete}
       \dWAsy_{11}(\bk)
=
 \frac{2\chi_{[k_{1,n_1}, k_{1,n_1+1}]}(k_1)\chi_{[k_{3,n_3}, k_{3,n_3+1}]}(k_3)}
      {|k_1|^3[|k_1|^2+(k_3)^2]^2}     
\notag\\&\hskip17mm\times
     \sum_{n_2=1}^{N_2-1} 
 \int_{k_{2,1}}^{k_2} dk'_2
           \Expal(\bk;k'_2;0)
           \big(F(\bk)-F(\bk')\big)^2
           |\bk'|^4\chi_{[k_{2,n_2}, k_{2,n_2+1}]}(k'_2)
\notag\\&\hskip34mm\times 
\bigg(
  \frac{k'_2 k'_l}{|\bk'|^2}\,\pdContractedtocAsy_{l2}(\bk',n_1,n_2,n_3)
 -\pdContractedtocAsy_{22}(\bk',n_1,n_2,n_3)
\bigg)
\notag\\&\hskip16mm
-\frac{2\chi_{[k_{1,n_1}, k_{1,n_1+1}]}(k_1)\chi_{[k_{3,n_3}, k_{3,n_3+1}]}(k_3)}
      {|k_1|^2[|k_1|^2+(k_3)^2]}   
\notag\\&\hskip20mm\times
    \sum_{n_2=1}^{N_2-1} 
 \int_{k_{2,1}}^{k_2} dk'_2
                      \Expal(\bk;k'_2;0)
                      \big(F(\bk)-F(\bk')\big)
                      \chi_{[k_{2,n_2}, k_{2,n_2+1}]}(k'_2)
\notag\\&\hskip38mm\times 
\bigg[
     k'_l
       \Big(
             k_1\pdContractedtocAsy_{l2}(\bk',n_1,n_2,n_3)
            +k'_2\pdContractedtocAsy_{l1}(\bk',n_1,n_2,n_3)
       \Big)       
\notag\\&\hskip44mm
-|\bk'|^2\Big(\pdContractedtocAsy_{12}(\bk',n_1,n_2,n_3)
              +\pdContractedtocAsy_{21}(\bk',n_1,n_2,n_3)\Big)
\bigg] 
\notag\\&\hskip16mm
+\frac{2 \chi_{[k_{1,n_1}, k_{1,n_1+1}]}(k_1)\chi_{[k_{3,n_3}, k_{3,n_3+1}]}(k_3)}{|k_1|}
\notag\\&\hskip20mm\times
     \sum_{n_2=1}^{N_2-1} 
 \int_{k_{2,1}}^{k_2} dk'_2
          \Expal(\bk;k'_2;0)\chi_{[k_{2,n_2}, k_{2,n_2+1}]}(k'_2)
\bigg( 
       \frac{k_1 k'_l}{|\bk'|^2}\,\pdContractedtocAsy_{l1}(\bk',n_1,n_2,n_3)
      -\pdContractedtocAsy_{11}(\bk',n_1,n_2,n_3)
\bigg).
\end{align}
\end{subequations}
The expressions may be recast further with the help of
\begin{align}
\Expal(\bk;k'_2;\sigma)=\Expal(\bk;k_{2,n_2};\sigma)\Expal((k_1,k_{2,n_2},k_3);k'_2;\sigma),
\label{DcomposibityOfE}
\end{align}
which is used in coding for the calculation of 
$\tWAsy_{11}$, $\tWAsy_{12}$, and $\tWAsy_{22}$
 at the collocation points specified in Appendix~\ref{Appensec:DiscretizedModel}.

In the evaluation of the objective
and the second order correlations in the physical space,
  three types of integrals are computed,
 owing to the primary  components
 $\tWAsy_{11}$, $\tWAsy_{12}$, and $\tWAsy_{22}$.
They are
\begin{subequations}
\label{Intdbk_PsiijdWAsyij_GeneralCase}
\begin{align}
\label{Intdbk_Psi22dWAsy22_GeneralCase}
&
\int_{-\infty}^{0}dk_1\int_{0}^{+\infty}dk_3
\int_{\mathbb{R}}dk_2\Psi_{22}(\bk,\br)\dWAsy_{22}(\bk)
\notag\\&
=
\sum\iiiint
           \frac{2\Psi_{22}(\bk,\br)\Expal(\bk;k'_2;0)|\bk'|^4}{|k_1||\bk|^4}
\bigg( \frac{k'_2 k'_l}{|\bk'|^2}\pdContractedtocAsy_{l2}(\bk',n_1,n_2,n_3)
         -\pdContractedtocAsy_{22}(\bk',n_1,n_2,n_3)\bigg),
\end{align}
\vspace{-.5em}
\begin{align}
\label{Intdbk_Psi12dWAsy12_GeneralCase}
&
\int_{-\infty}^{0}dk_1\int_{0}^{+\infty}dk_3
\int_{\mathbb{R}}dk_2\Psi_{12}(\bk,\br)\dWAsy_{12}(\bk)
\notag\\&
=
\sum\iiiint
\frac{-2\Psi_{12}(\bk,\br)\Expal(\bk;k'_2;0)
\big(F(\bk)-F(\bk')\big)|\bk'|^4}{|k_1|^2[|k_1|^2+(k_3)^2]|\bk|^2}
\bigg( \frac{k'_2 k'_l}{|\bk'|^2}\pdContractedtocAsy_{l2}(\bk',n_1,n_2,n_3)
                       -\pdContractedtocAsy_{22}(\bk',n_1,n_2,n_3)\bigg)
\notag\\&\hskip3mm
+\sum\iiiint
\frac{\Psi_{12}(\bk,\br)\,\Expal(\bk;k'_2;0)|\bk'|^2}{|k_1||\bk|^2}
\bigg[
 \frac{k'_l}{|\bk'|^2}\Big(k_1\,\pdContractedtocAsy_{l2}(\bk',n_1,n_2,n_3)
                +k'_2\pdContractedtocAsy_{l1}(\bk',n_1,n_2,n_3)\Big)
\notag\\&\hskip63.5mm
    -\pdContractedtocAsy_{12}(\bk',n_1,n_2,n_3)
    -\pdContractedtocAsy_{21}(\bk',n_1,n_2,n_3)
\bigg],
\end{align}
\vspace{-1.5em}
\begin{align}
\label{Intdbk_Psi11dWAsy11_GeneralCase}
&
\int_{-\infty}^{0}dk_1\int_{0}^{+\infty}dk_3
\int_{\mathbb{R}}dk_2\Psi_{11}(\bk,\br)\,\dWAsy_{11}(\bk)
\notag\\&
=
\sum\iiiint
\frac{2\Psi_{11}(\bk,\br)\Expal(\bk;k'_2;0)
\big(F(\bk)-F(\bk')\big)^2|\bk'|^4}{|k_1|^3[|k_1|^2+(k_3)^2]^2}
\bigg( \frac{k'_2 k'_l}{|\bk'|^2}\pdContractedtocAsy_{l2}(\bk',n_1,n_2,n_3)
                           -\pdContractedtocAsy_{22}(\bk',n_1,n_2,n_3)\bigg)
\notag\\&\hskip3mm
+\sum\iiiint
\frac{-2\Psi_{11}(\bk,\br)\,\Expal(\bk;k'_2;0)
\big(F(\bk)-F(\bk')\big)}{|k_1|^2[|k_1|^2+(k_3)^2]}
\bigg[
        k'_l\Big(k_1\pdContractedtocAsy_{l2}(\bk',n_1,n_2,n_3)
              +k'_2\pdContractedtocAsy_{l1}(\bk',n_1,n_2,n_3)\Big)
\notag\\&\hskip84.5mm
-|\bk'|^2\Big(\pdContractedtocAsy_{12}(\bk',n_1,n_2,n_3)
              +\pdContractedtocAsy_{21}(\bk',n_1,n_2,n_3)\Big)
\bigg] 
\notag\\&\hskip3mm
+\sum\iiiint
\frac{2\Psi_{11}(\bk,\br)\Expal(\bk;k'_2;0)}{|k_1|}
 \bigg( \frac{k_1 k'_l}{|\bk'|^2}\pdContractedtocAsy_{l1}(\bk',n_1,n_2,n_3)  
       -\pdContractedtocAsy_{11}(\bk',n_1,n_2,n_3)
\bigg).
\end{align}
\end{subequations}
Here,
\begin{align*}
 \sum\iiiint=
 \sum_{n_1=1}^{N_1-1}\sum_{n_3=1}^{N_3-1}\sum_{n_2=1}^{N_2-1}
   \int_{k_{1,n_1}}^{k_{1,n_1+1}} dk_1
   \int_{k_{3,n_3}}^{k_{3,n_3+1}} dk_3
\bigg[
\int_{k_{2,n_2}}^{k_{2,n_2+1}} dk_2\int_{k_{2,n_2}}^{k_2} dk'_2
+\int_{k_{2,n_2}}^{k_{2,n_2+1}} dk'_2\int_{k_{2,n_2+1}}^{k_{2,M_2}} dk_2
\bigg],
\end{align*}
and $\Psi_{ij}$  denote the relevant functions.
The proof of \eqref{Intdbk_PsiijdWAsyij_GeneralCase} is outlined below.

In the derivations of Eqs.~\eqref{Intdbk_PsiijdWAsyij_GeneralCase},
the following operations are carried out,
together with Eqs.~\eqref{dWAsyij_Discrete}.
We start with the set of equalities,
\begin{subequations}
\label{GobalIntegrFormuProofStep1}
\begin{align}
\label{GobalIntegrFormuProofStep1a}
&
 \int_{-\infty}^{0}dk_1\int_{0}^{\infty}dk_3
=\sum_{n_1=1}^{N_1-1}\sum_{n_3=1}^{N_3-1}
   \int_{k_{1,n_1}}^{k_{1,n_1+1}} dk_1
   \s \int_{k_{3,n_3}}^{k_{3,n_3+1}} dk_3,
\quad
\int_{\mathbb{R}} dk_2
=\int_{k_{2,1}}^{k_{2,N_2}} dk_2
+\int_{k_{2,N_2}}^{k_{2,M_2}} dk_2,
\notag\\[-5pt]&
\\[-5pt]&
\int_{k_{2,1}}^{k_{2,N_2}} dk_2
=\sum_{m_2=1}^{N_2-1} \int_{k_{2,m_2}}^{k_{2,m_2+1}}dk_2,
\quad
\int_{k_{2,N_2}}^{k_{2,M_2}} dk_2
\int_{k_{2,1}}^{k_{2}} dk'_2
=
\int_{k_{2,N_2}}^{k_{2,M_2}} dk_2 
  \int_{k_{2,1}}^{k_{2,N_2}} dk'_2,
 \notag
\end{align}
\vspace{-1.5em}
\begin{align}
\label{GobalIntegrFormuProofStep1b}
\int_{k_{2,1}}^{k_{2,N_2}} dk'_2\, 
\CharacteristicFunction_{(k_{2,n_2},k_{2,n_2+1}]}(k'_2)
=\int_{k_{2,n_2}}^{k_{2,n_2+1}}dk'_2\,
       \CharacteristicFunction_{(k_{2,n_2},k_{2,n_2+1}]}(k'_2), 
\end{align}
\vspace{-1.5em}
\begin{align}
\label{GobalIntegrFormuProofStep1c}
&
\sum_{m_2=1}^{N_2-1} 
   \int_{k_{2,m_2}}^{k_{2,m_2+1}}dk_2
    \int_{k_{2,1}}^{k_{2}} dk'_2
     \sum_{n_2=1}^{N_2-1}\CharacteristicFunction_{(k_{2,n_2},k_{2,n_2+1}]}(k'_2)
           \Phi(n_2\ldots)
\notag\\&
=
  \sum_{m_2=1}^{N_2-1}\int_{k_{2,m_2}}^{k_{2,m_2+1}}dk_2
    \sum_{n_2=1}^{m_2}
     \int_{k_{2,n_2}}^{k_{2}} dk'_2\,\CharacteristicFunction_{(k_{2,n_2},k_{2,n_2+1}]}(k'_2)
                  \Phi(n_2\ldots)
\notag\\&
=\sum_{m_2=1}^{N_2-1}
\int_{k_{2,m_2}}^{k_{2,m_2+1}}dk_2
\int_{k_{2,m_2}}^{k_2} dk'_2\,
      \CharacteristicFunction_{(k_{2,m_2},k_{2,m_2+1}]}(k'_2)
                 \Phi(m_2\ldots)
\notag\\&\quad
+\sum_{m_2=2}^{N_2-1}
\int_{k_{2,m_2}}^{k_{2,m_2+1}}dk_2
\sum_{n_2=1}^{m_2-1}
\bigg(\int_{k_{2,n_2}}^{k_{2,m_2}} dk'_2+\int_{k_{2,m_2}}^{k_2} dk'_2\bigg)
      \CharacteristicFunction_{(k_{2,n_2},k_{2,n_2+1}]}(k'_2)
                 \Phi(n_2\ldots)
\notag\\&
=\sum_{m_2=1}^{N_2-1}\,
\int_{k_{2,m_2}}^{k_{2,m_2+1}}dk_2
\int_{k_{2,m_2}}^{k_2} dk'_2\,
      \CharacteristicFunction_{(k_{2,m_2},k_{2,m_2+1}]}(k'_2)
                  \Phi(m_2\ldots)
\notag\\&\quad
+\sum_{m_2=2}^{N_2-1}\,
\int_{k_{2,m_2}}^{k_{2,m_2+1}}dk_2
\sum_{n_2=1}^{m_2-1}
\int_{k_{2,n_2}}^{k_{2,n_2+1}} dk'_2
      \CharacteristicFunction_{(k_{2,n_2},k_{2,n_2+1}]}(k'_2)
                 \Phi(n_2\ldots).
\end{align}
\end{subequations}
Equalities~\eqref{GobalIntegrFormuProofStep1a}
and \eqref{GobalIntegrFormuProofStep1b} are self-explanatory, 
Eqs.~\eqref{GobalIntegrFormuProofStep1c}
may be verified directly with the use of properties
of characteristic functions.
Applying Eqs.~\eqref{GobalIntegrFormuProofStep1} to Eq.~\eqref{dWAsy22_Discrete}
step-by-step results in
\begin{align}
\label{GobalIntegrFormuProofOutcome1}
&
\int_{k_{1,n_1}}^{k_{1,n_1+1}}dk_1
\int_{k_{3,n_3}}^{k_{3,n_3+1}}dk_3
\int_{\mathbb{R}}dk_2\Psi_{22}(\bk,\br)\dWAsy_{22}(\bk)
\notag\\&
=\sum_{m_2=2}^{N_2-1}
\sum_{n_2=1}^{m_2-1} 
\int_{k_{1,n_1}}^{k_{1,n_1+1}}dk_1
\int_{k_{3,n_3}}^{k_{3,n_3+1}}dk_3
\int_{k_{2,m_2}}^{k_{2,m_2+1}}dk_2
 \int_{k_{2,n_2}}^{k_{2,n_2+1}} dk'_2\,
      \CharacteristicFunction_{(k_{2,n_2},k_{2,n_2+1}]}(k'_2)
    \frac{2\Psi_{22}(\bk,\br)\Expal(\bk;k'_2;\sigma)|\bk'|^4\big[n_2\cdots\big]}{|k_1||\bk|^4}
\notag\\&\quad
+\sum_{m_2=1}^{N_2-1}
\int_{k_{1,n_1}}^{k_{1,n_1+1}}dk_1
\int_{k_{3,n_3}}^{k_{3,n_3+1}}dk_3
\int_{k_{2,m_2}}^{k_{2,m_2+1}}dk_2
\int_{k_{2,m_2}}^{k_2} dk'_2\,
      \CharacteristicFunction_{(k_{2,m_2},k_{2,m_2+1}]}(k'_2)
           \frac{2\Psi_{22}(\bk,\br)\Expal(\bk;k'_2;\sigma)|\bk'|^4\big[m_2\cdots\big]}{|k_1||\bk|^4}
\notag\\&\quad
+\sum_{n_2=1}^{ N_2-1}
\int_{k_{1,n_1}}^{k_{1,n_1+1}}dk_1
\int_{k_{3,n_3}}^{k_{3,n_3+1}}dk_3
\int_{k_{2,N_2}}^{k_{2,M_2}}dk_2
 \int_{k_{2,n_2}}^{k_{2,n_2+1}} dk'_2\,
      \CharacteristicFunction_{(k_{2,n_2},k_{2,n_2+1}]}(k'_2)
        \frac{2\Psi_{22}(\bk,\br)\Expal(\bk;k'_2;\sigma)|\bk'|^4\big[n_2\cdots\big]}{|k_1||\bk|^4},
\end{align}
where $\big[n_2\cdots\big]$ denotes the big bracketed quantity in Eq.~\eqref{dWAsy22_Discrete}.

With regard to Eq.~\eqref{GobalIntegrFormuProofOutcome1},
the characteristic functions are redundant and may be removed.
For the first term on its right-hand side,
we have
\begin{align}
\label{GobalIntegrFormuProofStep2}
&
\sum_{m_2=2}^{N_2-1}
\sum_{n_2=1}^{m_2-1} \s
\int_{k_{1,n_1}}^{k_{1,n_1+1}}dk_1\s
\int_{k_{3,n_3}}^{k_{3,n_3+1}}dk_3\s
\int_{k_{2,m_2}}^{k_{2,m_2+1}}dk_2
\int_{k_{2,n_2}}^{k_{2,n_2+1}} dk'_2
\Phi(n_2\ldots)
\notag\\&
=\sum_{n_2=1}^{N_2-2} 
\sum_{m_2=n_2+1}^{N_2-1}\s
\int_{k_{1,n_1}}^{k_{1,n_1+1}}dk_1\s
\int_{k_{3,n_3}}^{k_{3,n_3+1}}dk_3\s
\int_{k_{2,m_2}}^{k_{2,m_2+1}}dk_2
\int_{k_{2,n_2}}^{k_{2,n_2+1}} dk'_2
\Phi(n_2\ldots).
\end{align}
This equality is proved via mathematical induction:
It is simple to verify that the relation holds for $N_2=3$; 
It is straight-forward to check that if the relation
holds for $N_2\geq 3$, it holds for $(N_2+1)$.
Further, the right-hand side of Eq.~\eqref{GobalIntegrFormuProofStep2}
is recast in the form of
\begin{align*}
\sum_{n_2=1}^{N_2-2}
\sum_{m_2=n_2+1}^{N_2-1}
\int_{k_{2,m_2}}^{k_{2,m_2+1}}dk_2
\int_{k_{2,n_2}}^{k_{2,n_2+1}} dk'_2
=
\sum_{n_2=1}^{N_2-2} 
\int_{k_{2,n_2+1}}^{k_{2,N_2}}dk_2
\int_{k_{2,n_2}}^{k_{2,n_2+1}} dk'_2
=
\sum_{n_2=1}^{N_2-1} 
\int_{k_{2,n_2+1}}^{k_{2,N_2}}dk_2
\int_{k_{2,n_2}}^{k_{2,n_2+1}} dk'_2.
\end{align*}
Applying the above results to Eq.~\eqref{GobalIntegrFormuProofOutcome1} 
yields Eq.~\eqref{Intdbk_Psi22dWAsy22_GeneralCase}.

The same procedure is used to prove  Eqs.~\eqref{Intdbk_Psi12dWAsy12_GeneralCase}
and \eqref{Intdbk_Psi11dWAsy11_GeneralCase}.

\section{\label{Appensec:DiscretizedModel}Discretized model}

The main parts of the  discretized second-order model 
and their derivations and productions
are described in this appendix.

Objective~\eqref{ObjectiveFunctionMaxIntWkkbrAsy}
 is represented 
 in terms of $\dContracted_{ij}$,
 with the help of 
 Eqs.~\eqref{tWkk_tW112212} and  \eqref{Intdbk_PsiijdWAsyij_GeneralCase}. 
Algorithms `{\small Cuhre}' and `{\small Divonne}'
of the open source software package {\small CUBA} library
 \cite{Hahn2005, Hahn2006, httpcuba} are used to compute
the four-dimensional integrals in Eqs.~\eqref{Intdbk_PsiijdWAsyij_GeneralCase}.

The discrete control variables at the nodes
 are directly constrained by 
  Eqs.~\eqref{CVgamma_TransformedAsy_SymmetryAboutk3}, \eqref{tocBounds},
  and  the support~\eqref{tCVgammaAsyij_SupportEstimate},
\begin{align}
\label{LocalLinearConstraintsCV}
&
\dContracted_{ij}(n_1,n_2,n_3)=0,\ \
n_1\in\{1,N_1\}\ \
\text{or}\ n_2\in\{1,N_2\}
\ \ 
\text{or}\ n_3=N_3,\ \ \forall i,\ j;
\notag\\&
\dContracted_{ij}(n_1,n_2,2)=\dContracted_{ij}(n_1,n_2,1),\ \ ij=11,21,12,22;
\quad
\dContracted_{ij}(n_1,n_2,1)=0,\ \ ij=31,32;
\\&
\big|\dContracted_{ij}(n_1,n_2,n_3)\big|\leq 1,\ \
 2\leq n_1\leq N_1-1,\ 
 2\leq n_2\leq N_2-1,\ 
 1\leq n_3\leq N_3-1,\ \ \forall i,\ j.
 \notag
\end{align}
Substituting Eqs.~\eqref{TriLinearApproximation} 
into Eqs.~\eqref{dCVgamma_Constraint_Integral_Asy}
 and integrating analytically the integrals over the elements
 in $\EstimatedSupportForgammaij(0)$ give 
  four global linear equality constraints for
  $\dContracted_{ij}$.

To implement the constraints of inequality~\eqref{ElementaryInequalitiesFortwiwj_Asy}, 
we select three types of collocation points in $\EstimatedSupportForWij(0)$,
according to the integral structures of the discretized
solutions~\eqref{dWAsyij_Discrete} with respect to $k'_2$.

\paragraph{Collocation points for constraints 1C}
The points are located in the middle of the element edges parallel to the $k_2$-axis,
\begin{align}
 k_1=k_{1,n_1},\ 
k_3=k_{3,n_3},\
k_2=(k_{2,n_2}+k_{2,n_2+1})/2.
\label{Collocation_1C}
\end{align}
Equations~\eqref{dWAsyij_Discrete} indicate
  that $\tWAsy_{ij}(\bk)$ 
  and Eqs.~\eqref{ElementaryInequalitiesFortwiwj_Asy}
 evaluated at these collocation points involve only 
\begin{align*}
 \dContracted_{ij}(n_1,n'_2,n_3),\  
n'_2\in\!\{1,\ldots, n_2+1\},
\end{align*}
like 1-column parallel to the $k_2$-axis. Accordingly,
the collocation points~\eqref{Collocation_1C} 
are denoted as {1C}. This {1C}
is also used to identify the collection of constraints \eqref{ElementaryInequalitiesFortwiwj_Asy}
evaluated at the {1C} collocation points; A similar practice is  adopted 
for the other two types, {2C}\,({2C$k_1$}, {2C$k_3$}) and {4C}, below.

\paragraph{Collocation points for constraints 2C}
The points are located in the center of the element facets parallel to the $k_2$-axis,
\begin{subequations}
\label{Collocation_2C}
\begin{align}
\label{Collocation_2Ck1}
 k_1=k_{1,n_1},\ \
k_3=(k_{3,n_3}+k_{3,n_3+1})/2,\ \ 
k_2=(k_{2,n_2}+k_{2,n_2+1})/2;
\end{align}
\vspace{-2.5em}
\begin{align}
\label{Collocation_2Ck3}
k_1=(k_{1,n_1}+k_{1,n_1+1})/2,\ \
 k_3=k_{3,n_3},\ \ 
k_2=(k_{2,n_2}+k_{2,n_2+1})/2.
\end{align}
\end{subequations}
Constraints~\eqref{ElementaryInequalitiesFortwiwj_Asy}
 evaluated at the collocation points \eqref{Collocation_2C} involve 
 either
\begin{align*}
 \dContracted_{ij}(n_1,n'_2,n'_3),\ 
 n'_3\in\!\{n_3,n_3+1\},\ 
n'_2\in\!\{1,\ldots, n_2+1\},
\end{align*}
or
\begin{align*}
 \dContracted_{ij}(n'_1,n'_2,n_3), \ 
 n'_1\in\!\{n_1, n_1+1\},\ 
 n'_2\in\!\{1,\ldots, n_2+1\},
\end{align*}
 like 2-columns parallel to the $k_2$-axis,  these collocation points 
 are denoted as {2C}. 
 Specifically, the set associated with
 Eqs~\eqref{Collocation_2Ck1} is denoted as {2C$k_1$}
 and  the set associated with 
 Eqs.~\eqref{Collocation_2Ck3} is denoted as {2C$k_3$}.

\paragraph{Collocation points for constraints 4C}
The points are located in the center of the elements,
\begin{align}
\label{Collocation_4C}
k_1=(k_{1,n_1}+k_{1,n_1+1})/2, \ \ 
k_3=(k_{3,n_3}+k_{3,n_3+1})/2,\ \ 
k_2=(k_{2,n_2}+k_{2,n_2+1})/2.
\end{align}
Constraints~\eqref{ElementaryInequalitiesFortwiwj_Asy}
evaluated at these collocation points involve
\begin{align*}
  \dContracted_{ij}(n'_1,n'_2,n'_3), \ &
n'_1\in\!\{n_1, n_1+1\},\  n'_3\in\!\{n_3,n_3+1\},\ 
 n'_2\in\!\{1,\ldots, n_2+1\},
\end{align*}
 like 4-columns parallel to the $k_2$-axis. Accordingly
 the collocation points~\eqref{Collocation_4C} are referred to as {4C}.

The evaluation of Eqs.~\eqref{dWAsyij_Discrete}
 at collocation points {1C}, {2C}, and {4C}
 results in the need to compute 
 a collection of  1-dimensional integrals with respect to $k'_2$ over the
 mesh elements.
Function `{\small NIntegrate}' of the commercial software {\small MATHEMATICA}
is used to compute these integrals 
(with the default ``Global Adaptive Strategy" and global error tolerance).

\section{\label{Appensec:Algorithm}Algorithm for parallel computing}

This appendix outlines the major procedural steps
in the solution of the discretized second-order model, 
including  details essential
for the development of the ADMM consensus algorithm
and the use of the open-source SCS solver.

In Appendix~\ref{Appensec:DiscretizedModel},
the discretized model is presented in its natural form
in terms of $\dContracted_{ij}$.
For convenience to employ the ideas
developed and the notations commonly adopted
in ADMM algorithm \cite{Boydetal2010, ParikhBoyd2013},
   the above discretized model is recast in the ADMM form,
\begin{align}
\label{SOM_Original}
&
\minimize\ \ \  a^T Z
\notag\\[-6.5pt]&
\\[-6.5pt]&
\subjectto\ \ Z\in\cC.
\notag
\end{align}
Here, 
$\bZ$ denotes the vector associated with
\begin{align*}
&
\Big\{\dContracted_{ij}(n_1,n_2,n_3):\ 
            n_1=1,\ldots, N_1,\
  n_2=1,\ldots, N_2, \
     n_3=1,\ldots, N_3;\ i, j\Big\}\subset\bbR^{6 \NCV},
     \ \ \NCV=N_1 N_2  N_3,
\end{align*}
through the one-to-one linear  mapping
 specified by the  pseudocode,
\begin{align}
\label{ADMMAlgorithmSubSOcpXvsCVs}
&
\text{$n=0$} 
\notag\\&
\text{FOR $n_1$ $=$ 1 to $N_1$}
\notag\\&
\text{\hskip10mm      FOR $n_2$ $=$ 1 to $N_2$}
\notag\\&
\text{\hskip20mm           FOR $n_3$ $=$ 1 to $N_3$}
\notag\\&
\text{\hskip30mm            $n= n+1$}
\notag\\&
\text{\hskip30mm            $Z[n] =\dContracted_{11}(n_1,n_2,n_3)$}
\notag\\&
\text{\hskip20mm           END FOR}
\notag\\&
\text{\hskip10mm      END FOR}
\notag\\&
\text{END FOR}
\notag\\&
\text{    \hskip10mm  $\vdots$  \{21,31,12,22\} }
\notag\\&
\text{FOR $n_1$ $=$ 1 to $N_1$}
\notag\\&
\text{\hskip10mm      FOR $n_2$ $=$ 1 to $N_2$}
\notag\\&
\text{\hskip20mm           FOR $n_3$ $=$ 1 to $N_3$}
\notag\\&
\text{\hskip30mm            $n=n+1$}
\notag\\&
\text{\hskip30mm            $Z[n] =  \dContracted_{32}(n_1,n_2,n_3)$}
\notag\\&
\text{\hskip20mm           END FOR}
\notag\\&
\text{\hskip10mm      END FOR}
\notag\\&
\text{END FOR}
\end{align}
The quantity, $-a^T Z$, represents
the discretized  objective functional~\eqref{ObjectiveFunctionMaxIntWkkbrAsy}
 to be tested,
\begin{align}
-a^T\bZ
=32
   \int_{-\infty}^{0}\s dk_1
    \int_{0}^{+\infty} dk_3 \int_{-\infty}^{+\infty} dk_2\,
    \tWAsy_{kk}(\bk)
    \frac{\sin(L_1\,k_1) \sin(L_2\,k_2) \sin(L_3\,k_3)}{k_1 k_2 k_3},
\label{SOM_EqnsForab}
\end{align}
whose right-hand side is evaluated with the help of Eqs.~\eqref{tWkk_tW112212} 
and  \eqref{Intdbk_PsiijdWAsyij_GeneralCase}.
Equality~\eqref{SOM_EqnsForab}
is used to determine 
the coefficient vector $a$ with the aid of
      Eq.~\eqref{ADMMAlgorithmSubSOcpXvsCVs}.
The set of constraints $\cC$ in 
Eqs.~\eqref{SOM_Original} consists of
   the local linear constraints~\eqref{LocalLinearConstraintsCV}, 
the four global linear constraints of equality,
and the constraints of {1C} or \{1C, 2C, 4C\} or some other choices,
as discussed in Appendix~\ref{Appensec:DiscretizedModel}.

Next, for parallelization,
 the global consensus problem formulation \cite{Boydetal2010, ParikhBoyd2013}
 is employed to write Eqs.~\eqref{SOM_Original} in an equivalent form,
\begin{align}
\label{SOM_GlobalConsensus}
&
 \minimize\ \ \, \sum_{i=1}^N f_i(X_i) 
\notag\\[-13pt]&
\\[-2pt]&
\subjectto\ \ X_i-Z=0,\ \ i=1,\ldots,N.
\notag
\end{align}
$X_i\in\bbR^{6 \NCV}$, $i=1,\ldots,N$,
denote the local variables
associated, respectively, with the $N$ computing processing elements (or processes),
  and $Z$ is the common global variable.
The function $f_i$ is defined as
\begin{align}
\label{SOM_ithFunction}
f_i(X_i)=a^TX_i+\Indicator_{\cC_i}(X_i).
\end{align}
$\Indicator_{\cC_i}$ is the indicator function of the convex set $\cC_i$;
the collection $\{\cC_i: i=1,\ldots, N\}$ is a partition of $\cC$.
The size of each $\cC_i$ is controlled by the computer memory available
for the generation of standard conic form.

To apply ADMM to solve Eqs.~\eqref{SOM_GlobalConsensus} iteratively,
 the augmented Lagrangian \cite{Boydetal2010, ParikhBoyd2013} is employed,
\begin{align}
 L_{\rho}(X,Z,Y)
=\sum_{i=1}^N \big[
f_i(X_i)+(Y_i)^T(X_i-Z)
  +(\rho/2)||X_i-Z||_2^2\big],
\end{align}
where
\begin{align*}
X=(X_1,\ldots,X_N)\in \big(\bbR^{6 \NCV}\big)^N,\ \ \
Y=(Y_1,\ldots, Y_N)\in \big(\bbR^{6 \NCV}\big)^N,
\end{align*}
$Y_i$ is the dual variable associate with the consensus equality constraint
$X_i-Z=0$ in  Eqs.~\eqref{SOM_GlobalConsensus},
 and $\rho \, (>0)$ is the penalty parameter.
The ADMM algorithm
  consists of the following iterations \cite{Boydetal2010, ParikhBoyd2013},
\begin{align}
\label{SOM_ADMMAlgorithm}
 &
 X^{k+1}:=\underset{X}{\argmin}\, L_{\rho}(X,Z^k,Y^k),
 \quad
 Z^{k+1}:=\underset{Z}{\argmin}\, L_{\rho}(X^{k+1},Z,Y^k),
 \notag\\[-6pt]&
 \\[-6pt]&
Y^{k+1}:=Y^k+\rho\big[X^{k+1}-\big(Z^{k+1},\ldots,Z^{k+1}\big)\big],
\ \ 
k=0,1,\ldots,
\notag
\end{align}
which may be manipulated in a straight-forward fashion \cite{Boydetal2010}
to obtain the ADMM consensus algorithm in the numerical simulation,
\begin{subequations}
\label{SOM_ADMMAlgorithmProjection}
\begin{align}
 Z^{k}=\overX^{k}:=\frac{1}{N}\sum_{i=1}^N X_i^k,
\label{SOM_ADMMAlgorithmProjectionZ}
\end{align}
\vspace{-1.5em}
\begin{align}
 X_i^{k+1}:=\underset{X_i\in\cC_i}{\argmin}\,
             \big|\big|X_i-\Lik\big|\big|_2^2
           =\Projection_{\cC_i}\big(\Lik\big),
\label{SOM_ADMMAlgorithmProjection2}
\end{align}
\vspace{-2.0em}
\begin{align}
\tilde{Y}_i^{k+1}:=\tilde{Y}_i^k+X_i^{k+1}-\overX^{k+1},\quad
\frac{1}{N}\sum_{i=1}^N \tilde{Y}_i^0=0.
\label{SOM_ADMMAlgorithmProjectionY}
\end{align}
\end{subequations}
Here,
\begin{align}
 \Lik=\overX^k-\big(\tilde{a}+\tilde{Y}_i^k\big),
 \label{LikDef}
\end{align}
and $\Projection_{\cC_i}\big(\Lik\big)$ denotes Euclidean projection of vector $\Lik$
onto $\cC_i$, $i=1,\ldots,N$.
In the above derivation, we have used Eq.~\eqref{SOM_ithFunction} and the scaled,
\begin{align*}
Y_i^k=\rho\,\tilde{Y}_i^k,\quad 
 a=\rho\,\tilde{a}.
\end{align*}
The key part of the consensus algorithm is 
 the $X_i$-update~\eqref{SOM_ADMMAlgorithmProjection2}
whose computation is discussed  below.

The convergence of the iterative process~\eqref{SOM_ADMMAlgorithmProjection}
is guaranteed analytically
 \cite{Boydetal2010, EcksteinBertsekas1992}.
To terminate the iterative process,
 the conventional stopping criteria \cite{Boydetal2010} are adopted.
First, the primal and dual residuals, $R^k$ and $S^k$, and their squared norms
at the $k$-th iteration are calculated through
\begin{align}
\label{PriDualResidues}
&
 R^k=\big(X_1^k-\overX^k,\ldots, X_N^k-\overX^k\big),\ \ \
 S^k=-\rho \big(\overX^k-\overX^{k-1},\ldots, \overX^k-\overX^{k-1}\big);
\notag\\[-4pt]&
\\[-12pt]&
\big|\big| R^k\big|\big|_2^2=\sum_{i=1}^N \big|\big|X_i^k-\overX^k\big|\big|_2^2,\ \ \
\big|\big| S^k\big|\big|_2^2=N \rho^2 \big|\big|\overX^k-\overX^{k-1}\big|\big|_2^2.
\notag
\end{align}
These residuals converge to zero as the iterative process proceeds,
according to the convergence theory \cite{Boydetal2010}.
The stopping criteria are 
\begin{align}
\label{StoppingCriteria}
 \big|\big| R^k\big|\big|_2\leq \epsilon^{\text{pri}},\quad
\big|\big| S^k\big|\big|_2\leq \epsilon^{\text{dual}},
\end{align}
where $\epsilon^{\text{pri}}$ and $\epsilon^{\text{dual}}$ denote, respectively, the feasibility
tolerances for the primal and dual feasibility conditions \cite{Boydetal2010}.
Further, the tolerances are expressed in terms of 
the absolute and relative criteria,
$\epsilon^{\text{abs}}$ and $\epsilon^{\text{rel}}$ \cite{Boydetal2010},
\begin{align}
 \epsilon^{\text{pri}}
=\sqrt{N}\big( \epsilon^{\text{abs}}
              +\big|\big|\overX^k\big|\big|_2\epsilon^{\text{rel}}\big),\quad
\epsilon^{\text{dual}}
=\rho \sqrt{N} \big( \epsilon^{\text{abs}}
                    +\big|\big|\overX^k\big|\big|_2\epsilon^{\text{rel}}\big).
\end{align}
The values of $\rho=1$, $\epsilon^{\text{abs}}=10^{-4}$, and 
$\epsilon^{\text{rel}}=10^{-3}$ are employed and are adequate for the present
exploration of the second-order model.

The open source conic solver, {\small SCS in C} \cite{ODonoghueetal2016},
is used to compute the $X_i$-update~\eqref{SOM_ADMMAlgorithmProjection2}
in the $i$-th processing element. 
The solver employs a first-order method to solve large convex cone programs,
suitable to the large-scale nature of the present problem.
Because the solver accepts only standard forms,
 we resort to the open source package 
 {\small CVXPY} \cite{DiamondBoyd2016}
   with its default solver {\small SCS} (denoted as 
   {\small CVXPY/SCS} hereafter) to generate
   such forms.
To this end,
 the  $X_i$-update~\eqref{SOM_ADMMAlgorithmProjection2}
is first cast in the equivalent ADMM form,
\begin{align}
\label{SOM_ADMMAlgorithmSubSOCPNaturalForm}
&
\minimize\ \ \ 
             \big|\big|X_i-\Lik\big|\big|_2^2
\notag\\[-6pt]&
\\[-6pt]&
 \subjectto\ \  X_i \in\cC_i.
\notag
\end{align}
This subproblem has an equivalent representation
in the natural form in terms of $\dContracted_{ij}$ as presented 
in Appendix~\ref{Appensec:DiscretizedModel}.
Since this natural form is much easier to code and acceptable to 
{\small CVXPY}, 
{\small CVXPY/SCS} is applied to this natural form, 
while $\cC_i$ is coded 
according to the disciplined convex programming (DCP)
  ruleset \cite{GrantBoyd2014} as required.
The standard form associated with
Eqs.~\eqref{SOM_ADMMAlgorithmSubSOCPNaturalForm} is then obtained,
\begin{align}
\label{SOM_ADMMAlgorithmSubStandardForm}
&
\minimize\ \ \ \,
             c_i^T x_i
\notag\\[-6pt]&
\\[-6pt]&
\subjectto\ \ \, A_ix_i + s_i = b_i^k,\ \ s_i \in\cK_i.
\notag
\end{align}
Here, $\cK_i$ denotes the convex cone composed of $\cC_i$, 
described by a group of dimension parameters and dimension array
(defined in \cite{ODonoghueetal2016} and {\small README.md} of 
        {\small SCS} package  \cite{ODonoghueetalPackage2016}).
Independent of $\Lik$,
 vector $c_i$, matrix $A_i$, 
  and cone $\cK_i$ are fixed completely by $\cC_i$,
 i.e., they remain constant for all the iterations $k$.
The value of vector $b_i^k$ depends on both $\cC_i$ and $\Lik$,
 the dependence on the latter is characterized by the (underneath stacked) pattern of
\begin{align}
\label{bLinkedToLambdaik}
&
 b_i^k
=
 \Big(
  \ldots,\underset{\text{$3\NCV$ components}}{\underbrace{1,1,-2\Lik[1],\ldots, 1,1,-2\Lik[\NCV]}},
  \underset{\text{$3\NCV$ components}}{\underbrace{1,1,-2\Lik[\NCV+1],\ldots, 1,1,-2\Lik[2\NCV]}},
\notag\\&\hskip12mm
\underset{\text{$3\s\times\s 3\NCV$ components}}{\underbrace{\ldots}},
  \underset{\text{$3\NCV$ components}}{\underbrace{1,1,-2\Lik[5\NCV+1],\ldots, 1,1,-2\Lik[6\NCV]}}, \ldots
\Big),
\end{align}
while the components outside the (underneath stacked) pattern are fixed by  $\cC_i$ itself.
It then follows from Eqs.~\eqref{LikDef} and \eqref{bLinkedToLambdaik}
that $b_i^k$ evolves as iteration $k$ proceeds and
its update is  straight-forward.

The above scheme of standard form generations leads to
 a mapping between $x_i$ of the standard form 
 and $\dContracted_{ij}$ of the natural form
 specified by the  pseudocode,
\begin{align}
\label{SOM_ADMMAlgorithmSubSOCPxvsCVs}
&
\text{$n=0$} 
\notag\\&
\text{FOR $n_1$ $=$ 1 to $N_1$}
\notag\\&
\text{\hskip10mm      FOR $n_3$ $=$ 1 to $N_3$}
\notag\\&
\text{\hskip20mm           FOR $n_2$ $=$ 1 to $N_2$}
\notag\\&
\text{\hskip30mm            $n = n+1$}
\notag\\&
\text{\hskip30mm             $\dContracted_{11}(n_1,n_2,n_3) = x_i[n]$}
\notag\\&
\text{\hskip20mm           END FOR}
\notag\\&
\text{\hskip10mm      END FOR}
\notag\\&
\text{    \hskip10mm  $\vdots$  \{21,31,12,22\} }
\notag\\&
\text{\hskip10mm      FOR $n_3$ $=$ 1 to $N_3$}
\notag\\&
\text{\hskip20mm           FOR $n_2$ $=$ 1 to $N_2$}
\notag\\&
\text{\hskip30mm            $n = n+1$}
\notag\\&
\text{\hskip30mm             $\dContracted_{32}(n_1,n_2,n_3)= x_i[n]$}
\notag\\&
\text{\hskip20mm           END FOR}
\notag\\&
\text{\hskip10mm      END FOR}
\notag\\&
\text{END FOR}
\end{align}
The relationship between $x_i$ 
of the standard form~\eqref{SOM_ADMMAlgorithmSubStandardForm}
and $X_i$ of the ADMM form~\eqref{SOM_ADMMAlgorithmSubSOCPNaturalForm}
is determined through the composition of mappings~\eqref{ADMMAlgorithmSubSOcpXvsCVs}
and \eqref{SOM_ADMMAlgorithmSubSOCPxvsCVs},
\begin{align}
 \label{MappingFromxiToXi}
x_i
\overset{\text{Eq.~\eqref{SOM_ADMMAlgorithmSubSOCPxvsCVs}}}{\rightarrow}
\Big\{\dContracted_{ij}(n_1,n_2,n_3)\Big\}
\overset{\text{Eq.~\eqref{ADMMAlgorithmSubSOcpXvsCVs}}}{\rightarrow}
Z=X_i.
\end{align}

To implement the ADMM consensus scheme~\eqref{SOM_ADMMAlgorithmProjection},
 the constraint set $\cC$ of Eqs.~\eqref{SOM_Original}
 in its natural form
 is first partitioned  into
$N$ subsets $\cC_i$, $i=1,\ldots,N$; the value of $N$
is fixed on the basis that the computer memory required
to generate the standard form~\eqref{SOM_ADMMAlgorithmSubStandardForm}
be less than the $64$GB available.
Then, {\small CVXPY/SCS} is applied to the natural form 
 associated with Eqs.~\eqref{SOM_ADMMAlgorithmSubSOCPNaturalForm}
 to determine both 
 $\{c_i, A_i, \cK_i\}$ and the pattern of $b_i^k$ that specify the standard form.

The solver, {\small SCS}, solves Eqs.~\eqref{SOM_ADMMAlgorithmSubStandardForm}
and returns the optimal 
solution ($x_i^*, y_i^*, s_i^*$)
where $x_i^*$ is the solution of 
the primal problem~\eqref{SOM_ADMMAlgorithmSubStandardForm}
 and $y_i^*$ is the solution of its dual problem.
 The parameters controlling the computational procedure are taken
 from the examples of  \cite{ODonoghueetalPackage2016}.

Suppose that all the quantities, 
\begin{align*}
 \big\{X_i^k,\ \tilde{Y}_i^k:\ i=1,\ldots,N\big\},
\end{align*}
contained in Eqs.~\eqref{SOM_ADMMAlgorithmProjection} 
are known at the $k$-th iteration,
then  
\begin{align*}
 \big\{X_i^{k+1},\, \tilde{Y}_i^{k+1}:\ i=1,\ldots,N\big\}
\end{align*}
 are determined in parallel as follows:
\begin{enumerate}
\item[1:]
Compute $\Lik$ from Eq.~\eqref{LikDef}
and $b_i^k$ from Eq.~\eqref{bLinkedToLambdaik} for all $i$;

\item[2:]
Solve Eqs.~\eqref{SOM_ADMMAlgorithmSubStandardForm} with 
{\small SCS} to obtain optimal solutions 
 $\big(x_i^{\ast (k+1)}, y_i^{\ast (k+1)}, s_i^{\ast (k+1)}\big)$
for all $i$, 
with the previously obtained
  $\big(x_i^{\ast (k)}, y_i^{\ast (k)}, s_i^{\ast (k)}\big)$ to warm start the solver.

\item[3:]
Compute $\big\{X_i^{k+1}$, $\overX^{k+1}$, $\tilde{Y}_i^{k+1}\big\}$
with the help of the optimal solutions obtained in Step 2 and
Eqs.~\eqref{MappingFromxiToXi},
\eqref{SOM_ADMMAlgorithmProjectionZ},
and \eqref{SOM_ADMMAlgorithmProjectionY}.

\item[4:]
Check the stopping criteria~\eqref{StoppingCriteria}; 
If not satisfied, go to Step 1.

\end{enumerate}

Following conventional practice, we usually start with the initial iteration condition,
\begin{align}
\label{InitialIterationCondition}
&
 X_i^{0}=Y_i^{0}=0,\ \
 x_i^{\ast (0)}=0,\  \ 
 y_i^{\ast (0)}=s_i^{\ast (0)}=0,\  \ 
 i=1,\ldots,N.
 \notag
\end{align}
In  case it is required to improve a solution obtained,
\begin{align*}
 \{X_i, Y_i, x_i, y_i, s_i: i=1,\ldots,N\},
\end{align*}
or to test whether such a solution may be further improved 
or to study how it varies under the modification of
some parameter values, the available solution is  taken as the initial
iteration condition to warm start the iteration process.

We code the idea and procedure outlined above in {\small C}
by modifying the {\small MPI} code of \cite{Boydetal2011}
and the examples of  \cite{ODonoghueetalPackage2016}
and we implement the parallel computing with 
{\small MPICH} \cite{AmerMPICH2015}.

\section{\label{Appensec:Turbulentenergyasobjectivefunction}Turbulent energy as an objective}

In the case of the second-order model in the asymptotic state with $\sigma=0$,
we have tested the possibility of $\WAsy_{kk}(\mathbf{0})$ 
as an objective functional which is maximized.
Its numerical and computational implementation is 
the same as that of $\int_{\mathbb{R}^3} d\br\, \WAsy_{kk}(\br)$,
the only modification is the replacement of 
$\int_{\mathbb{R}^3} d\br\, \WAsy_{kk}(\br)$
with $\WAsy_{kk}(\mathbf{0})$.

Table~\ref{OSmaxUkk0Data} lists
the predicted  nontrivial components of 
the Reynolds stress tensor $\tau^{(\infty)}_{ij}=\WAsy_{ij}(\mathbf{0})$ and
the predicted viscous dissipation
$\epsilon^{(\infty)}=\overline{w_{j,k}\,w_{j,k}}^{(\infty)}(\bo)$
under various combinations of constraints and mesh sizes.
The meanings of the elements in the column `Constraints' 
are the same as those of Table~\ref{OSDataVsExperimental}.
\begin{table}
\caption{The predicted nontrivial components of Reynolds stress tensor and viscous dissipation
  under $\max\WAsy_{kk}(\mathbf{0})$.}
\label{OSmaxUkk0Data}
\begin{tabular}{ll}
\hline\hline
Constraints &$\big(\tau^{(\infty)}_{11}$,  $\tau^{(\infty)}_{22}$,\ \, $\tau^{(\infty)}_{33}$,\ \
        \s $\tau^{(\infty)}_{12},\ \ \epsilon^{(\infty)}\big)$\\[2pt]
\hline
L$\,\&\,$1C\,(I) &($3.519, 1.091, 1.015, -1.559, 1.560$)\\
L$\,\&\,\text{1C}\,\&\,\text{4C}$\,(I) & ($3.476, 1.070, 1.014, -1.539, 1.539$)\\
L$\,\&\,\text{1C}\,\&\,\text{2C}\,\&\,\text{4C}$\,(I) &($3.469, 1.069,  1.014,  -1.536, 1.537$)\\
\hline
L$\,\&\,\text{1C}$\,(II) &($4.625, 1.160,  1.155,  -1.768, 1.769$)\\
\hline\hline
\end{tabular}
\end{table}
The adequacy of the solutions
is partially indicated by separate computations of 
$\tau^{(\infty)}_{12}$ and $\epsilon^{(\infty)}$
and the ratio,
% % \vspace{-0.5em}
\begin{align*}
 -\tau^{(\infty)}_{12}/\epsilon^{(\infty)}\in[0.9994,0.9996],
\end{align*}
consistent with the exact value of $1$ dictated 
by Eq.~\eqref{IntrinsicRelationForUkk(0)Asy} under $\sigma=0$.

Though the predicted values of $\WAsy_{11}(\mathbf{0})$ change significantly 
when the mesh size is reduced from $0.1$ of (I) to $0.05$ of (II),
all these discretized model versions 
produce the same numerical order pattern,
$\WAsy_{11}(\mathbf{0})>\WAsy_{22}(\mathbf{0})>\WAsy_{33}(\mathbf{0})$
\big(with $\WAsy_{33}(\mathbf{0})$ very close to $\WAsy_{22}(\mathbf{0})$\big);
this pattern is incompatible with the experimental result that  $\WAsy_{33}(\mathbf{0})$
is significantly higher than $\WAsy_{22}(\mathbf{0})$,
(see Section 4.5 of \cite{Piquet1999} 
or infer the relative values from the experimental values presented
in Table~\ref{OSDataVsExperimental}).

\bibliography{HSTasSOCParXivv6.bib}

\end{document}